\documentclass[%
reprint,
amsmath,amssymb,
aps,
pra,
floatfix,
longbibliography,
]{revtex4-2}

\usepackage{graphicx}
\usepackage{dcolumn}
\usepackage{bm}

\usepackage{array}
\usepackage{diagbox}
\usepackage[percent]{overpic}
\usepackage{xcolor}
\usepackage{makecell}

\usepackage[
unicode=true,
bookmarks=true,
bookmarksnumbered=true,
bookmarksopen=true,
bookmarksopenlevel=2
]{hyperref}
\usepackage{bookmark}

\begin{document}
	\raggedbottom
	\setlength{\parskip}{0pt}
	\preprint{APS/123-QED}
	
	\title{Vector Akhmediev Breathers
		and State Transitions in the Degenerate and Nondegenerate Regimes for the
		Coupled Sasa--Satsuma  System}
	
	\author{Mengyuan Lan}
	\affiliation{%
		School of Mathematics and Physics, North China Electric Power University,
		Beijing 102206, China
	}
	
	\author{Lei Wang}
	\email{Corresponding author: 50901924@ncepu.edu.cn}
	\affiliation{%
		School of Mathematics and Physics, North China Electric Power University,
		Beijing 102206, China
	}
	
	\author{Yinchuan Zhao}
	\affiliation{%
		School of Mathematics and Physics, North China Electric Power University,
		Beijing 102206, China
	}
	
\begin{abstract}
		We investigate vector Akhmediev breathers (ABs) in the degenerate and nondegenerate regimes for the
		coupled Sasa--Satsuma  system describing two coupled ultrashort
		optical pulse envelopes. Based on the carrier-wavenumber relations
of two wave-component backgrounds, we examine three configurations: a single vanishing wavenumber,
a pair of opposite wavenumbers, and two nonzero wavenumbers of unequal magnitudes. These
configurations admit distinct maximal degrees of nondegeneracy. For each configuration, we derive
the reduced spectral equations, construct the existence diagrams, and compare the resulting branches
with the modulation-instability gain spectra. Owing to the distinct carrier-wavenumber symmetries,
nondegenerate-type modes can occur in spectrally degenerate regimes, whereas degenerate-type ones may also
appear in spectrally nondegenerate regimes. Moreover, all admissible branches in the degenerate regimes obey the state-transition condition and correspond to periodic-wave states, while such branches also exist in nondegenerate regimes. Direct numerical simulations
of representative solutions further confirm the analytical predictions.
	\end{abstract}
	\maketitle
	
\section{INTRODUCTION}

Breathers constitute a broad class of nonlinear coherent structures
supported by finite-amplitude continuous-wave backgrounds
\cite{AkhmedievAnkiewicz1997,
	DudleyDiasErkintaloGenty2014,
	ChabchoubKiblerDudleyAkhmediev2014}.
For the scalar nonlinear Schr\"odinger (NLS) equation, the fundamental
breather solutions include Akhmediev breathers (ABs), Kuznetsov--Ma
solitons (KMs), Tajiri--Watanabe (TW) breathers, and Peregrine rogue
waves
\cite{AkhmedievEleonskiiKulagin1987,
	AkhmedievKorneev1986,
	Kuznetsov1977,
	Ma1979,
	TajiriWatanabe1998,
	Peregrine1983,
	KedzioraAnkiewiczAkhmediev2012}.
Among these solutions, the AB is spatially periodic
and temporally localized
\cite{AkhmedievKorneev1986,AkhmedievEleonskiiKulagin1987}.
It provides an exact nonlinear description of modulation instability
(MI), in which a weak periodic perturbation on a continuous-wave
background grows to its maximum amplitude and subsequently decays toward
the background
\cite{DudleyGentyDiasKiblerAkhmediev2009,
	HammaniWetzelKiblerFatomeFinotMillotAkhmedievDudley2011}.
The associated spectral expansion and contraction are closely related
to Fermi--Pasta--Ulam recurrence and have been observed experimentally
in optical fibers
\cite{VanSimaeysEmplitHaelterman2001,
	HammaniWetzelKiblerFatomeFinotMillotAkhmedievDudley2011}.
	
As the integrable two-component extension of the NLS equation,
the Manakov system serves as a fundamental model for coupled nonlinear
waves in various physical settings, including multi-component
Bose--Einstein condensates, two orthogonally polarized modes in optical
fibers, and interacting waves in hydrodynamic systems
\cite{Manakov1974,
	KevrekidisFrantzeskakisCarreteroGonzalez2008,
	Agrawal2013,
	OnoratoOsborneSerio2006}. The coupling between the components introduces
additional degrees of freedom and allows vector localized waves to
display intensity and polarization structures without the scalar setting. Representative studies have shown the vector breathers, vector rogue
waves and established a close connection between vector rogue-wave
formation and baseband MI
\cite{BaronioDegasperisConfortiWabnitz2012,
	BaronioConfortiDegasperisLombardoOnoratoWabnitz2014,
	BaronioChenGreluWabnitzConforti2015,GelashRaskovalov2023}, to name a few.
Polarization MI in the Manakov system has also been
observed experimentally in optical fibers
\cite{FrisquetKiblerFatomeMorinBaronioConfortiMillotWabnitz2015}.
These results indicate that vector models may support richer
dynamical behaviors than their scalar counterparts.

Recently, Liu~\textit{et al.} investigated the nondegenerate ABs in the focusing Manakov system and related their existence to the multiple eigenvalue branches associated with MI \cite{LiuChenYaoAkhmediev2022}. For the same background and modulation frequency, different eigenvalue branches are responsible for distinct vector ABs with the same growth rate but different spatiotemporal profiles. Physically, these solutions can exhibit contrasting bright--dark intensity modes in the two components, together with tilted propagation directions. Moreover, a single-frequency modulation can simultaneously excite different nondegenerate AB branches, leading to more complex growth--return dynamics \cite{LiuChenYaoAkhmediev2022}. These results were subsequently extended to the nondegenerate KMs and higher-order dynamics involving multiple AB branches \cite{CheChenLiuZhaoAkhmediev2022, ChenLiuAkhmediev2023}. In addition, the recurrence and spectral evolution of two-component ABs, including the nondegenerate structures, have also been observed experimentally in optical fibers \cite{LiuLiChenYaoYangAkhmediev2025}.

Higher-order effects become important for femtosecond optical pulses, for which third-order dispersion and nonlinear derivative corrections can no longer be neglected. For example, the coupled Hirota (CH) system incorporates these higher-order effects into the Manakov system~\cite{TasgalPotasek1992}. It was shown that these additional terms give rise to a state-transition mechanism absent in the Manakov system \cite{PanWangLiuSunRen2024}. Under the corresponding parameter conditions, various nondegenerate localized waves are transformed into soliton or periodic-wave states. This is regarded as the state transition of nonlinear waves. These behaviors occur only in the nondegenerate regime, whereas no corresponding transition is found in the degenerate regime of the CH system. Such result suggests that higher-order effects can modify wave states in the nondegenerate regime and introduce dynamical mechanisms beyond those of the Manakov system.

The coupled Sasa--Satsuma (CSS) system is another integrable higher-order extension of the Manakov system~\cite{PorsezianSundaramMahalingam1994}. It incorporates a different combination of third-order dispersion and nonlinear derivative terms and describes the coupled propagation of two ultrashort optical pulse envelopes, such as two polarization components in nonlinear optical fibers. These higher-order effects provide a natural setting for examining how the dynamics of vector localized waves differ from those in the Manakov and CH systems. Such system takes the form
\begin{equation}
	\begin{aligned}
		\mathrm{i}q_{j,z}
		+\frac{1}{2}q_{j,\tau\tau}
		+S q_j
		;&+\mathrm{i}\epsilon
		\left(
		q_{j,\tau\tau\tau}
		+6S q_{j,\tau}
		+3q_jS_{\tau}
		\right)
		=0,
	\end{aligned}
	\label{eq:css-system}
\end{equation}
for $j=1,2$, where $S=|q_1|^2+|q_2|^2$
denotes the total intensity. Here, $q_j=q_j(z,\tau)$ stands for the envelope of the $j$th component, $z$ is the propagation coordinate, $\tau$ is the retarded time, and $\epsilon$ characterizes the strength of the higher-order contributions~\cite{NakkeeranEtAl1998}.
The CSS system has been considered from several perspectives. Previous studies have developed its inverse-scattering formulation, Darboux transformations and derived various soliton solutions~\cite{Nandy2004,XuXu2013,ZhangWangMa2017}. Subsequent works investigated the localized waves on continuous-wave backgrounds, including different types of solitons, breathers and rogue waves~\cite{ZhaoYangLing2014,LiuTianYuanDu2018,ZhangShiWuFeng2025Dark,ZhangChenFengWu2025Rogue}. 
Although the CSS system shares several physical features with the Manakov and CH systems, its spectral structure exhibits substantial differences. The Manakov and CH systems share the same 3×3 spatial spectral matrix, whereas the CSS system is related to a 5×5 spatial spectral matrix \cite{Manakov1974,PanWangLiuSunRen2024,Nandy2004}. This enlarged spectral structure may make the determination and classification of admissible characteristic roots considerably more complex and provide room for localized-wave behaviors beyond those found in the Manakov and CH systems.

In this work, we investigate the vector AB dynamics in the CSS system, with particular
emphasis on the degenerate and nondegenerate regimes. Owing to the complicated structure of the
associated spectral problem, we classify the background carrier wavenumbers into
three configurations: a single vanishing wavenumber, a pair of opposite
wavenumbers, and two nonzero wavenumbers of unequal magnitudes. These
configurations have different maximal degrees of nondegeneracy, determined by
the number of distinct admissible eigenvalue-branch pairs. For each case, we
derive the reduced spectral equations, construct the existence diagrams, and
calculate the branch-resolved MI gain. The three configurations display different existence diagrams and MI
characteristics. The eigenvalue branches and mode classifications do not always
coincide: nondegenerate-type modes may occur in degenerate regimes, whereas degenerate-type
ones may appear in nondegenerate regimes. A common condition for state transition is
obtained from the exact solutions and is shown to coincide with the zero-gain
condition of the corresponding MI branch. In particular, all admissible branches
in the degenerate regimes obey this condition and correspond to periodic-wave
states, and nondegenerate regimes also cover state-transition branches.
Moreover, the degree of nondegeneracy does not uniquely determine the
intensities of vector breathers, and distinct admissible branches can also generate
symmetry-related profiles. Finally, direct numerical simulations reproduce the
predicted spatiotemporal evolutions and support the analytical results.

The remainder of this paper is organized as follows.
Section~\ref{sec:lax-darboux} presents a preliminary analysis of the
spectral problem for the CSS system.
In Sec.~\ref{sec:spectral-classification}, we analytically classify the
degenerate and nondegenerate vector AB branches under different
carrier-wavenumber configurations.
Section~\ref{sec:existence-mi} gives the corresponding exact
solutions, existence diagrams, and MI gain and discusses their
connections.
Section~\ref{sec:numerical} performs the direct numerical simulations.
Finally, Sec.~\ref{sec:conclusions} summarizes the main conclusions and
discusses several open problems.
\section{\label{sec:lax-darboux}Lax pair and eigenvalue structure}

We start from the CSS system~\eqref{eq:css-system}.
Using the  transformation introduced in
Ref.~\cite{NakkeeranEtAl1998},
\begin{equation*}
	\begin{aligned}
		u_j(x,t)
		&=
		q_j(\tau,z)
		\exp\left[
		-\frac{i}{6\epsilon}
		\left(
		\tau-\frac{z}{18\epsilon}
		\right)
		\right],
		\\
		x&=\tau-\frac{z}{12\epsilon},
		\qquad
		t=z,
		\qquad j=1,2,
	\end{aligned}
\end{equation*}
the CSS system is transformed into
\begin{equation}
	\begin{aligned}
		u_{j,t}
		&+\epsilon\Bigl[
		u_{j,xxx}
		+6\left(|u_1|^2+|u_2|^2\right)u_{j,x}
		\\
		&+3u_j
		\left(|u_1|^2+|u_2|^2\right)_x
		\Bigr]
		=0,
		\qquad j=1,2 .
	\end{aligned}
	\label{eq:reduced-system}
\end{equation}
Following Ref.~\cite{ZhaoYangLing2014}, we rewrite the Lax pair for the
system~\eqref{eq:reduced-system} as
\begin{equation}
	\Phi_x=U\Phi,
	\qquad
	\Phi_t=V\Phi.
	\label{eq:lax-pair}
\end{equation}
The matrices $U$ and $V$ are given by
\begin{equation*}
	U=i\lambda J+Q,
	\qquad
	V=\lambda^3V_0+\lambda^2V_1+\lambda V_2+V_3,
\end{equation*}
with
\begin{equation*}
	J=\operatorname{diag}(1,0,0,0,0)
\end{equation*}
and
\begin{equation*}
	Q=i
	\begin{pmatrix}
		0&u_1&u_1^\ast&u_2&u_2^\ast\\
		u_1^\ast&0&0&0&0\\
		u_1&0&0&0&0\\
		u_2^\ast&0&0&0&0\\
		u_2&0&0&0&0
	\end{pmatrix}.
\end{equation*}
The remaining matrices are
\begin{align*}
	V_0&=i\epsilon J,
	V_1=\epsilon Q,
	V_2=-i\epsilon
	\left([J,Q_x]+[Q,J]Q\right),\\
	V_3&=-\epsilon
	\left(Q_{xx}+[Q,Q_x]-2Q^3\right),
\end{align*}
where $[A,B]=AB-BA$.
 The compatibility condition $U_t-V_x+[U,V]=0$
reproduces the system~\eqref{eq:reduced-system}.

To construct exact solutions on nonzero backgrounds, we choose the plane-wave
seeds
\begin{equation}
	u_1^{[0]}=c_1e^{i\theta_1},\qquad
	u_2^{[0]}=c_2e^{i\theta_2},
		\label{eq:seed-solution}
\end{equation}
where $c_1$ and $c_2$ are real constants. The phase functions are
\begin{equation*}
	\theta_j
	=\beta_j x
	+\epsilon\left[
	\beta_j^3-6\beta_j(c_1^2+c_2^2)
	\right]t,\qquad j=1,2.
\end{equation*}
The parameters $\beta_1$ and $\beta_2$ are the carrier wavenumbers of the
two backgrounds. 

Substituting Eq.~\eqref{eq:seed-solution} into
Eq.~\eqref{eq:lax-pair} and applying a gauge transformation, the spatial part of the Lax pair reduces to a
constant-coefficient system
\begin{equation*}
	\widetilde U_0=iM,
\end{equation*}
where
\begin{equation}
	M=
	\begin{pmatrix}
		\lambda&c_1&c_1&c_2&c_2\\
		c_1&\beta_1&0&0&0\\
		c_1&0&-\beta_1&0&0\\
		c_2&0&0&\beta_2&0\\
		c_2&0&0&0&-\beta_2
	\end{pmatrix}.
	\label{eq:M-def}
\end{equation}
Details of the gauge reduction are provided in
Appendix~\ref{app:spectral-reduction}.

For a fixed spectral parameter $\lambda$, let $\chi$ be an eigenvalue
of $M$. The corresponding characteristic equation reads as
\begin{equation}
	\begin{aligned}
		P(\chi)
		&={}
		(\chi^2-\beta_1^2)(\chi^2-\beta_2^2)(\chi-\lambda)
		\\
		&\quad-2c_1^2\chi(\chi^2-\beta_2^2)
		-2c_2^2\chi(\chi^2-\beta_1^2)=0 .
	\end{aligned}
	\label{eq:char-compact}
\end{equation}
Its algebraic structure depends directly on the relation between the two
carrier wavenumbers. 

When a single carrier wavenumber vanishes, we set
$\beta_1=0$ without loss of generality, since $\beta_2=0$ follows by
interchanging the two components. Equation~\eqref{eq:char-compact} then factorizes as
\begin{equation}
	P(\chi)
	=
	\chi
	\left\{
	(\chi^2-\beta_2^2)
	\left[\chi(\chi-\lambda)-2c_1^2\right]
	-2c_2^2\chi^2
	\right\}
	=0 .
	\label{eq:char-beta1-zero}
\end{equation}
Thus, the vanishing carrier wavenumber yields the explicit root
$\chi=0$, while the remaining four roots are determined by a quartic
equation.

For the equal-magnitude carrier wavenumbers, namely
$\beta_1^2=\beta_2^2=\beta^2$, the characteristic polynomial takes
the factorized form
\begin{equation}
		P(\chi)
		={}
		(\chi^2-\beta^2)
		\left[
		(\chi^2-\beta^2)(\chi-\lambda)
		-2\rho\chi
		\right]
		=0,
	\label{eq:char-equal-square}
\end{equation}
where $\rho=c_1^2+c_2^2$.
Hence, the two carrier-related roots $\chi=\pm\beta$ are obtained
explicitly,
while the other three ones verify a cubic equation.
Although the equal- and opposite-wavenumber configurations 
($\beta_1=\beta_2$ and $\beta_1=-\beta_2$) lead to the same
characteristic polynomial, they correspond to different plane-wave
backgrounds. In the equal-wavenumber case, a fixed-polarization reduction to the
scalar Sasa--Satsuma equation is also possible when the two components
remain proportional. In the following analysis, we use the
opposite-wavenumber configuration as the representative
equal-magnitude case; its relation to the equal-wavenumber
configuration is discussed in
Sec.~\ref{subsec:classification-beta-opposite}. 

Finally, for two nonzero carrier wavenumbers of unequal magnitudes, i.e., 
$
	\beta_1\beta_2\neq0,
	\beta_1^2\neq\beta_2^2,
$
no carrier-related factor can be extracted from
Eq.~\eqref{eq:char-compact}, and the characteristic equation remains quintic. The choice $\beta_1=\beta_2=0$ is included as the $\beta=0$ limit of
the opposite-wavenumber configuration and requires no separate
treatment. 

Accordingly, the subsequent analysis is organized around three
representative carrier-wavenumber configurations: a single vanishing
carrier wavenumber, a pair of opposite carrier wavenumbers, and two
nonzero carrier wavenumbers of unequal magnitudes.
\section{\label{sec:spectral-classification}
	Admissible eigenvalue branches}

\subsection{\label{subsec:classification-beta1-zero}
	Single vanishing carrier wavenumber}

For $\beta_1=0$, Eq.~\eqref{eq:char-beta1-zero} yields the special
eigenvalue $\chi_e=0$, while the remaining four,
$\chi_a$, $\chi_b$, $\chi_c$, and $\chi_d$, satisfy the quartic
equation
\begin{equation}
	\chi^4-\lambda\chi^3
	-\left(\beta_2^2+2\rho\right)\chi^2
	+\lambda\beta_2^2\chi
	+2c_1^2\beta_2^2
	=0.
	\label{eq:quartic-beta1-zero}
\end{equation}
For a nonspecial eigenvalue, the spectral curve is
\begin{equation}
	\lambda(\chi)
	=
	\chi-\frac{2c_1^2}{\chi}
	-\frac{c_2^2}{\chi-\beta_2}
	-\frac{c_2^2}{\chi+\beta_2}.
	\label{eq:lambda-chi-beta1-zero}
\end{equation}

To identify the admissible eigenvalue branches, we select two roots
$\chi_a$ and $\chi_b$ of Eq.~\eqref{eq:quartic-beta1-zero} and impose
the separation relation
\begin{equation}
	\chi_b=\chi_a+K,
	\qquad
	K=i\alpha+\gamma,
	\label{eq:chi-branch-separation}
\end{equation}
where $\alpha$ and $\gamma$ are real parameters. For the AB solutions
considered in this work, we set $\alpha=0$, so that
$K=\gamma\in\mathbb R$. The selected roots must also
correspond to the same spectral parameter
\begin{equation}
	\lambda(\chi_a)=\lambda(\chi_b).
	\label{eq:same-lambda-condition}
\end{equation}
Introducing the auxiliary variable
\begin{equation}
	R=\chi_a\chi_b,
	\label{eq:R-def-general-principle}
\end{equation}
the condition~\eqref{eq:same-lambda-condition} becomes
\begin{equation}
	R^3+a_2R^2+a_1R+a_0=0,
	\label{eq:R-cubic-beta1zero}
\end{equation}
where
\begin{equation}
	\begin{aligned}
		a_0&=2c_1^2\beta_2^2
		\left(\beta_2^2-K^2\right),
		a_1=\beta_2^2
		\left(
		\beta_2^2-K^2-4c_1^2+2c_2^2
		\right),\\
		a_2&=2(\rho-\beta_2^2).
	\end{aligned}
	\label{eq:R-cubic-coefficients}
\end{equation}

Let $R_j$ $(j=1,2,3)$ denote the three roots of
Eq.~\eqref{eq:R-cubic-beta1zero}. Combining
Eq.~\eqref{eq:R-def-general-principle} with
Eq.~\eqref{eq:chi-branch-separation}, we obtain six candidate values
of $\chi_a$
\begin{equation}
	\chi_a^{(j,\sigma)}
	=
	\frac{-K+\sigma\sqrt{K^2+4R_j}}{2},
	\quad
	j=1,2,3,\quad
	\sigma=\pm1.
	\label{eq:chi-branches-beta1zero}
\end{equation}
For each candidate value of $\chi_a$, Eq.~\eqref{eq:chi-branch-separation} directly gives the associated eigenvalue
$\chi_b$. Once an admissible value of $\chi_a$ has been selected, the remaining ones $\chi_c$ and $\chi_d$ are fixed by applying the Vieta's
relations to Eq.~\eqref{eq:quartic-beta1-zero}.

The eigenvectors associated with the nonspecial eigenvalues can be
chosen as
\begin{equation*}
	\begin{aligned}
		\xi_j
		&=
		\Biggl(
		1,\,
		\frac{c_1}{\chi_j},\,
		\frac{c_1}{\chi_j},\,
		\frac{c_2}{\chi_j-\beta_2},\,
		\frac{c_2}{\chi_j+\beta_2}
		\Biggr)^{\mathrm T},
		\quad
		j\in\{a,b,c,d\}.
	\end{aligned}
\end{equation*}
For the special eigenvalue $\chi_e=0$, a corresponding eigenvector is
\begin{equation*}
	\xi_e=(0,1,-1,0,0)^{\mathrm T}.
\end{equation*}

The structure of admissible eigenvalue-branch is determined by the roots
of Eq.~\eqref{eq:R-cubic-beta1zero}. Its discriminant is
\begin{equation}
	\Delta_{\mathrm c}
	=
	18a_2a_1a_0-4a_2^3a_0+a_2^2a_1^2-4a_1^3-27a_0^2.
	\label{eq:Q-cubic-discriminant}
\end{equation}
When $\Delta_{\mathrm c}>0$,
Eq.~\eqref{eq:R-cubic-beta1zero} has three distinct real roots $R_1$, $R_2$ and $R_3$. For real $K$, each real root
gives rise to an admissible complex-conjugate eigenvalue-branch pair if
and only if
\begin{equation}
	K^2+4R_j<0.
	\label{eq:complex-condition-beta1zero}
\end{equation}
Thus, the number of admissible eigenvalue-branch pairs equals the number of negative quantities among
\begin{equation*}
	K^2+4R_1,\qquad
	K^2+4R_2,\qquad
	K^2+4R_3.
\end{equation*}

When $\Delta_{\mathrm c}<0$, the cubic equation has one real root and
one complex-conjugate pair of roots. We write
\begin{equation*}
	R_1\in\mathbb R,\quad
	R_2=A+iB,\quad
	R_3=A-iB,
	\quad B\neq0.
\end{equation*}
The real root $R_1$ is subject to the same condition
$K^2+4R_1<0$. For the complex roots,
\begin{equation}
	K^2+4R_3=(K^2+4R_2)^\ast,
	\label{eq:conjugate-R23}
\end{equation}
and the conjugate eigenvalue branches are paired across the two complex $R$ branches. The curves $\Delta_{\mathrm c}=0$ and
$K^2+4R_j=0$ define the analytical boundaries of the existence diagram for this carrier-wavenumber configuration.

\subsection{\label{subsec:classification-beta-opposite}
	Opposite carrier wavenumbers}

The condition $\beta_1^2=\beta_2^2$ admits two possible
carrier-wavenumber configurations, namely $\beta_1=\beta_2$ and
$\beta_1=-\beta_2$. For the equal-amplitude background $c_1=c_2$ used in the intensity
analysis, the two configurations are related through complex
conjugation of the second component. More precisely, for the
same eigenvalue branches and Darboux coefficients,
\begin{equation}
	u_{1,-}=u_{1,+},
	\qquad
	u_{2,-}=u_{2,+}^{\ast}.
	\label{eq:beta-sign-conjugation-sec2}
\end{equation}
Hence, under identical background amplitude and other parameters, the equal- and opposite‑wavenumber configurations are equivalent in terms of intensity of breather. We take the
opposite-wavenumber configuration as the representative
equal-magnitude case, one has
\begin{equation}
	\beta_1=-\beta_2=\beta.
	\label{eq:opposite-wavenumbers}
\end{equation}
The proof of the equivalence relation
\eqref{eq:beta-sign-conjugation-sec2} is given in
Appendix~\ref{app:conjugation-equivalence}. The case $\beta=0$ is
included as the corresponding zero-wavenumber limit and is therefore
not treated separately.

Specializing Eq.~\eqref{eq:char-equal-square} to
\eqref{eq:opposite-wavenumbers} yields the carrier-locked eigenvalues
$
	\chi_d=\beta,
	\chi_e=-\beta,
$
while the other ones hold
\begin{equation}
	\chi^3-\lambda\chi^2
	-\left(\beta^2+2\rho\right)\chi
	+\lambda\beta^2=0.
	\label{eq:cubic-opposite}
\end{equation}
Solving Eq.~\eqref{eq:cubic-opposite} for $\lambda$ yields the spectral
curve
\begin{equation*}
	\lambda(\chi)
	=
	\chi
	-\frac{\rho}{\chi-\beta}
	-\frac{\rho}{\chi+\beta},
	\quad
	\chi\neq\pm\beta.
\end{equation*}

For two distinct eigenvalues $\chi_a$ and $\chi_b$ satisfying
$\lambda(\chi_a)=\lambda(\chi_b)$, we combine
$R=\chi_a\chi_b$ with $\chi_b-\chi_a=K$ to reduce the
same spectral-parameter condition to
\begin{equation}
	R^2
	+2(\rho-\beta^2)R
	+\beta^2\left(\beta^2-K^2+2\rho\right)=0.
	\label{eq:R-quadratic-betaop}
\end{equation}
We write the two roots of Eq.~\eqref{eq:R-quadratic-betaop} as $R_1$ and $R_2$. Its discriminant is
\begin{equation}
	\Delta_R
	=
	4\left[\rho^2+\beta^2\left(K^2-4\rho\right)\right].
	\label{eq:discriminant-betaop}
\end{equation}
The candidate values of $\chi_a$ are
\begin{equation}
	\chi_a^{(j,\sigma)}
	=
	\frac{-K+\sigma\sqrt{K^2+4R_j}}{2},
	\quad
	j=1,2,
	\quad
	\sigma=\pm1.
	\label{eq:chi-branches-opposite}
\end{equation}
Once $\chi_a$ is fixed, $\chi_b$ is recovered from
Eq.~\eqref{eq:chi-branch-separation}. Vieta's relation then gives
the third nonspecial eigenvalue
\begin{equation*}
	\chi_c=\lambda-\chi_a-\chi_b.
\end{equation*}
For the nonspecial eigenvalues $\chi_j$ $(j=a,b,c)$, we take
\begin{equation*}
	\xi(\chi_j)=
	\left(
	1,\
	\frac{c_1}{\chi_j-\beta},\
	\frac{c_1}{\chi_j+\beta},\
	\frac{c_2}{\chi_j+\beta},\
	\frac{c_2}{\chi_j-\beta}
	\right)^{\rm T}.
\end{equation*}
The special eigenvalues admit the eigenvectors
\begin{equation*}
		\xi_d=(0,-c_2,0,0,c_1)^{\rm T},
		\quad
		\xi_e=(0,0,-c_2,c_1,0)^{\rm T}.
\end{equation*}

The structure of admissible eigenvalue-branch is governed by the two
roots of Eq.~\eqref{eq:R-quadratic-betaop}. If both roots are real,
each yields an admissible eigenvalue-branch pair if
\begin{equation}
	K^2+4R_j<0,
	\qquad j=1,2.
	\label{eq:betaop-admissible-condition}
\end{equation}
If the two roots form a complex-conjugate pair, complex conjugation
links the eigenvalue branches generated by the two $R$ branches.
The conditions $\Delta_R=0$ and $K^2+4R_j=0$ then provide
the analytical boundaries in the $(\beta,\gamma)$ plane.

\subsection{\label{subsec:classification-general}
	Nonzero carrier wavenumbers of unequal magnitudes}

We finally turn to the general configuration
$
	\beta_1 \beta_2\neq0,
	\beta_1^2\neq\beta_2^2.
$
In this case, the four carrier-related values
$\pm\beta_1$ and $\pm\beta_2$ are mutually distinct and cannot serve
as admissible eigenvalues. No carrier-related factor can be
extracted from the characteristic polynomial, which therefore
remains its quintic form
\begin{equation}
	\begin{aligned}
		&\chi^5-\lambda\chi^4
		-\left(\beta_1^2+\beta_2^2+2\rho\right)\chi^3
		+\lambda\left(\beta_1^2+\beta_2^2\right)\chi^2
		\\
		&\quad
		+\left[
		\beta_1^2\beta_2^2
		+2\left(c_1^2\beta_2^2+c_2^2\beta_1^2\right)
		\right]\chi
		-\lambda\beta_1^2\beta_2^2=0.
	\end{aligned}
	\label{eq:char-poly-general-case}
\end{equation}
Equivalently, the spectral curve is
\begin{equation}
	\lambda(\chi)
	=
	\chi-\frac{c_1^2}{\chi-\beta_1}
	-\frac{c_1^2}{\chi+\beta_1}
	-\frac{c_2^2}{\chi-\beta_2}
	-\frac{c_2^2}{\chi+\beta_2}.
	\label{eq:lambda-chi-general}
\end{equation}

Imposing
$\lambda(\chi_a)=\lambda(\chi_b)$ on two distinct eigenvalues and
using $R=\chi_a\chi_b$ together with $\chi_b-\chi_a=K$ leads to the
quartic equation
\begin{equation}
	R^4+b_3R^3+b_2R^2+b_1R+b_0=0,
	\label{eq:R-quartic-general}
\end{equation}
where
\begin{equation*}
	\begin{aligned}
		b_3
		&=-2\left(\beta_1^2+\beta_2^2\right)+2\rho,\\
		b_2
		&=\mathcal E_1+\mathcal E_2+4\beta_1^2\beta_2^2
		+2c_1^2\left(\beta_1^2-2\beta_2^2\right)\\
		&\quad
		+2c_2^2\left(\beta_2^2-2\beta_1^2\right),\\
		b_1
		&=-2\left(\beta_2^2\mathcal E_1+\beta_1^2\mathcal E_2\right)
		+2c_1^2\left(\mathcal E_2-2\beta_1^2\beta_2^2\right)\\
		&\quad
		+2c_2^2\left(\mathcal E_1-2\beta_1^2\beta_2^2\right),\\
		b_0
		&=\mathcal E_1\mathcal E_2
		+2c_1^2\mathcal E_2\beta_1^2
		+2c_2^2\mathcal E_1\beta_2^2,
	\end{aligned}
\end{equation*}
with
\begin{equation*}
	\mathcal E_1
	=
	\beta_1^2(\beta_1^2-K^2),
	\qquad
	\mathcal E_2
	=
	\beta_2^2(\beta_2^2-K^2).
\end{equation*}

We label the four roots of Eq.~\eqref{eq:R-quartic-general} as $R_j$ $(j=1\sim4)$. These four roots generate eight candidate values of $\chi_a$,
\begin{equation*}
	\chi_a^{(j,\sigma)}
	=
	\frac{-K+\sigma\sqrt{K^2+4R_j}}{2},
	\quad
	j=1,2,3,4,\quad
	\sigma=\pm1.
\end{equation*}
Once $\chi_a$ is specified,
Eq.~\eqref{eq:chi-branch-separation} immediately fixes $\chi_b$. The remaining three eigenvalues
$\chi_c$, $\chi_d$, and $\chi_e$ follow from the cubic
equation that remains after dividing
Eq.~\eqref{eq:char-poly-general-case} by
$(\chi-\chi_a)(\chi-\chi_b)$.
A convenient eigenvector representation for any eigenvalue $\chi_l$ $(l=a,b,c,d,e)$ is 
\begin{equation*}
	\xi_l=
	\left(
	1,\
	\frac{c_1}{\chi_l-\beta_1},\
	\frac{c_1}{\chi_l+\beta_1},\
	\frac{c_2}{\chi_l-\beta_2},\
	\frac{c_2}{\chi_l+\beta_2}
	\right)^{\rm T}.
\end{equation*}

In the general configuration, admissibility depends jointly on the roots of
Eq.~\eqref{eq:R-quartic-general} and the quantities
\begin{equation*}
	K^2+4R_j,
	\qquad j=1,2,3,4.
\end{equation*}
For a real root $R_j$, the associated eigenvalue-branch pair is admissible if
\begin{equation}
	K^2+4R_j<0.
	\label{eq:complex-condition-general}
\end{equation}
If two roots meet $R_m=R_n^\ast$, then
\begin{equation*}
	K^2+4R_n=(K^2+4R_m)^\ast,
\end{equation*}
so the conjugate eigenvalue branches lie on the two associated
$R$ branches. Multiple roots of
Eq.~\eqref{eq:R-quartic-general}, together with the additional
condition
\begin{equation*}
	K^2+4R_j=0,
\end{equation*}
mark the critical boundaries separating the different admissible-branch regimes. Compared with the two reduced cases above, the quartic equation can
provide up to four distinct $R$ branches and hence a larger set of
candidate eigenvalue-branch pairs whose admissibility must be
determined in the parameter space.
\section{\label{sec:existence-mi}
Breathers, existence diagrams, state transitionS and MI}
The wave profiles of vector breathers discussed below are obtained by evaluating the
closed-form Darboux solutions. For an admissible eigenvalue branch
$\chi_a$ and its associated eigenvalue $\chi_b$, we set
\begin{equation*}
	\chi_b=\chi_a+\gamma,
	\qquad
	\gamma\in\mathbb R,
	\qquad
	\chi_a=\chi_a^r+i\chi_a^i .
\end{equation*}
The corresponding exponential phases are
\begin{equation*}
	\eta_\mu
	=
	i\chi_\mu x
	+i\epsilon\left(\chi_\mu^3-6\rho\chi_\mu\right)t,
	\quad
	\phi_\mu=e^{\eta_\mu},
	\quad
	\mu=a,b,
\end{equation*}
where $\rho=c_1^2+c_2^2$.

We introduce
\begin{equation*}
	\delta
	=
	\eta_b-\eta_a
	=
	\Psi+i\Theta,
\end{equation*}
where
\begin{equation}
	\Psi
	=
	-3\epsilon\chi_a^i\gamma
	\left(2\chi_a^r+\gamma\right)t ,
	\label{eq:Psi-main}
\end{equation}
and
\begin{equation}
	\Theta
	=
	\gamma x
	+\epsilon\gamma
	\left[
	3(\chi_a^r)^2
	-3(\chi_a^i)^2
	+3\chi_a^r\gamma
	+\gamma^2
	-6\rho
	\right]t .
	\label{eq:Theta-main}
\end{equation}

When $\lambda+\lambda^\ast\neq0$, the two reduction-related Darboux
spectral points $\lambda$ and $-\lambda^\ast$ are distinct, and the
solution is obtained through the binary Darboux transformation,
\begin{equation}
	u_s
	=
	u_{s0}
	\frac{\mathcal N_s}{\mathcal D},
	\qquad
	s=1,2,
	\label{eq:binary-DT-nine-main}
\end{equation}
where $u_{10}=c_1e^{i\theta_1}$ and
$u_{20}=c_2e^{i\theta_2}$. The numerator is
\begin{align}
	\mathcal N_s
	&={}
	\mathcal K^{(s)}_{22}e^{2\Psi}
	+\mathcal K^{(s)}_{21}e^{\Psi+i\Theta}
	+\mathcal K^{(s)}_{12}e^{\Psi-i\Theta}
	+\mathcal K^{(s)}_{20}e^{2i\Theta}
	\nonumber\\
	&\quad+\mathcal K^{(s)}_{11}
	+\mathcal K^{(s)}_{02}e^{-2i\Theta}
	+\mathcal K^{(s)}_{10}e^{-\Psi+i\Theta}
	+\mathcal K^{(s)}_{01}e^{-\Psi-i\Theta}
	\nonumber\\
	&\quad+\mathcal K^{(s)}_{00}e^{-2\Psi},
	\qquad s=1,2 ,
	\label{eq:Ns-nine-term-main}
\end{align}
and the denominator reads as
\begin{align}
	\mathcal D
	&={}
	\mathcal K^{(0)}_{22}e^{2\Psi}
	+\mathcal K^{(0)}_{21}e^{\Psi+i\Theta}
	+\mathcal K^{(0)}_{12}e^{\Psi-i\Theta}
	+\mathcal K^{(0)}_{20}e^{2i\Theta}
	\nonumber\\
	&\quad+\mathcal K^{(0)}_{11}
	+\mathcal K^{(0)}_{02}e^{-2i\Theta}
	+\mathcal K^{(0)}_{10}e^{-\Psi+i\Theta}
	+\mathcal K^{(0)}_{01}e^{-\Psi-i\Theta}
	\nonumber\\
	&\quad+\mathcal K^{(0)}_{00}e^{-2\Psi}.
	\label{eq:D-nine-term-main}
\end{align}

The binary Darboux solution is invariant under interchange of the
reduction-related spectral points $\lambda$ and $-\lambda^\ast$.
For
\begin{equation}
	\widehat{\chi}=-\gamma-\chi^\ast,
	\label{eq:chi-reduction-equivalence}
\end{equation}
one has $\lambda(\widehat{\chi})=-\lambda(\chi)^\ast$ and hence
\begin{equation*}
	u_s(\widehat{\chi})=u_s(\chi),
	\qquad s=1,2.
\end{equation*}
We thus regard two admissible branches related by
Eq.~\eqref{eq:chi-reduction-equivalence} as one eigenvalue-branch pair.

When
\begin{equation*}
	\lambda+\lambda^\ast=0,
\end{equation*}
the two reduction-related Darboux spectral points coalesce,
$-\lambda^\ast=\lambda$, and the binary construction reduces to a
one-fold Darboux transformation. For the same exponential modes
$\phi_a$ and $\phi_b$, the resulting solution takes the form
\begin{equation}
	\begin{aligned}
		u_s
		&=
		u_{s0}
		\frac{
			h^{(s)}_{11}e^{-\Psi}
			+h^{(s)}_{12}e^{\Psi}
			+h^{(s)}_{13}e^{i\Theta}
			+h^{(s)}_{14}e^{-i\Theta}
		}{
			h^{(0)}_{11}e^{-\Psi}
			+h^{(0)}_{12}e^{\Psi}
			+h^{(0)}_{13}e^{i\Theta}
			+h^{(0)}_{14}e^{-i\Theta}
		},
		\\
		s&=1,2 .
	\end{aligned}
	\label{eq:single-DT-four-main}
\end{equation}
In this case, the conjugate branches associated with the
same admissible real root yield identical modulus-squared profiles
and are likewise regarded as one eigenvalue-branch pair.

The detailed derivations of
Eqs.~\eqref{eq:Ns-nine-term-main}--\eqref{eq:D-nine-term-main}
and~\eqref{eq:single-DT-four-main}, together with the explicit
coefficients $\mathcal K_{mn}^{(s)}$ and $h_{1n}^{(s)}$, are given in
Appendix~\ref{app:coefficients-darboux}. The binary/one-fold
distinction concerns the Darboux construction and is independent of
the degenerate/nondegenerate classification introduced in
Sec.~\ref{sec:spectral-classification}.
\subsection{\label{subsec:existence-beta1-zero}
	Case I: $\beta_1=0$}

\begin{figure}[htbp]
	\centering
	\includegraphics[width=0.55\columnwidth]{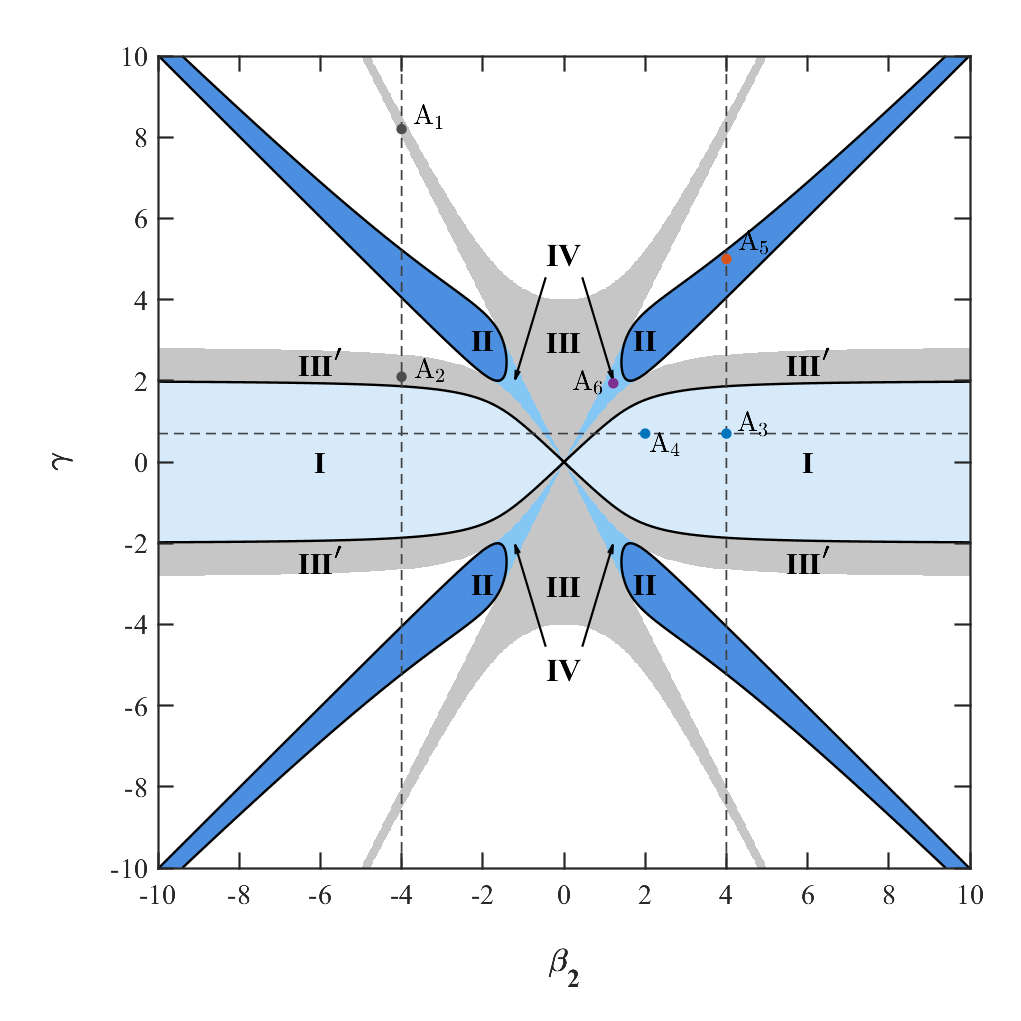}
	\caption{
		Existence diagram in the $(\beta_2,\gamma)$ plane for
		$\beta_1=0$, with $c_1=c_2=1$ and $\alpha=0$.
		Gray and blue regions denote degenerate and nondegenerate regimes,
		respectively.
		The annotated coordinate points
		$\mathrm{A}_1=(-4,8.2)$,
		$\mathrm{A}_2=(-4,2.1)$,
		$\mathrm{A}_3=(4,0.7)$,
		$\mathrm{A}_4=(2,0.7)$,
		$\mathrm{A}_5=(4,5)$, and
		$\mathrm{A}_6=(1.21402,1.93992)$
		are used in the analyses below.
		Dashed vertical and horizontal lines indicate common values of
		$\beta_2$ and $\gamma$, respectively.
	}
	\label{fig:existence-beta1-zero}
\end{figure}

For $\beta_1=0$, the admissible eigenvalue branches are governed by
the cubic equation~\eqref{eq:R-cubic-beta1zero}. With
$c_1=c_2=1$ and $\alpha=0$, the existence diagram in the
$(\beta_2,\gamma)$ plane is shown in
Fig.~\ref{fig:existence-beta1-zero}. Its boundaries are determined by
the cubic discriminant together with the admissibility
condition~\eqref{eq:complex-condition-beta1zero}.

The gray region contains only one admissible eigenvalue-branch pair,
whereas each of the remaining ones harbors at least two.
We define the degree of nondegeneracy as
\begin{equation*}
	d_{\mathrm{nd}}=N_{\mathrm p}-1,
\end{equation*}
where $N_{\mathrm p}$ denotes the number of distinct admissible
eigenvalue-branch pairs. Accordingly, $d_{\mathrm{nd}}=0$
corresponds to the degenerate regime, whereas
$d_{\mathrm{nd}}\geq1$ identifies a nondegenerate regime.
Here, $d_{\mathrm{nd}}$ counts distinct admissible
eigenvalue-branch pairs rather than inequivalent intensity structures,
since symmetry-related branch pairs may generate equivalent
modulus-squared profiles.

\begin{table*}[tbp]
	\caption{
		Maximum degree of nondegeneracy $d_{\mathrm{nd}}^{\max}$ for
		different integrable vector systems and carrier-wavenumber
		configurations. Here, $d_{\mathrm{nd}}=N_{\mathrm p}-1$, where
		$N_{\mathrm p}$ denotes the number of distinct admissible
		eigenvalue-branch pairs. A dash indicates that the corresponding
		configuration has not been considered.
	}
	\label{tab:max-degree-nondegeneracy}
	\centering
	\small
	
	\renewcommand{\arraystretch}{1.10}
	\setlength{\tabcolsep}{3pt}
	
	\begin{ruledtabular}
		\begin{tabular}{@{}cccc@{}}
			
			\noalign{\vskip-\doublerulesep}
			
			\diagbox[
			width=4.2cm,
			height=3.0\baselineskip,
			trim=l,
			innerleftsep=4pt,
			innerrightsep=4pt
			]
			{\raisebox{-0.2ex}{Models}}
			{\raisebox{-0.2ex}{Wavenumbers}}
			&
			\parbox[c][3.0\baselineskip][c]{4.0cm}{
				\centering
				Single vanishing\\[-2pt]
				carrier wavenumber
			}
			&
			\parbox[c][3.0\baselineskip][c]{2.8cm}{
				\centering
				Opposite carrier\\[-2pt]
				wavenumbers
			}
			&
			\parbox[c][3.0\baselineskip][c]{4.0cm}{
				\centering
				Unequal nonzero\\[-2pt]
				carrier wavenumbers
			}
			\\
			
			\hline
			
			Manakov system
			& \textemdash
			& $1$
			& \textemdash
			\\
			
			CH system
			& \textemdash
			& $1$
			& \textemdash
			\\
			
			CSS system
			& $2$
			& $1$
			& $3$
			\\
			
		\end{tabular}
	\end{ruledtabular}
\end{table*}

\begin{figure*}[t]
	\centering
	\includegraphics[width=0.75\textwidth]{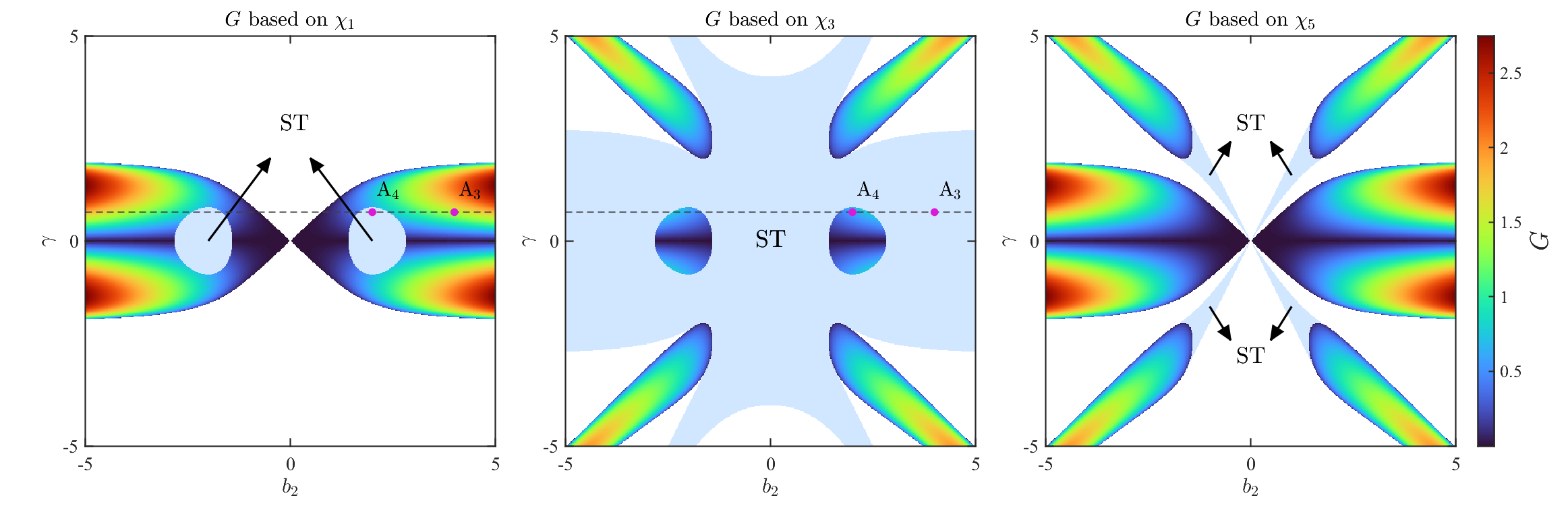}
	\caption{
		Branch-resolved MI gain in the $(\beta_2,\gamma)$ plane for
		$\beta_1=0$, with $c_1=c_2=1$ and $\alpha=0$.
		The three panels correspond to the eigenvalue-branch pairs
		represented by $\chi_1$, $\chi_3$, and $\chi_5$, respectively.
		Nonzero gain identifies AB-supporting branches, while ``ST''
		denotes admissible branches satisfying the state-transition
		condition and reducing to periodic-wave states.
		The annotated coordinate points
		$\mathrm{A}_3=(4,0.7)$ and
		$\mathrm{A}_4=(2,0.7)$
		are used in the representative analyses below, with the dashed
		horizontal line indicating their common $\gamma=0.7$.
	}
	\label{fig:MI-beta1-zero}
\end{figure*}

The maximum degree of nondegeneracy is set by the number of
admissible eigenvalue-branch pairs that can coexist. For the Manakov
system~\cite{LiuChenYaoAkhmediev2022} and the CH
system~\cite{PanWangLiuSunRen2024}, the configurations considered in
the corresponding studies support at most two admissible
eigenvalue-branch pairs, giving
$d_{\mathrm{nd}}^{\max}=1$. In the CSS system, up to four admissible
eigenvalue-branch pairs can coexist, yielding
$d_{\mathrm{nd}}^{\max}=3$. The results for the different
carrier-wavenumber configurations are summarized in
Table~\ref{tab:max-degree-nondegeneracy}.

For the Manakov and CH systems, other carrier-wavenumber
configurations can be related to the representative configurations
through Galilean transformations. Their degrees of nondegeneracy
have not been classified separately within the present framework,
and the corresponding entries are therefore left unspecified.
For the present configuration $\beta_1=0$, at most three admissible
eigenvalue-branch pairs coexist, so that
$d_{\mathrm{nd}}^{\max}=2$. Specifically, regions~I, II, and IV in
Fig.~\ref{fig:existence-beta1-zero} correspond to
$d_{\mathrm{nd}}=2$, $1$, and $1$.
The larger number of coexisting admissible eigenvalue-branch pairs
distinguishes the CSS system from the Manakov and CH cases considered
here and reflects its richer multi-branch structure.

The gray region admits a single eigenvalue-branch pair and therefore
belongs to the degenerate regime. This pair lies on the
state-transition locus and corresponds to a periodic wave. Region~I
is characterized by three admissible eigenvalue-branch pairs, one
associated with a periodic wave and the other two with ABs. Region~II
has two admissible pairs, both corresponding to the AB solutions, whereas
region~IV likewise admits two pairs, both located on the
state-transition locus and giving rise to the periodic waves.

The existence diagram determines the number of admissible
eigenvalue-branch pairs in each region but does not resolve the
behavior of individual branches. In particular, one of the three
pairs in region~I undergoes the state transition, while the existence
diagram alone does not specify which pair it is. We therefore
calculate the MI gain for each eigenvalue branch in the same
parameter plane. Linearization of the reduced CSS system around the
plane-wave background gives
\begin{equation*}
	G(\chi_a;\gamma)
	=
	3\left|
	\epsilon\gamma\chi_a^i
	\left(2\chi_a^r+\gamma\right)
	\right|,
\end{equation*}
with the detailed derivation provided in
Appendix~\ref{app:mi-analysis}. Among the admissible eigenvalue branches, those with nonzero gain
support the ABs, whereas those with zero gain correspond to the
state-transition periodic waves.

Comparing Figs.~\ref{fig:existence-beta1-zero} and
\ref{fig:MI-beta1-zero} resolves the branch structure within each
region. In the gray degenerate region, the unique admissible
eigenvalue-branch pair is $(\chi_3,\chi_4)$, whose vanishing MI gain
identifies it as a state-transition periodic-wave branch.

In the nondegenerate region~I, all three eigenvalue-branch pairs are
admissible, but only one lies on the state-transition locus. The
branch-resolved MI-gain maps reveal a small elliptical domain within
this region. Inside the ellipse, $(\chi_1,\chi_2)$ has zero gain and
thus corresponds to the periodic-wave state, while the other two
pairs remain on AB branches. Outside the ellipse, the state-transition
pair switches to $(\chi_3,\chi_4)$, whereas the remaining two pairs
continue to support ABs.

\begin{figure*}[hbpt]
	\centering
	
	\begin{minipage}[t]{0.49\textwidth}
		\centering
		\includegraphics[width=\linewidth]{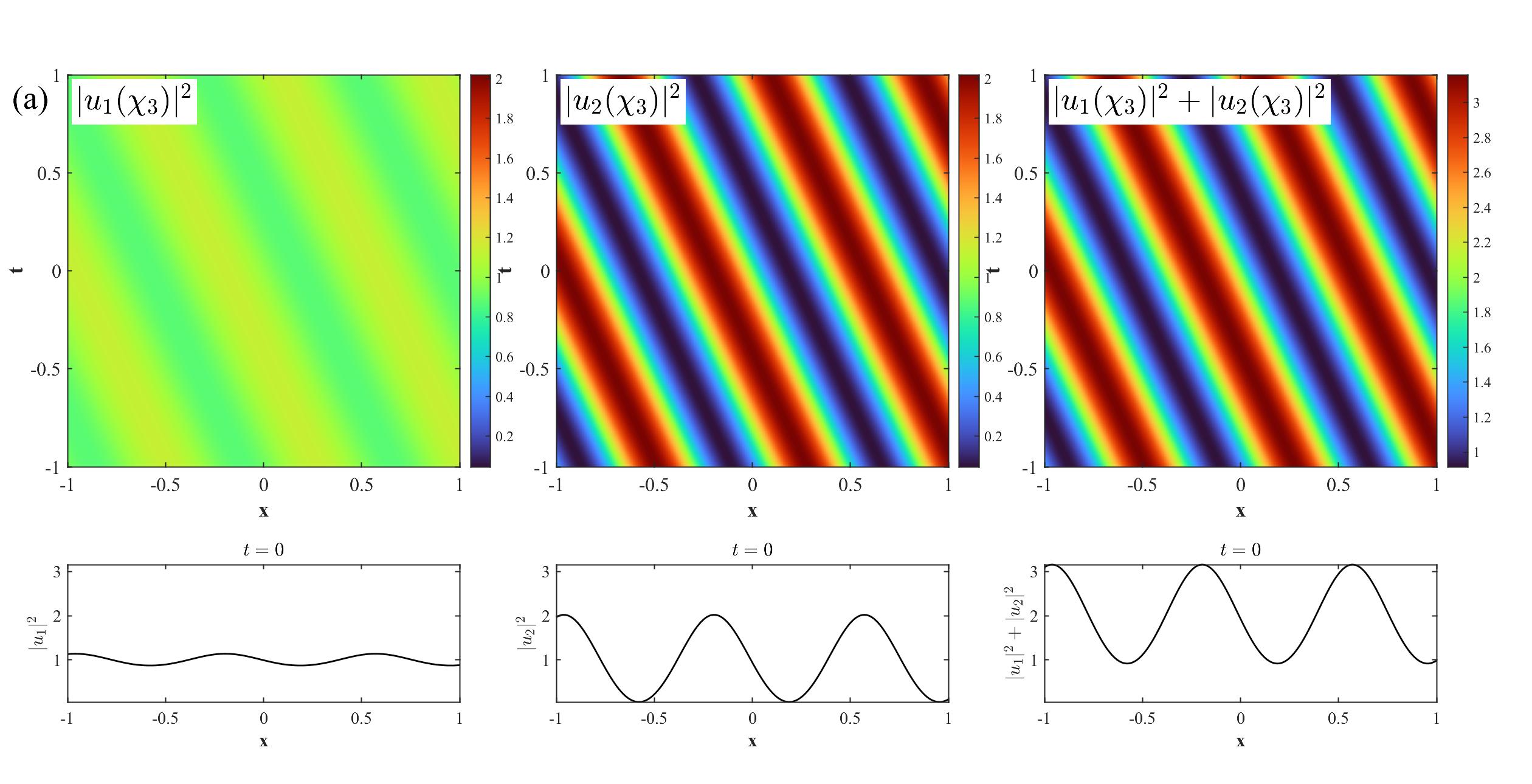}
	\end{minipage}
	\hfill
	\begin{minipage}[t]{0.49\textwidth}
		\centering
		\includegraphics[width=\linewidth]{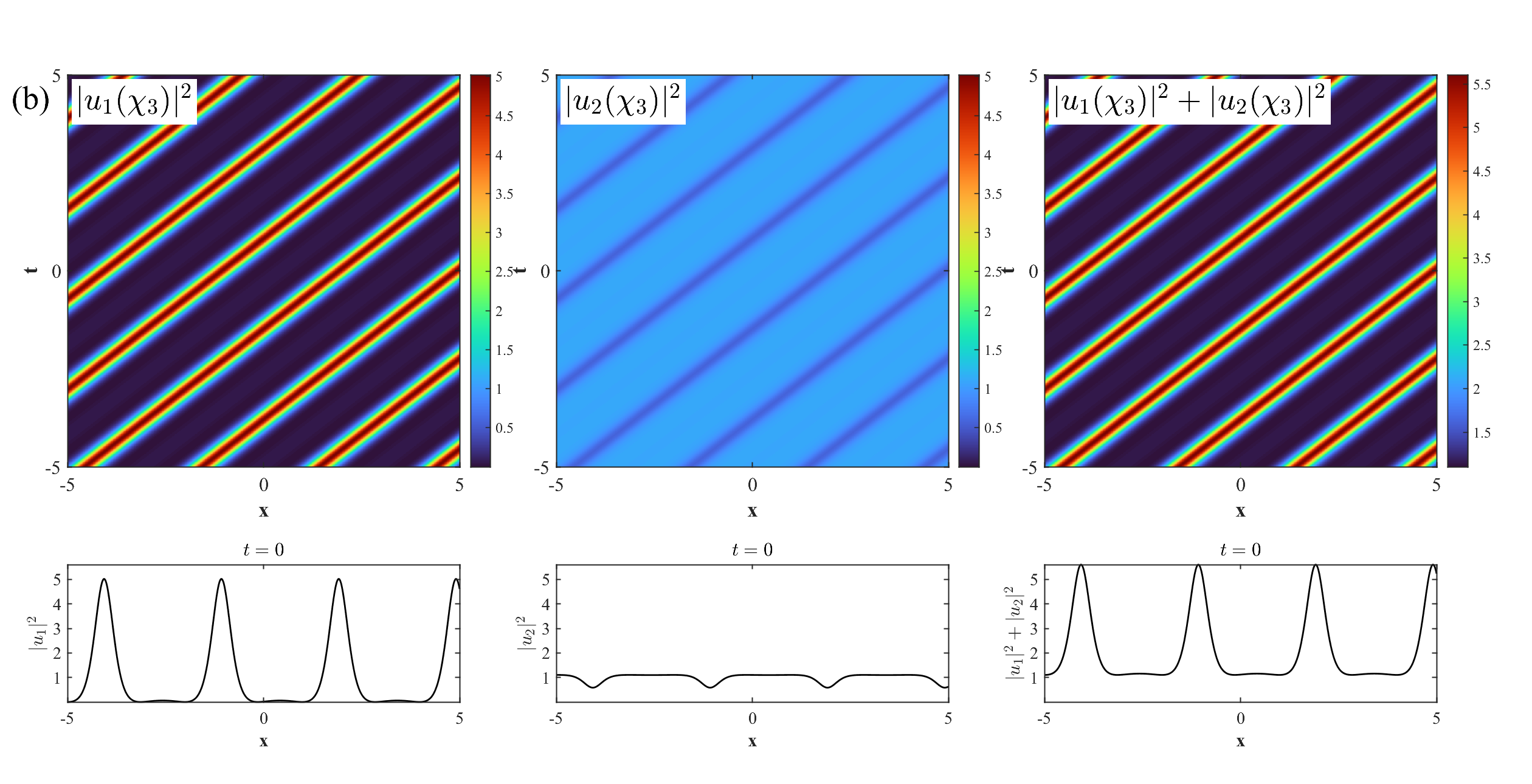}
	\end{minipage}
	
	\caption{
		State-transition periodic waves in the degenerate regime for
		$\beta_1=0$, corresponding to points
		$\mathrm{A}_1$ and $\mathrm{A}_2$ in
		Fig.~\ref{fig:existence-beta1-zero}.
		(a) Region~III at $\mathrm{A}_1=(-4,8.2)$, with
		$\chi_a=\chi_3=-4.1-0.1366i$ and
		$\chi_b=4.1-0.1366i$.
		(b) Region~III$^\prime$ at $\mathrm{A}_2=(-4,2.1)$, with
		$\chi_a=\chi_3=-1.0500-0.8309i$ and
		$\chi_b=1.0500-0.8309i$.
		In both cases,
		$\operatorname{Re}(\chi_a)=-\gamma/2$, with $\chi_4$ forming the other member of the unique
		admissible eigenvalue-branch pair.
		The intensity modulations are synchronous in region~III and
		opposite in region~III$^\prime$.
		For each case, the upper row shows
		$|u_1|^2$, $|u_2|^2$, and $|u_1|^2+|u_2|^2$ from left to right,
		and the lower row gives the corresponding $t=0$ cross sections.
	}
	\label{fig:degenerate-regionIII-comparison}
\end{figure*}

The MI-gain maps also distinguish regions~II and IV. Both admissible
eigenvalue-branch pairs in region~II have nonzero gain and support
ABs, whereas the remaining two pairs in region~IV have zero gain and correspond to
state-transition periodic waves. Consequently, the existence and MI-gain diagrams
 provide complementary information: the former determines
the number of admissible eigenvalue-branch pairs, and the latter
identifies their AB or state-transition character.

\subsubsection{\label{subsubsec:degenerate-beta1-zero}
	State-transition periodic waves in degenerate regions~III and III$^\prime$}

In regions~III and III$^\prime$, the cubic discriminant satisfies
$\Delta_{\mathrm c}>0$, so
Eq.~\eqref{eq:R-cubic-beta1zero} has three distinct real roots,
which we order as
\begin{equation*}
	R_1<R_2<R_3.
\end{equation*}
Among them, only $R_1$ satisfies the admissibility criterion,
\begin{equation*}
	K^2+4R_1<0,
\end{equation*}
whereas $K^2+4R_2>0$ and $K^2+4R_3>0$. Hence, $R_1$
generates the unique admissible eigenvalue-branch pair, and both
regions belong to the degenerate regime.

Writing
\begin{equation*}
	\kappa=\sqrt{-(K^2+4R_1)}>0,
\end{equation*}
the two admissible branches are
\begin{equation*}
	\chi_a^{(1,\pm)}
	=
	\frac{-K\pm i\kappa}{2}.
\end{equation*}
For $K=\gamma\in\mathbb R$, they share the real part
\begin{equation}
	\operatorname{Re}(\chi_a)
	=
	-\frac{\gamma}{2},
	\label{eq:state-transition-condition}
\end{equation}
which gives the state-transition condition. Substituting this
relation into Eq.~\eqref{eq:Psi-main} immediately yields
\begin{equation*}
	\Psi
	=
	-3\epsilon\chi_a^i\gamma
	(2\chi_a^r+\gamma)t
	=0.
\end{equation*}
Thus, the localizing exponential factors disappear and the
corresponding state is governed solely by the periodic phase.

The associated spectral parameter is purely imaginary,
$\lambda+\lambda^\ast=0$, so the one-fold Darboux solution
\eqref{eq:single-DT-four-main} applies. Using
Eq.~\eqref{eq:state-transition-condition},
Eq.~\eqref{eq:Theta-main} reduces to
\begin{equation*}
	\Theta
	=
	\gamma x
	+
	\epsilon\gamma
	\left[
	-3(\chi_a^i)^2
	+\frac{\gamma^2}{4}
	-6(c_1^2+c_2^2)
	\right]t.
\end{equation*}
The resulting solution shows a periodic-wave state with a
linear phase in the $(x,t)$ plane.

More generally, for $K=\gamma\in\mathbb{R}$, any admissible
pair generated by a real root $R_j$ satisfies
$\gamma^2+4R_j<0$ and therefore has
$\operatorname{Re}(\chi_a)=-\gamma/2$.
Hence, such a pair necessarily lies on the state-transition
locus. As shown below, the unique admissible pair in every
degenerate regime considered in the three carrier-wavenumber
configurations is of this type.

Although the regions~III and III$^\prime$ are both degenerate according
to the eigenvalue-branch classification and lie on the
state-transition locus, they exhibit different modes
between the two wave components. Here, the mode classification refers to
the relative intensity pattern of the two components within the same
solution. In region~III, the peaks and valleys of the two components
occur at the same spatiotemporal positions, corresponding to a
degenerate-type mode, as shown in
Fig.~\ref{fig:degenerate-regionIII-comparison}(a).
Region~III$^\prime$, by contrast, exhibits a nondegenerate-type mode, for
which a peak in one component coincides with a valley in the other,
as illustrated in
Fig.~\ref{fig:degenerate-regionIII-comparison}(b).
Thus, the eigenvalue-branch degeneracy does not uniquely determine
the mode character of the two-component solution.

The contrast between regions~III and III$^\prime$ shows that
degenerate- and nondegenerate-type modes can occur within the same
degenerate eigenvalue-branch classification. The
single vanishing-carrier configuration considered here is less
symmetric than the opposite-wavenumber case, since $\beta_2$
remains an independent carrier parameter rather than being
constrained by $\beta_1=-\beta_2$. This behavior differs from the
degenerate breathers of the Manakov system considered in
Ref.~\cite{LiuChenYaoAkhmediev2022}, where the two components
exhibit similar spatiotemporal intensity patterns. It also contrasts
with the opposite-wavenumber configuration discussed in
Sec.~\ref{subsec:phase-beta-opposite}, where degenerate-type modes can
instead occur within nondegenerate eigenvalue-branch regimes.
These comparisons highlight the distinctive mode behavior of the
single-vanishing-carrier configuration.
\subsubsection{\label{subsubsec:nondegenerate-beta1-zero}
	Nondegenerate solutions}

\paragraph*{Region~I.}
\begin{figure}[tbp]
	\centering
	\includegraphics[width=\columnwidth]{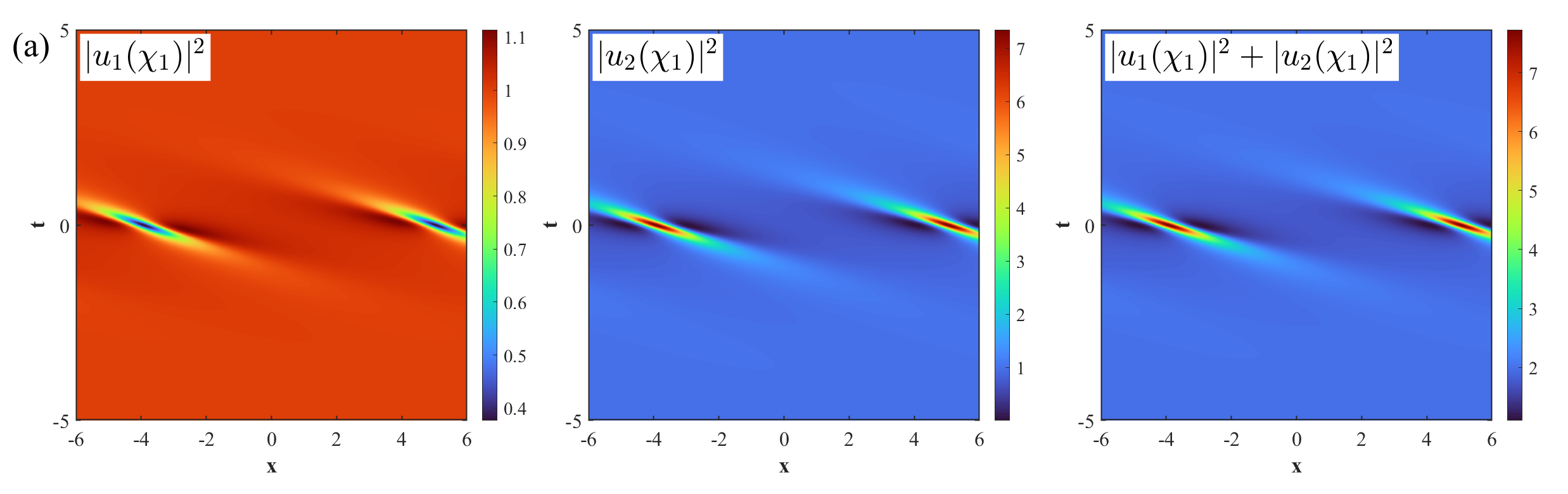}
	
	\includegraphics[width=\columnwidth]{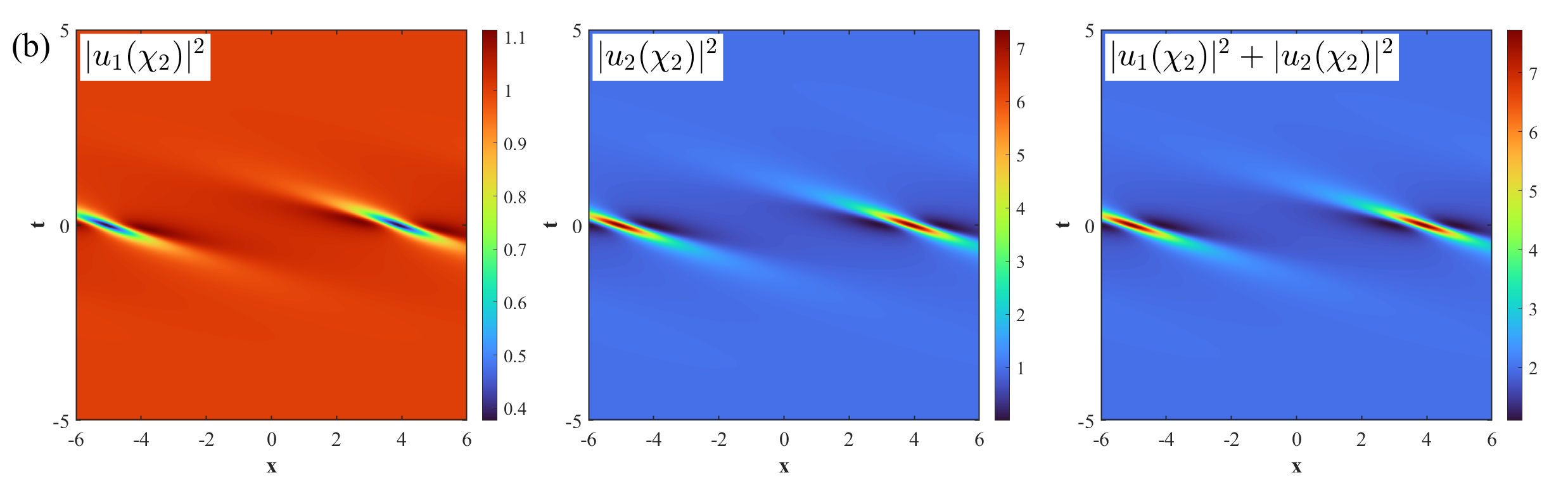}
	
	\includegraphics[width=\columnwidth]{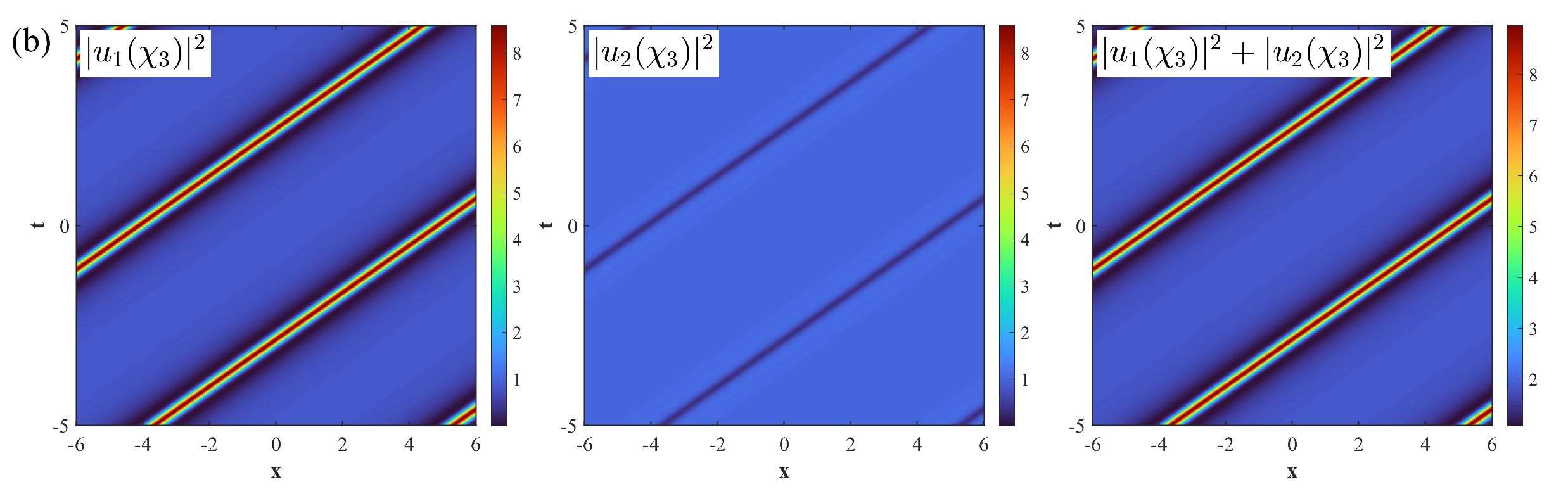}
	\caption{
		Representative solutions at $\mathrm{A}_3=(4,0.7)$ in the outer
		subregion of region~I for $\beta_1=0$.
		(a) $\chi_a=\chi_1=3.6254-0.8744i$,
		$\chi_b=4.3254-0.8744i$.
		(b) $\chi_a=\chi_2=-4.3254+0.8744i$,
		$\chi_b=-3.6254+0.8744i$.
		(c) $\chi_a=\chi_3=-0.3500-1.3077i$,
		$\chi_b=0.3500-1.3077i$.
		The branches represented by $\chi_1$ and $\chi_2$ support ABs,
		whereas $\chi_3$ satisfies
		$\operatorname{Re}(\chi_a)=-\gamma/2$ and lies on the
		state-transition locus, giving a periodic-wave state.
		In each row, the panels from left to right show
		$|u_1|^2$, $|u_2|^2$, and $|u_1|^2+|u_2|^2$.
	}
	\label{fig:kai123_ab}
\end{figure}
In region~I, $\Delta_{\mathrm c}<0$, and
Eq.~\eqref{eq:R-cubic-beta1zero} has one real root and one
complex-conjugate pair. The branches generated by the real root
also satisfy the admissibility condition, so all three
eigenvalue-branch pairs are admissible. This gives rise to the case $d_{\mathrm{nd}}=2$
 with $N_{\mathrm p}=3$, corresponding to the
nondegenerate regime.

\begin{figure}[tbp]
	\centering
	\includegraphics[width=\columnwidth]{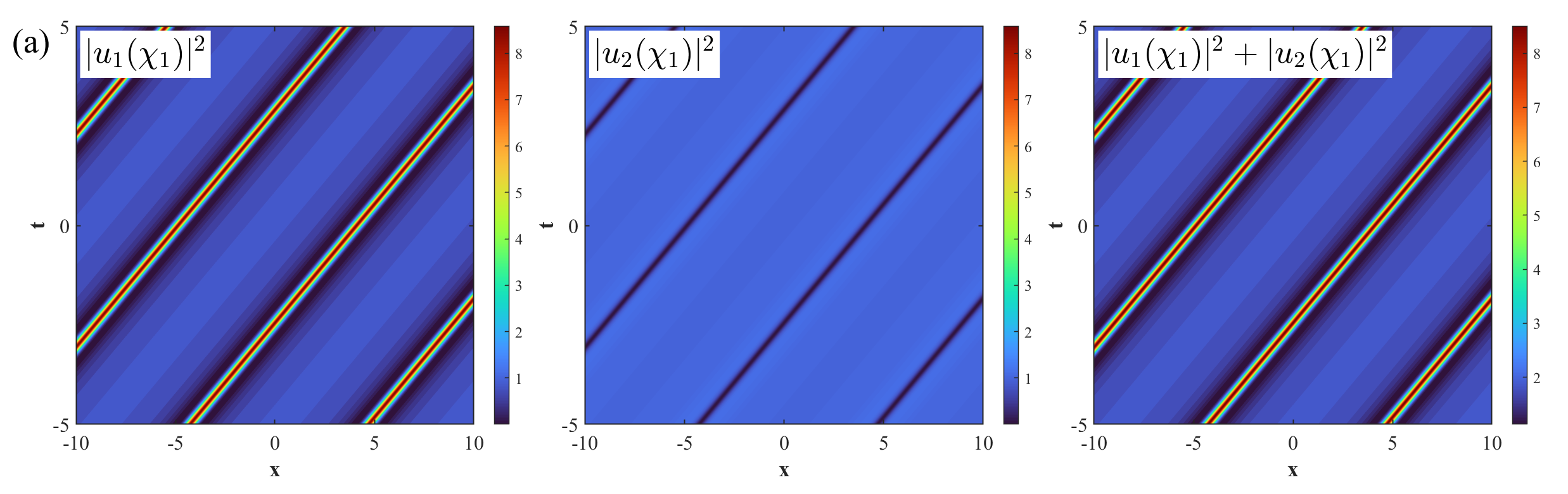}
	
	\includegraphics[width=\columnwidth]{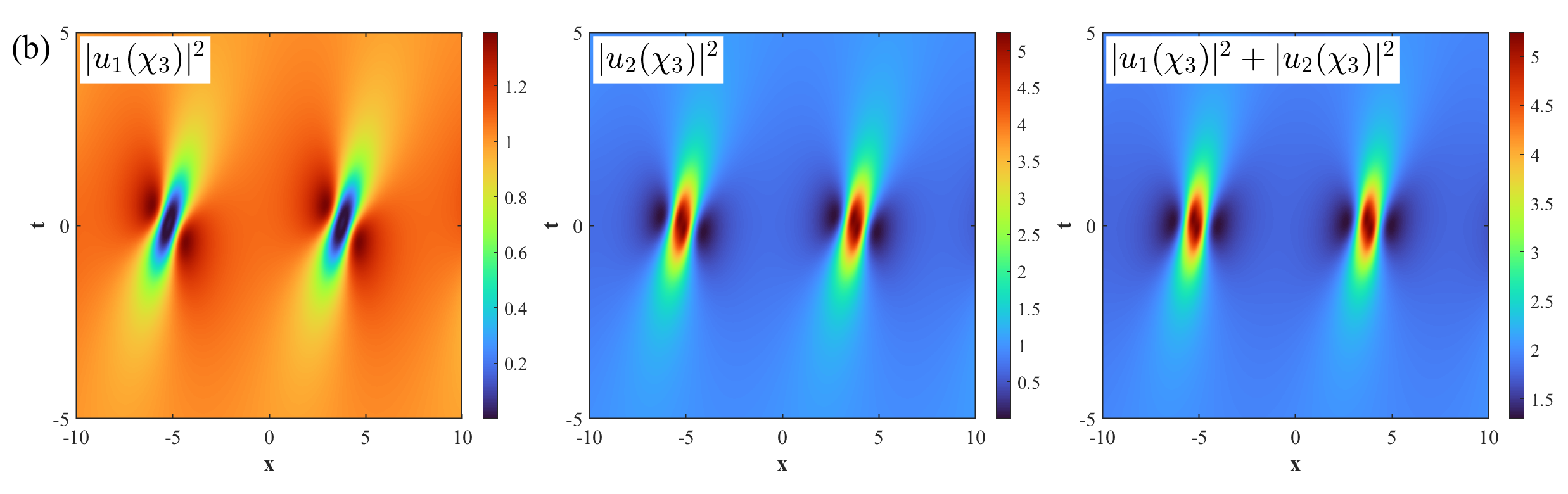}
	
	\includegraphics[width=\columnwidth]{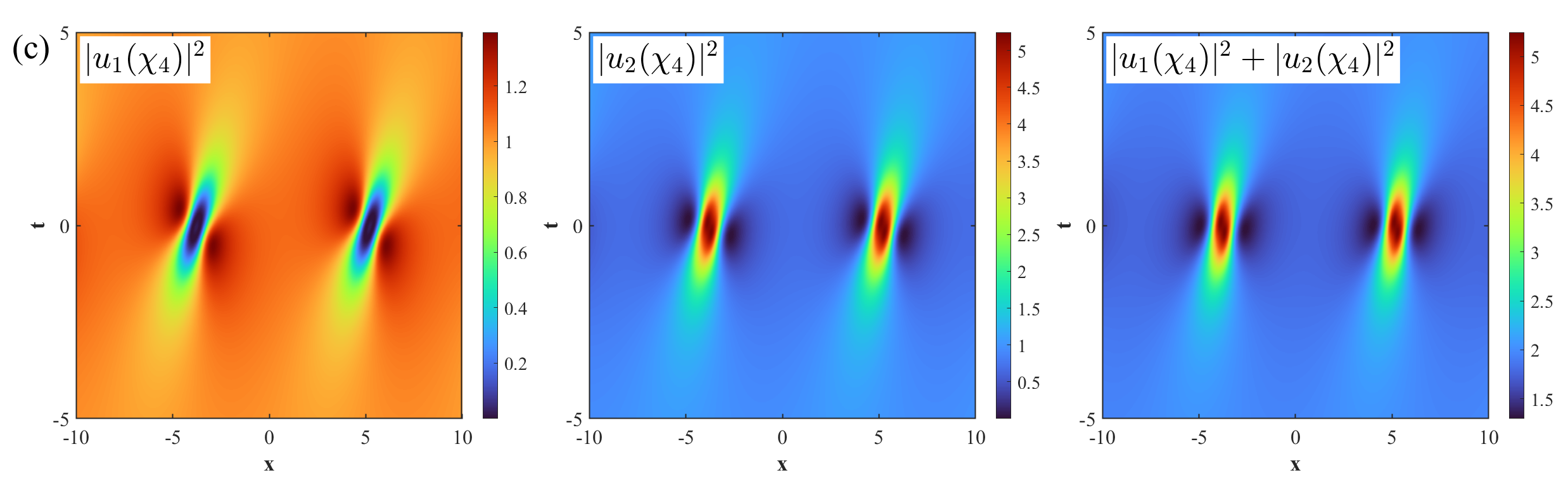}
	\caption{
		Representative solutions at $\mathrm{A}_4=(2,0.7)$ in an inner
		oval subregion of region~I for $\beta_1=0$.
		(a) $\chi_a=\chi_1=-0.3500+1.2744i$,
		$\chi_b=0.3500+1.2744i$.
		(b) $\chi_a=\chi_3=1.5334+0.7425i$,
		$\chi_b=2.2334+0.7425i$.
		(c) $\chi_a=\chi_4=-2.2334-0.7425i$,
		$\chi_b=-1.5334-0.7425i$.
		Here, the branch represented by $\chi_1$ satisfies
		$\operatorname{Re}(\chi_a)=-\gamma/2$ and lies on the
		state-transition locus, whereas those represented by
		$\chi_3$ and $\chi_4$ remain on AB branches.
		In each row, the panels from left to right show
		$|u_1|^2$, $|u_2|^2$, and $|u_1|^2+|u_2|^2$.
	}
	\label{fig:regionI-inner-134}
\end{figure}

The branch-resolved MI-gain maps in
Fig.~\ref{fig:MI-beta1-zero} further divide region~I into an outer
subregion and two symmetric inner oval domains. While the number of
admissible pairs remains unchanged, the state-transition pair
switches between these subregions: $(\chi_3,\chi_4)$ lies on the
state-transition locus in the outer part, whereas
$(\chi_1,\chi_2)$ does so inside the oval domains. The remaining
admissible pairs correspond to AB branches.
This branch switching is illustrated by the representative solutions
at $\mathrm{A}_3=(4,0.7)$ in the outer subregion and
$\mathrm{A}_4=(2,0.7)$ in an inner oval subregion, shown in
Figs.~\ref{fig:kai123_ab} and
\ref{fig:regionI-inner-134}, respectively.
At $\mathrm{A}_3$, the state-transition branch is represented by
$\chi_3$, whereas at $\mathrm{A}_4$ it switches to $\chi_1$.

Different admissible eigenvalue branches do not necessarily generate
inequivalent intensity structures. For example,
Figs.~\ref{fig:kai123_ab}(a) and
\ref{fig:kai123_ab}(b) are related via
\[
(x,t)\mapsto(-x,-t),
\]
corresponding to a $180^\circ$ rotation about the origin in the
$(x,t)$ plane. This equivalence can be traced to the symmetry of the
associated eigenvalue branches.

Let $\chi_{a,1}$ and $\chi_{a,2}$ denote the two representative
values of $\chi_a$. They obey
\begin{equation*}
	\chi_{a,1}+\chi_{a,2}=-\gamma.
\end{equation*}
Using $\chi_{b,j}=\chi_{a,j}+\gamma$, we obtain
\begin{equation*}
	\chi_{a,2}=-\chi_{b,1},
	\qquad
	\chi_{b,2}=-\chi_{a,1}.
\end{equation*}
For $\beta_1=0$, the spectral curve
\eqref{eq:lambda-chi-beta1-zero} satisfies
\begin{equation*}
	\lambda(-\chi)=-\lambda(\chi).
\end{equation*}
Together with
$\lambda(\chi_{a,j})=\lambda(\chi_{b,j})$, this yields
\begin{equation}
	\lambda_{a,2}
	=
	\lambda(\chi_{a,2})
	=
	\lambda(-\chi_{b,1})
	=
	-\lambda(\chi_{b,1})
	=
	-\lambda_{a,1}.
	\label{eq:lambda_eq}
\end{equation}
Thus, the two branches have opposite spectral parameters and the
same nonzero value of $|\operatorname{Im}\lambda|$, while their
modulus-squared profiles are related by the above spatiotemporal
transformation.

\begin{figure}[tbp]
	\centering
	\includegraphics[width=0.55\columnwidth]{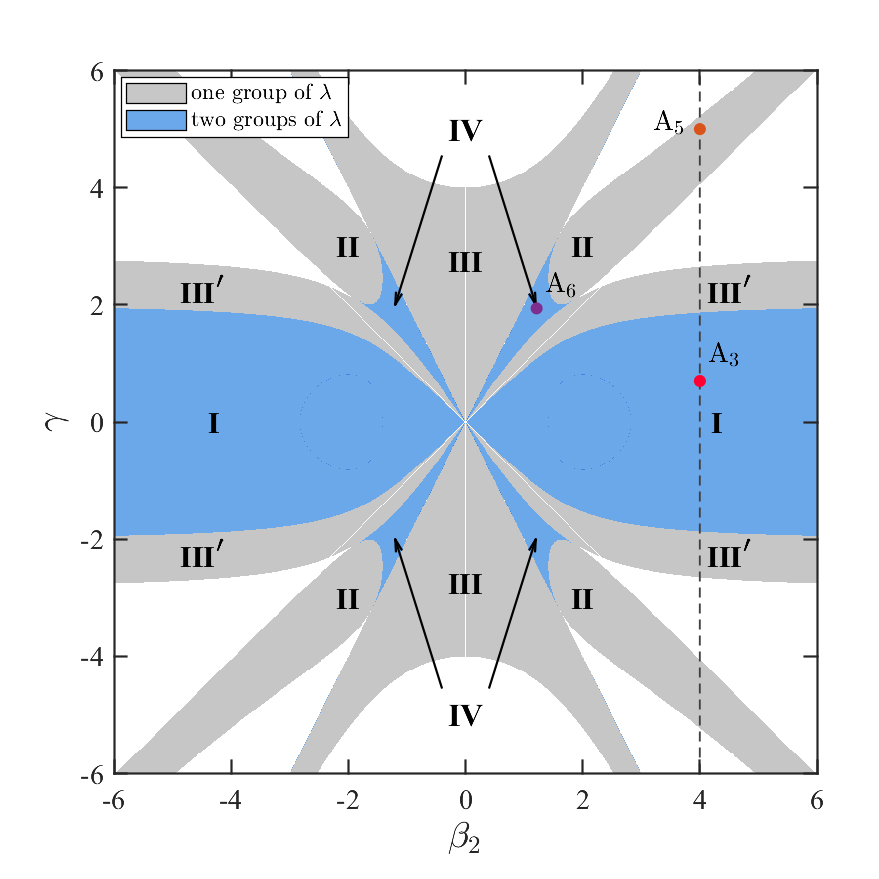}
	\caption{
		Structural-equivalence diagram in the $(\beta_2,\gamma)$ plane for
		$\beta_1=0$, with $c_1=c_2=1$ and $\alpha=0$.
		Symmetry-related branches sharing the same nonzero
		$|\operatorname{Im}\lambda|$ are grouped into
		structural-equivalence classes.
		Gray and blue denote one and two such classes, respectively;
		the color indicates only the number of classes, not the wave
		structure.
		The annotated coordinate points
		$\mathrm{A}_3=(4,0.7)$,
		$\mathrm{A}_5=(4,5)$, and
		$\mathrm{A}_6=(1.21402,1.93992)$
		are used in the representative analyses below.
		The dashed vertical line indicates the common value
		$\beta_2=4$ for $\mathrm{A}_3$ and $\mathrm{A}_5$.
	}
	\label{fig:lambda-group-beta1-zero}
\end{figure}

We therefore group branches connected by such transformations into
the same structural-equivalence class, conveniently labeled by their
common nonzero value of $|\operatorname{Im}\lambda|$. The resulting
classification in the $(\beta_2,\gamma)$ plane is shown in
Fig.~\ref{fig:lambda-group-beta1-zero}.
At $\mathrm{A}_3=(4,0.7)$, the branches
$\chi_1,\chi_2,\chi_5,\chi_6$ belong to one
structural-equivalence class, whereas $\chi_3$ and $\chi_4$
form the second class and lie on the state-transition locus,
producing the periodic-wave state exhibited in
Fig.~\ref{fig:kai123_ab}(c). Hence, although six admissible
$\chi$ branches are present at this point, they correspond to only
two inequivalent intensity structures: an AB structure and a
state-transition periodic-wave structure.

\paragraph*{Region~II.}
\begin{figure}[htbp]
	\centering
	\includegraphics[width=\columnwidth]{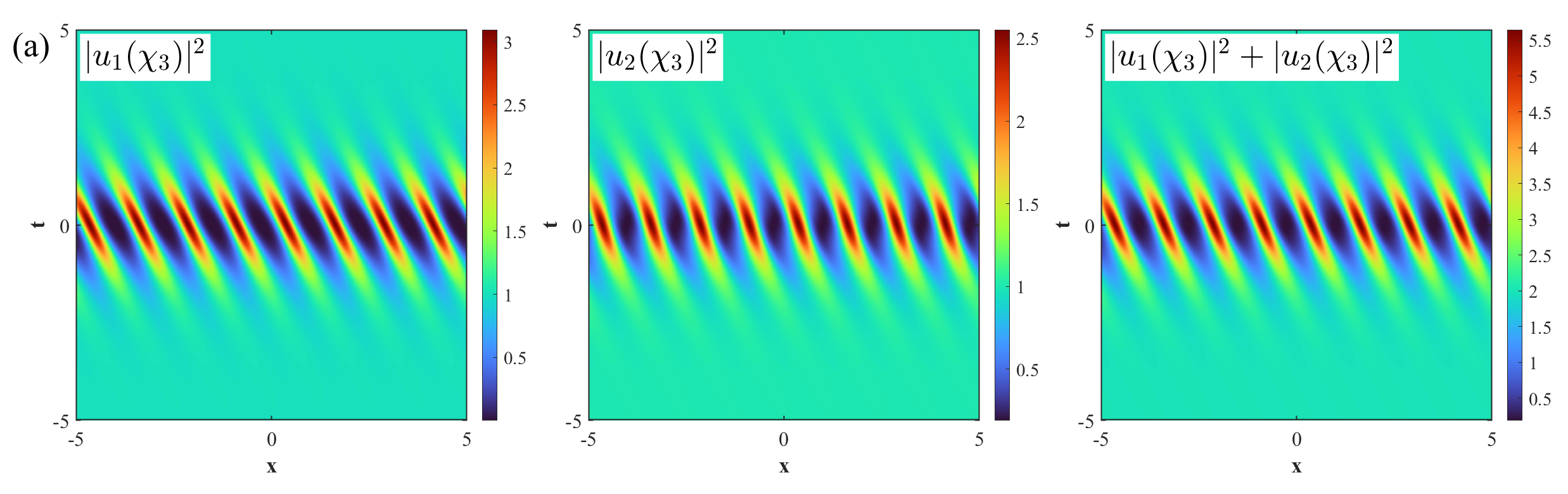}
	
	\includegraphics[width=\columnwidth]{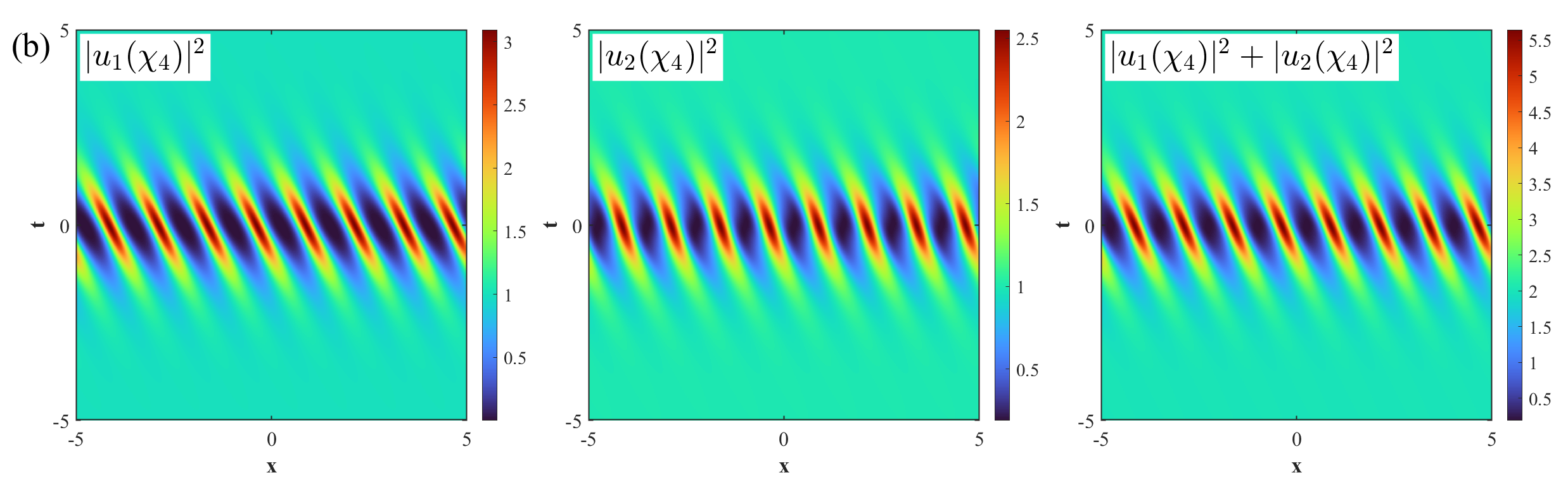}
	\caption{
		Vector ABs at $\mathrm{A}_5=(4,5)$ in region~II for
		$\beta_1=0$.
		(a) $\chi_a=\chi_3=-0.6116+0.2658i$,
		$\chi_b=4.3884+0.2658i$.
		(b) $\chi_a=\chi_4=-4.3884-0.2658i$,
		$\chi_b=0.6116-0.2658i$.
		Neither representative branch satisfies the state-transition
		condition, and both correspond to AB states.
		In each row, the panels from left to right show
		$|u_1|^2$, $|u_2|^2$, and $|u_1|^2+|u_2|^2$.
	}
	\label{fig:kai34_ab_regionII}
\end{figure}
Region~II likewise has $\Delta_{\mathrm c}<0$, but exhibits a
different admissibility pattern from region~I. The real root does not
produce an admissible branch, while the complex-conjugate roots
generate four admissible branches, $\chi_3$--$\chi_6$, grouped into
the two pairs $(\chi_3,\chi_6)$ and $(\chi_4,\chi_5)$.
Accordingly, $d_{\mathrm{nd}}=1$ with $N_{\mathrm p}=2$, and
region~II falls within the nondegenerate regime.

At the representative point $\mathrm{A}_5=(4,5)$, neither
admissible pair lies on the state-transition locus, so both support
AB solutions. Representative profiles for the branches $\chi_3$
and $\chi_4$ are shown in
Fig.~\ref{fig:kai34_ab_regionII}.

Although $\chi_3$ and $\chi_4$ represent different admissible
eigenvalue-branch pairs, the corresponding solutions are
structurally equivalent. Their spectral parameters differ only by a
sign, and the symmetry relation~\eqref{eq:lambda_eq} connects their
modulus-squared profiles through
$(x,t)\mapsto(-x,-t)$. The two pairs hence are grouped into the same
structural-equivalence class, consistent with
Fig.~\ref{fig:lambda-group-beta1-zero}.
\paragraph*{Region~IV.}

Region~IV, in contrast, lies in the $\Delta_{\mathrm c}>0$ sector,
where Eq.~\eqref{eq:R-cubic-beta1zero} possesses three distinct real
roots. Among them, two meet the admissibility condition and give the
eigenvalue-branch pairs $(\chi_3,\chi_4)$ and $(\chi_5,\chi_6)$.
The resulting branch count is $N_{\mathrm p}=2$, giving
$d_{\mathrm{nd}}=1$ and identifying region~IV as nondegenerate regime.

\begin{figure}[htbp]
	\centering
	\includegraphics[width=\columnwidth]{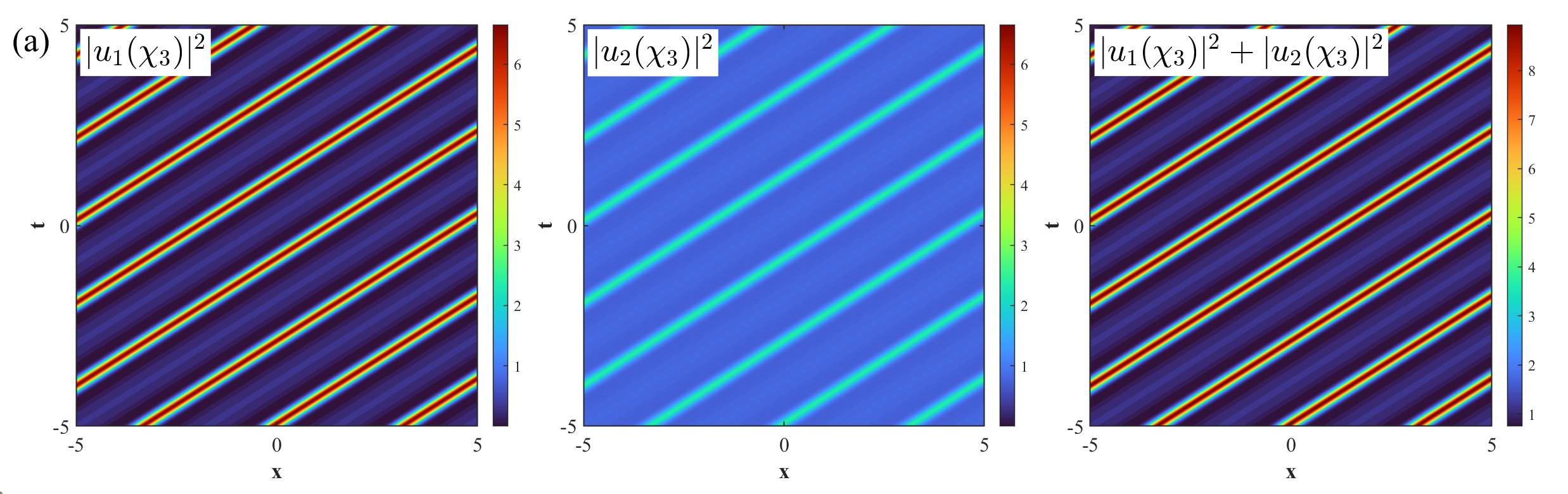}
	\includegraphics[width=\columnwidth]{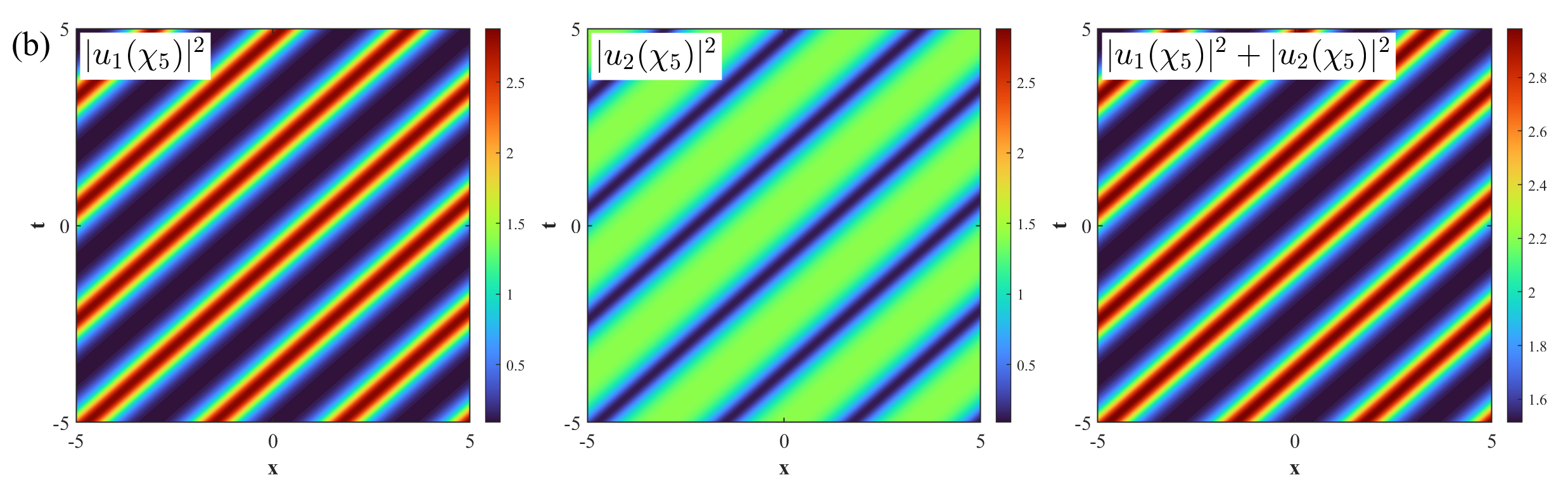}
	\caption{
		State-transition periodic waves at
		$\mathrm{A}_6=(1.21402,1.93992)$ in region~IV for
		$\beta_1=0$.
		(a) $\chi_a=\chi_3=-0.9700-1.2491i$,
		$\chi_b=0.9700-1.2491i$.
		(b) $\chi_a=\chi_5=-0.9700+0.3606i$,
		$\chi_b=0.9700+0.3606i$.
		The two representative branches both satisfy
		$\operatorname{Re}(\chi_a)=-\gamma/2$ and hence correspond to
		state-transition periodic waves.
		In each row, the panels from left to right show
		$|u_1|^2$, $|u_2|^2$, and $|u_1|^2+|u_2|^2$.
	}
	\label{fig:kai35_ab_regionIV}
	
\end{figure}

Unlike region~II, both admissible pairs lie on the
state-transition locus. At the representative point
$\mathrm{A}_6=(1.21402,1.93992)$, the localization variable
vanishes, $\Psi=0$, and both pairs thus cause periodic-wave
states. Their representative profiles are shown in
Fig.~\ref{fig:kai35_ab_regionIV}.

Despite both pairs correspond to state-transition periodic waves,
their intensity structures are distinct. As depicted in
Fig.~\ref{fig:lambda-group-beta1-zero}, the two admissible
eigenvalue-branch pairs belong to different nonzero
$|\operatorname{Im}\lambda|$ groups and hence to different
structural-equivalence classes. In consequence, they represent two
inequivalent periodic-wave structures.

\subsection{\label{subsec:phase-beta-opposite}
	Case II: $\beta_1=-\beta_2$}

\begin{figure}[htbp]
	\centering
	\includegraphics[width=0.55\columnwidth]{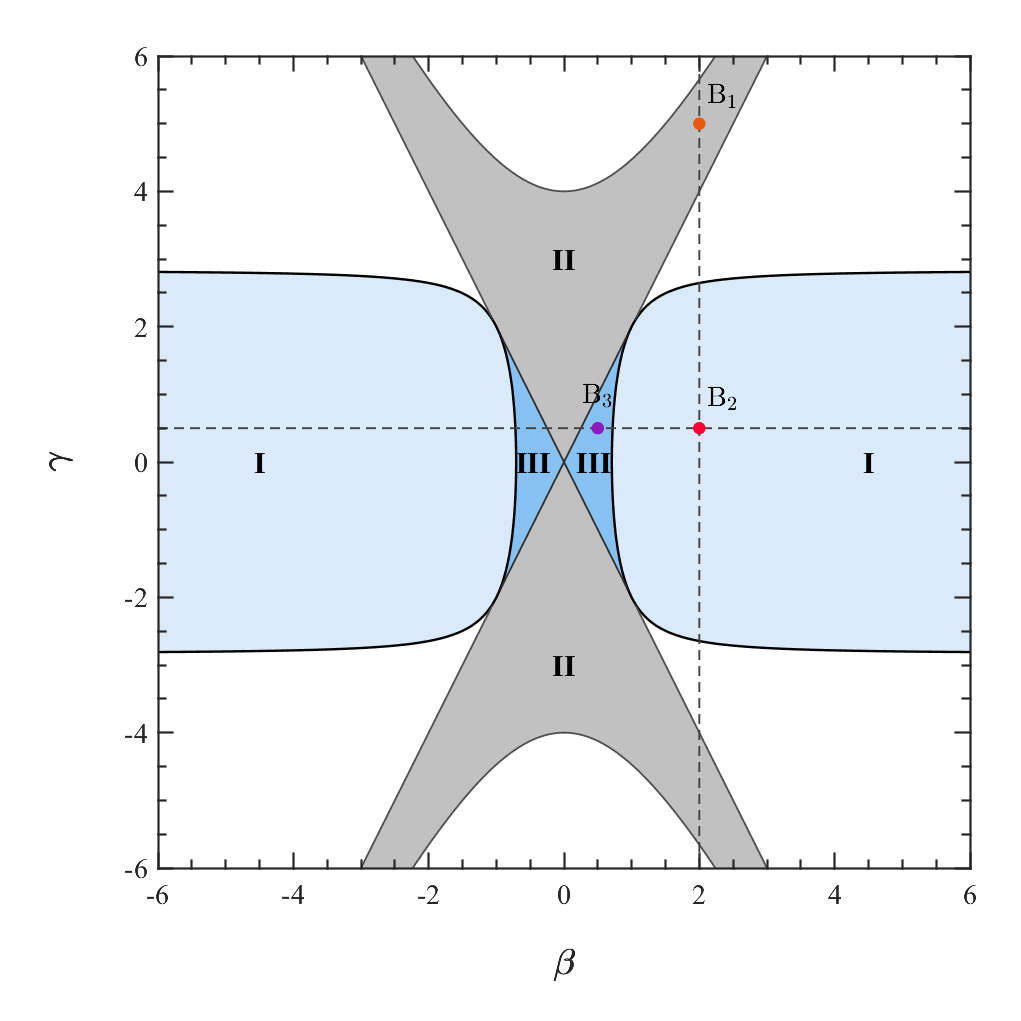}
	\caption{
		Existence diagram in the $(\beta,\gamma)$ plane for
		$\beta_1=-\beta_2=\beta$, with $c_1=c_2=1$ and $\alpha=0$.
		The same eigenvalue-branch classification applies to
		$\beta_1=\beta_2$; for $c_1=c_2$, the corresponding
		component intensities are equivalent as shown in Sec.~III B.
		Region~II (gray) admits one eigenvalue-branch pair, whereas
		regions~I and III (blue) admit two.
		The annotated coordinate points
		$\mathrm{B}_1=(2,5)$,
		$\mathrm{B}_2=(2,0.5)$, and
		$\mathrm{B}_3=(0.5,0.5)$
		are used in the analyses below.
	}
	\label{fig:existence-beta-opposite}
\end{figure}

We next specialize to the opposite-wavenumber case. Here, the admissible branches follow from
the quadratic reduction~\eqref{eq:R-quadratic-betaop} together with
the condition~\eqref{eq:betaop-admissible-condition}. For
$c_1=c_2=1$ and $\alpha=0$, their distribution in the
$(\beta,\gamma)$ plane is summarized in
Fig.~\ref{fig:existence-beta-opposite}.

\begin{figure}[tbp]
	\centering
	\includegraphics[width=0.95\columnwidth]
	{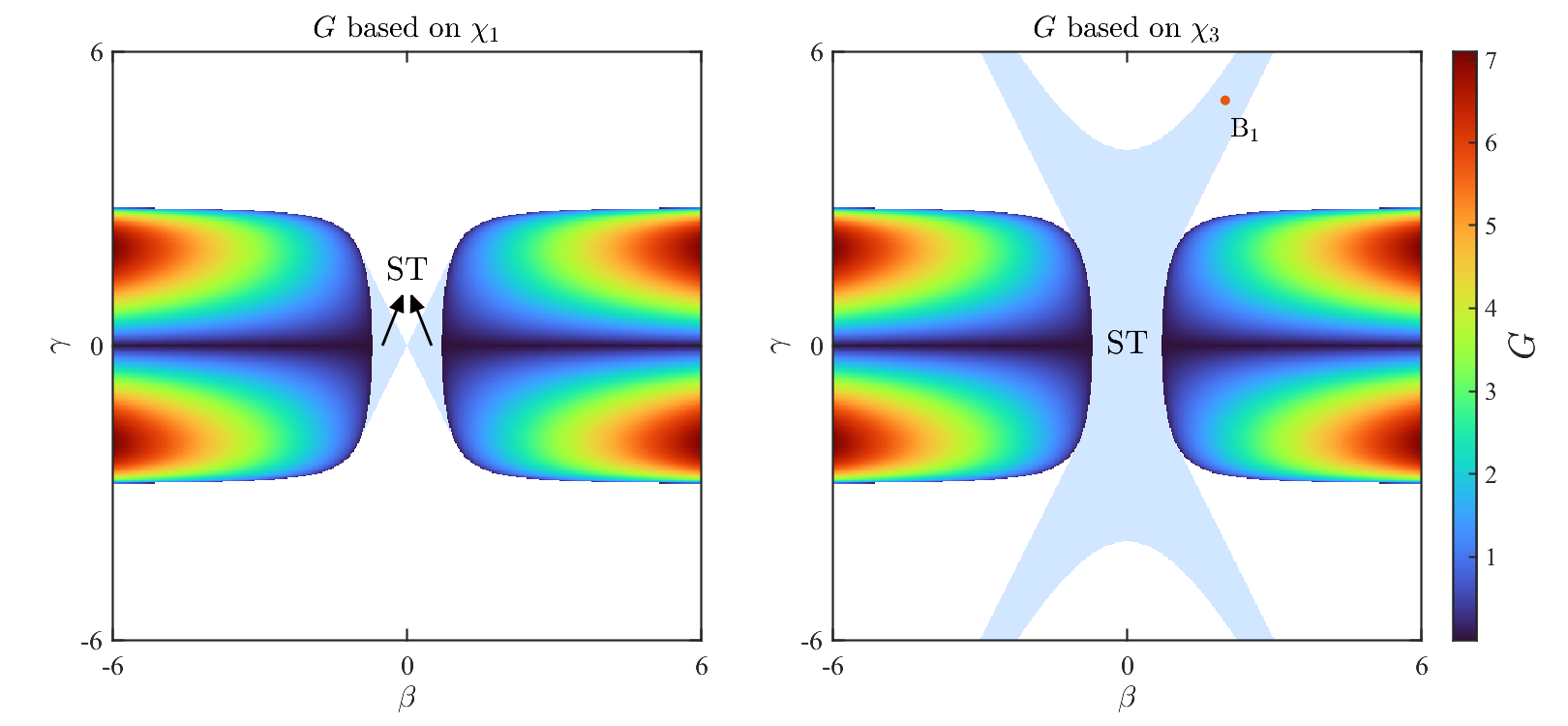}
	\caption{
		Branch-resolved MI gain in the $(\beta,\gamma)$ plane for the
		opposite-wavenumber configuration
		$\beta_1=-\beta_2=\beta$, with $c_1=c_2=1$ and $\alpha=0$.
		The left and right panels correspond to the representative
		branches $\chi_1$ and $\chi_3$, respectively.
		The regions labeled ``ST'' identify admissible branches satisfying
		the state-transition condition.
		The annotated coordinate point $\mathrm{B}_1=(2,5)$ in the $\chi_3$ panel
		corresponds to the state-transition periodic wave shown in
		Fig.~\ref{fig:betaop-regionII-kai3-ab}.
	}
	\label{fig:G-chi13-opposite-ST}
\end{figure}

Figure~\ref{fig:existence-beta-opposite} divides the
$(\beta,\gamma)$ plane into one degenerate region and two
nondegenerate ones. Region~II has a single admissible
eigenvalue-branch pair, giving $d_{\mathrm{nd}}=0$, whereas
regions~I and III each admit two pairs and hence have
$d_{\mathrm{nd}}=1$. Despite having the same degree of
nondegeneracy, regions~I and III differ in their branch character:
the former supports AB branches, while the latter is associated
with state-transition periodic waves.

The branch-resolved MI-gain maps in
Fig.~\ref{fig:G-chi13-opposite-ST} further specify the branch
assignment in each region, including the unique admissible pair
$(\chi_3,\chi_4)$ in region~II. We next use representative
 points to illustrate these distinct wave states and their
structural-equivalence properties.

\subsubsection{\label{subsubsec:degenerate-betaop-phase}
	State-transition periodic wave in region~II}

In region~II, Eq.~\eqref{eq:discriminant-betaop} gives $\Delta_R>0$, and Eq.~\eqref{eq:R-quadratic-betaop} consequently has two distinct real roots, ordered as \begin{equation*} R_1<R_2. \end{equation*} Only one of them meets the admissibility condition~\eqref{eq:betaop-admissible-condition}, leaving a single admissible eigenvalue-branch pair. The corresponding case $d_{\mathrm{nd}}=0$ with $N_{\mathrm p}=1$ identifies region~II as degenerate.

\begin{figure}[htbp]
	\centering
	\includegraphics[width=\columnwidth]{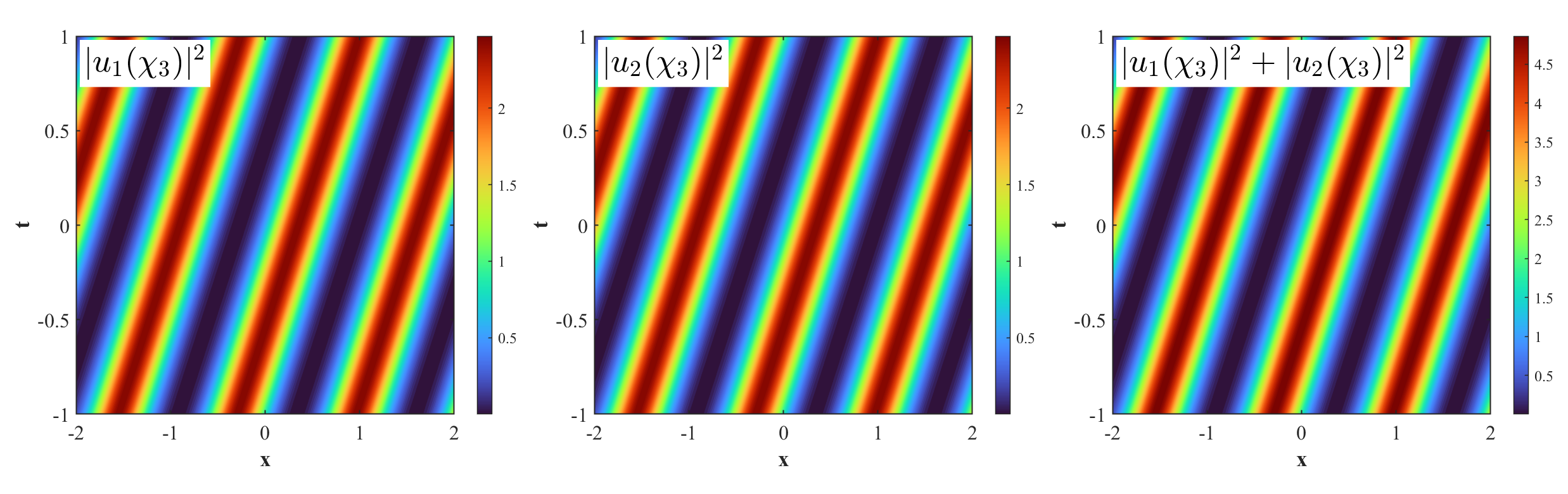}
	\caption{
		State-transition periodic wave at $\mathrm{B}_1=(2,5)$ in
		degenerate region~II for the opposite-wavenumber configuration.
		The selected branch is
		$\chi_a=\chi_3=-2.5000+0.4851i$, with
		$\chi_b=2.5000+0.4851i$.
		Since $\operatorname{Re}(\chi_a)=-\gamma/2$, this branch lies on
		the state-transition locus and reduces to a periodic-wave state.
		The three panels display, from left to right,
		$|u_1|^2$, $|u_2|^2$, and $|u_1|^2+|u_2|^2$.
	}
	\label{fig:betaop-regionII-kai3-ab}
\end{figure}

The branch-resolved MI-gain map in
Fig.~\ref{fig:G-chi13-opposite-ST} further identifies this pair as
$(\chi_3,\chi_4)$ and places it on the state-transition locus.
Consequently, $\Psi=0$, and the corresponding solution reduces to the
periodic-wave states. A representative example is shown in
Fig.~\ref{fig:betaop-regionII-kai3-ab}.

\subsubsection{\label{subsubsec:nondegenerate-betaop-phase}
	Nondegenerate solutions}

\paragraph*{Region~I.}

\begin{figure}[htbp]
	\centering
	\includegraphics[width=\columnwidth]{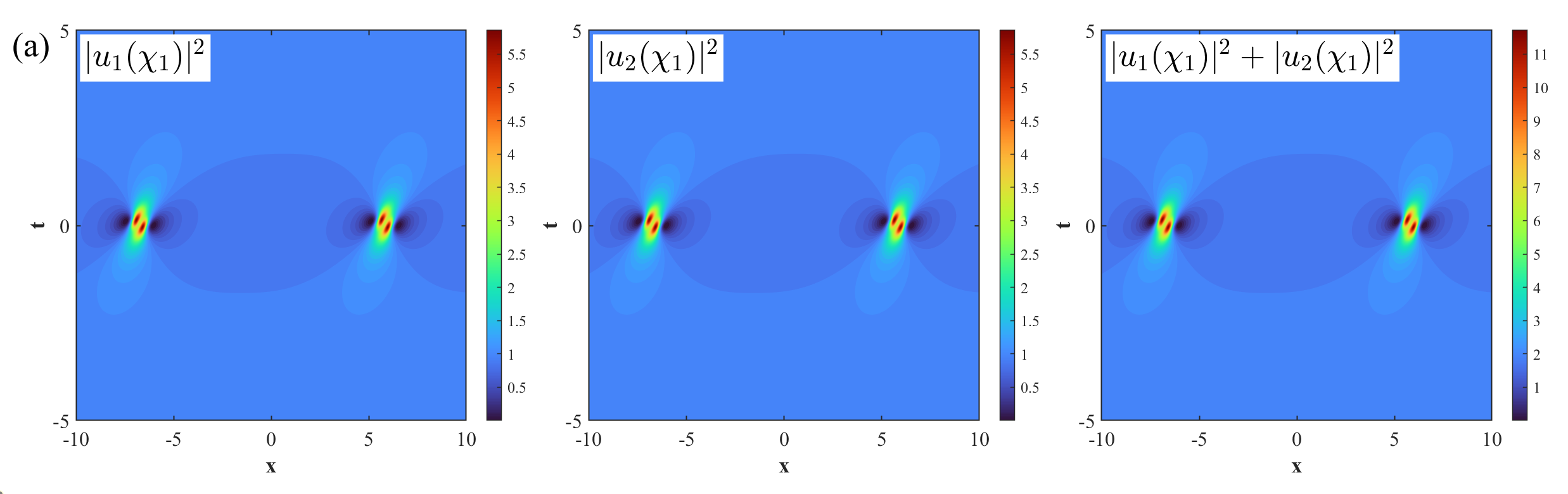}
	\includegraphics[width=\columnwidth]{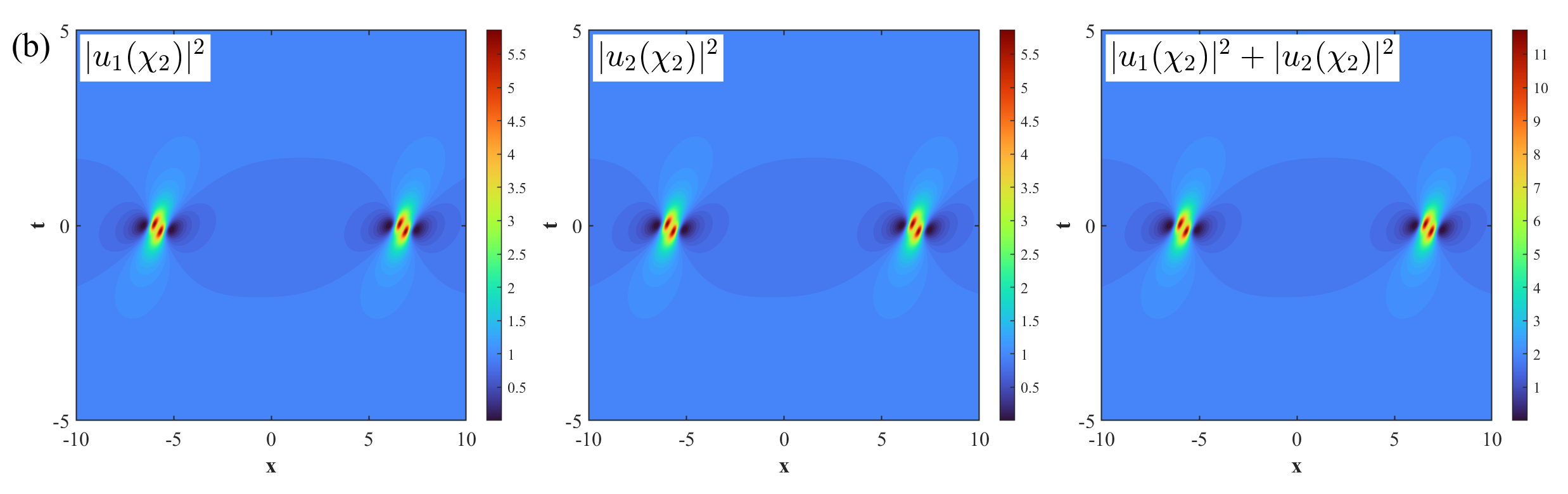}
	\caption{
		Vector ABs at $\mathrm{B}_2=(2,0.5)$ in region~I for the
		opposite-wavenumber configuration.
		(a) $\chi_a=\chi_1=1.7061+1.3282i$,
		$\chi_b=2.2061+1.3282i$.
		(b) $\chi_a=\chi_2=-2.2061-1.3282i$,
		$\chi_b=-1.7061-1.3282i$.
		Both branches retain nonzero localization and represent AB states
		away from the state-transition locus.
		In each row, the panels from left to right show
		$|u_1|^2$, $|u_2|^2$, and $|u_1|^2+|u_2|^2$.
	}
	\label{fig:betaop-regionI-kai12-ab}
	
\end{figure}

Region~I admits the two eigenvalue-branch pairs $(\chi_1,\chi_4)$ and $(\chi_2,\chi_3)$, yielding  $d_{\mathrm{nd}}=1$ with $N_{\mathrm p}=2$. We choose $\chi_1$ and $\chi_2$ as their representatives. At $\mathrm{B}_2=(2,0.5)$, both remain outside the state-transition locus and describe AB states, as illustrated in Fig.~\ref{fig:betaop-regionI-kai12-ab}. Despite this nondegenerate eigenvalue-branch structure, the two wave components display the same qualitative intensity pattern and thus describe a degenerate-type mode. This provides another example in which the eigenvalue-branch and mode classifications do not coincide.

\begin{figure}[htbp]
	\centering
	\includegraphics[width=0.55\columnwidth]{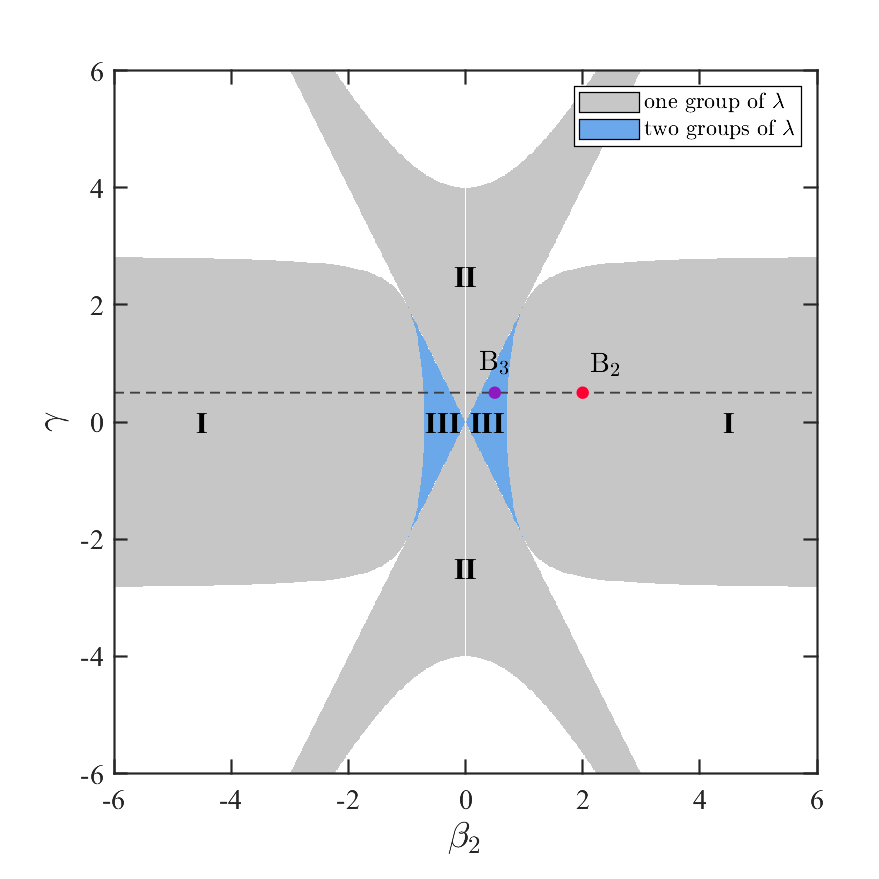}
	\caption{
		Structural-equivalence diagram in the $(\beta,\gamma)$ plane
		for the opposite-wavenumber configuration
		$\beta_1=-\beta_2=\beta$, with $c_1=c_2=1$ and $\alpha=0$.
		Gray and blue denote one and two structural-equivalence
		classes, respectively.
		The annotated coordinate points
		$\mathrm{B}_2=(2,0.5)$ and
		$\mathrm{B}_3=(0.5,0.5)$
		are used below.
	}
	\label{fig:lambda-group-betaop}
\end{figure}
The two representative solutions are nevertheless structurally
equivalent. For the opposite-wavenumber configuration, the spectral
curve obeys
\begin{equation*}
	\lambda(-\chi)=-\lambda(\chi),
\end{equation*}
so the symmetry relation~\eqref{eq:lambda_eq} derived above applies
directly. The corresponding spectral parameters differ only by a
sign, and their modulus-squared profiles are connected by
\begin{equation*}
	(x,t)\mapsto(-x,-t).
\end{equation*}
Hence, the two admissible eigenvalue-branch pairs at
$\mathrm{B}_2$ belong to the same structural-equivalence class.
The structural-equivalence classification over the
$(\beta,\gamma)$ plane is summarized in
Fig.~\ref{fig:lambda-group-betaop}.

\paragraph*{Region~III.}
\begin{figure}[htbp]
	\centering
	\includegraphics[width=\columnwidth]{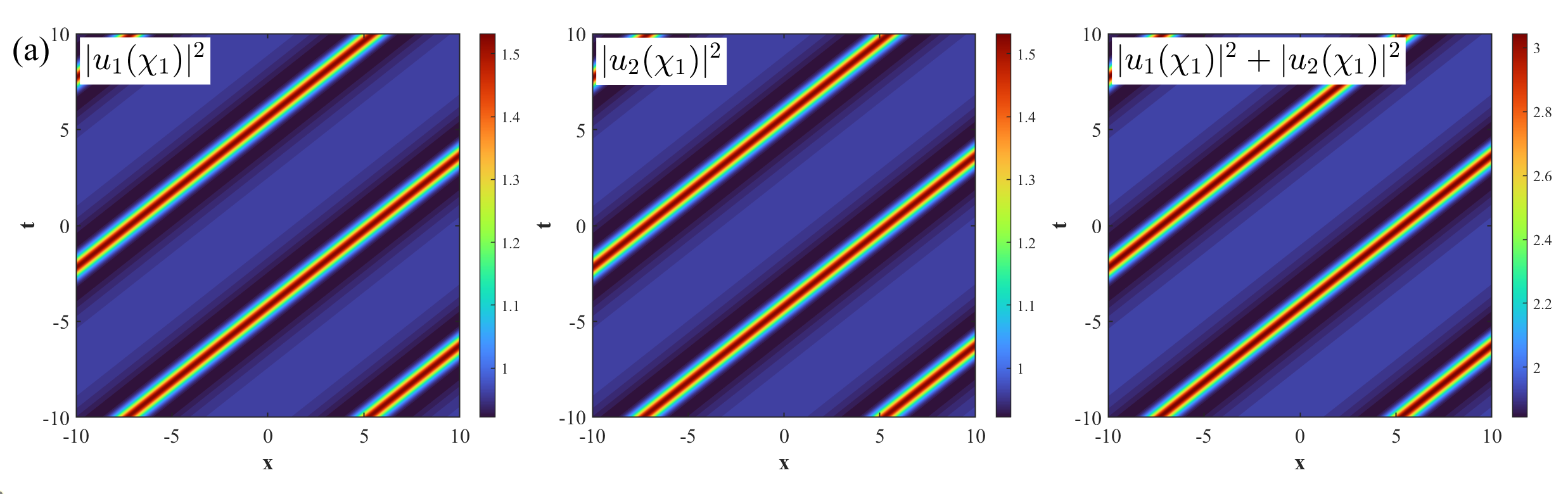}
	
	\includegraphics[width=\columnwidth]{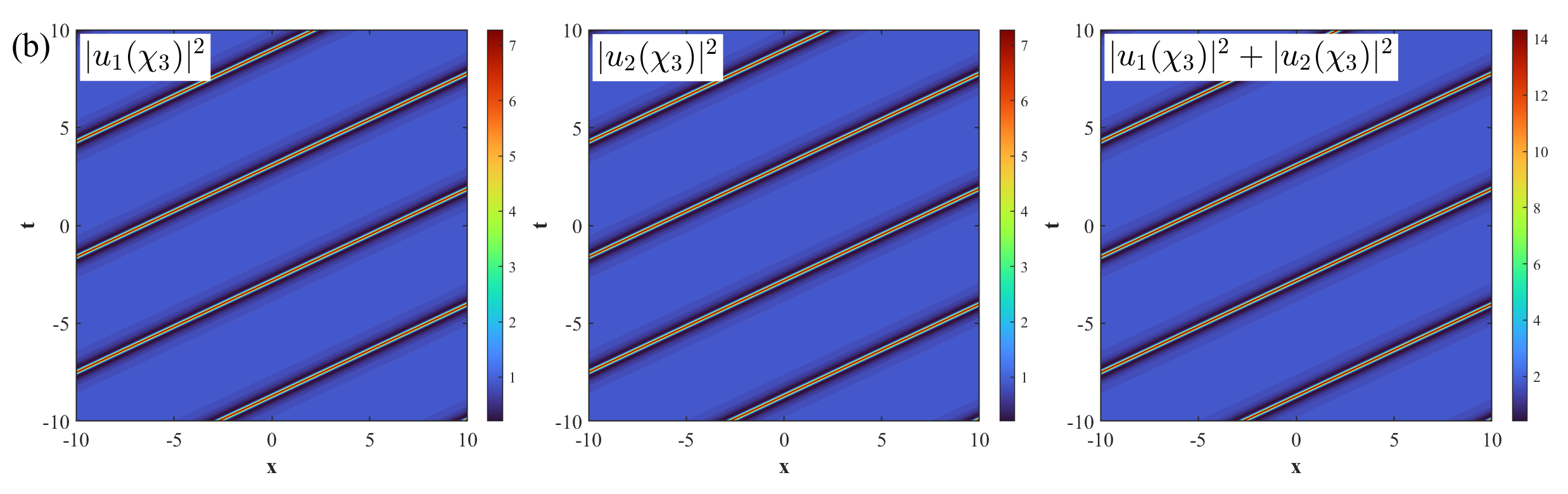}
	\caption{
		State-transition periodic waves at $\mathrm{B}_3=(0.5,0.5)$
		in region~III for the opposite-wavenumber configuration.
		(a) $\chi_a=\chi_1=-0.2500+0.5014i$,
		$\chi_b=0.2500+0.5014i$.
		(b) $\chi_a=\chi_3=-0.2500+1.7674i$,
		$\chi_b=0.2500+1.7674i$.
		The state-transition condition is satisfied by both representative
		branches.
		In each row, the panels from left to right show
		$|u_1|^2$, $|u_2|^2$, and $|u_1|^2+|u_2|^2$.
	}
	\label{fig:betaop-regionIII-kai13-ab}
\end{figure}
In region~III, both real roots of
Eq.~\eqref{eq:R-quadratic-betaop} meet the admissibility
condition and yield the two eigenvalue-branch pairs
$(\chi_1,\chi_2)$ and $(\chi_3,\chi_4)$.
Thus, one has $d_{\mathrm{nd}}=1$ with $N_{\mathrm p}=2$, as in region~I. The distinction is that both admissible pairs now
lie on the state-transition locus, so that $\Psi=0$ and the
corresponding states appear in the form of periodic waves. Representative profiles at
$\mathrm{B}_3=(0.5,0.5)$ are depicted in
Fig.~\ref{fig:betaop-regionIII-kai13-ab}.

Unlike the two admissible pairs at $\mathrm{B}_2$, which belong to
a single structural-equivalence class, those at $\mathrm{B}_3$
fall within two distinct nonzero $|\operatorname{Im}\lambda|$ groups
in Fig.~\ref{fig:lambda-group-betaop}. They represent two
structurally inequivalent periodic-wave states. Thus, the regions~I and
III have the same degree of nondegeneracy but differ in both their
wave character and structural-equivalence classification.

\subsection{\label{subsec:phase-general}
	Case III: $\beta_1\beta_2\neq0,\,\,
	\beta_1^2\neq\beta_2^2$}
	\begin{figure}[htbp]
		\centering
		\includegraphics[width=0.55\columnwidth]{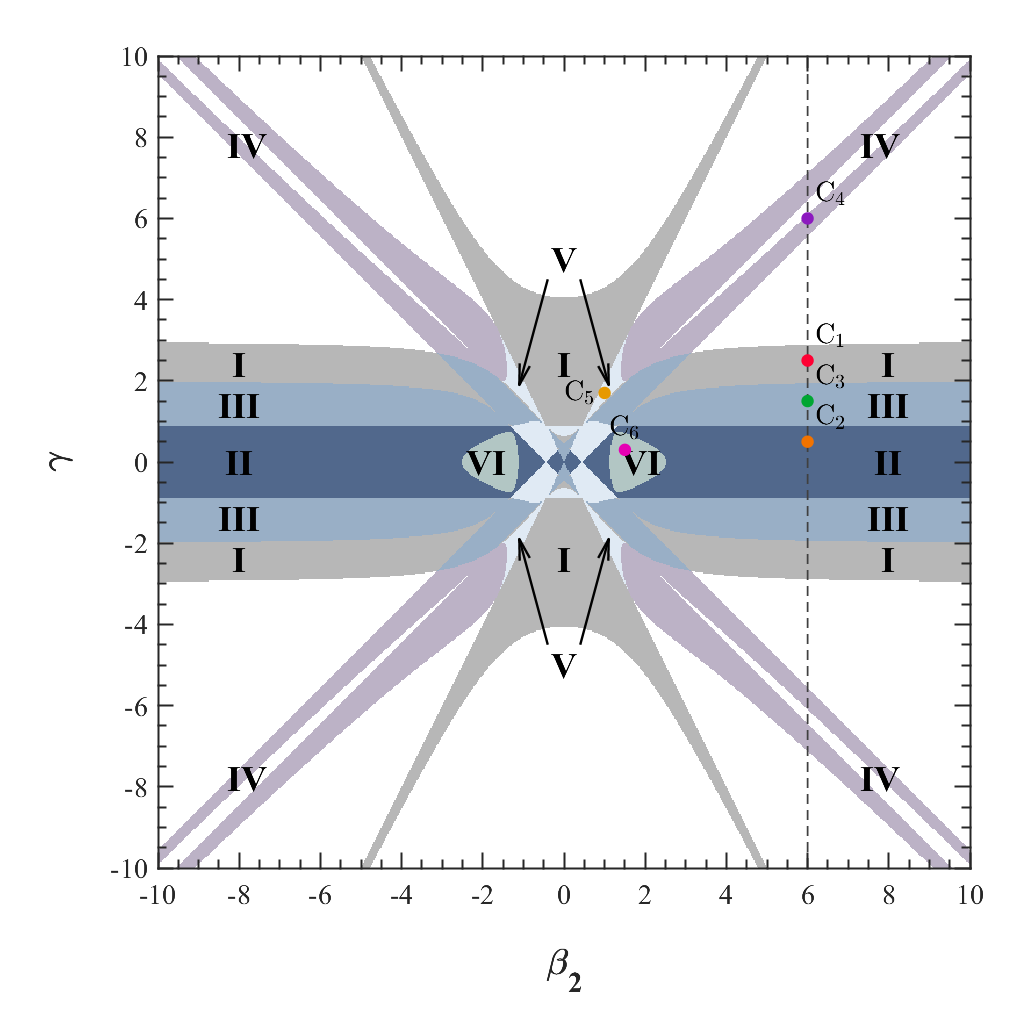}
		\caption{
			Existence diagram in the $(\beta_2,\gamma)$ plane for
			$\beta_1=0.45$, with $c_1=c_2=1$ and $\alpha=0$.
			Region~I (gray) is degenerate, whereas regions~II--VI are
			nondegenerate.
			The annotated coordinate points
			$\mathrm{C}_1=(6,2.5)$,
			$\mathrm{C}_2=(6,0.5)$,
			$\mathrm{C}_3=(6,1.5)$,
			$\mathrm{C}_4=(6,6)$,
			$\mathrm{C}_5=(1,1.7)$, and
			$\mathrm{C}_6=(1.5,0.3)$
			are used below.
			The dashed line marks $\beta_2=6$ for
			$\mathrm{C}_1$--$\mathrm{C}_4$.
		}
		\label{fig:phase-general-b1045}
	\end{figure}
	\begin{figure*}[!t]
		\centering
		\begin{tabular}{ccc}
			\includegraphics[width=0.23\textwidth]{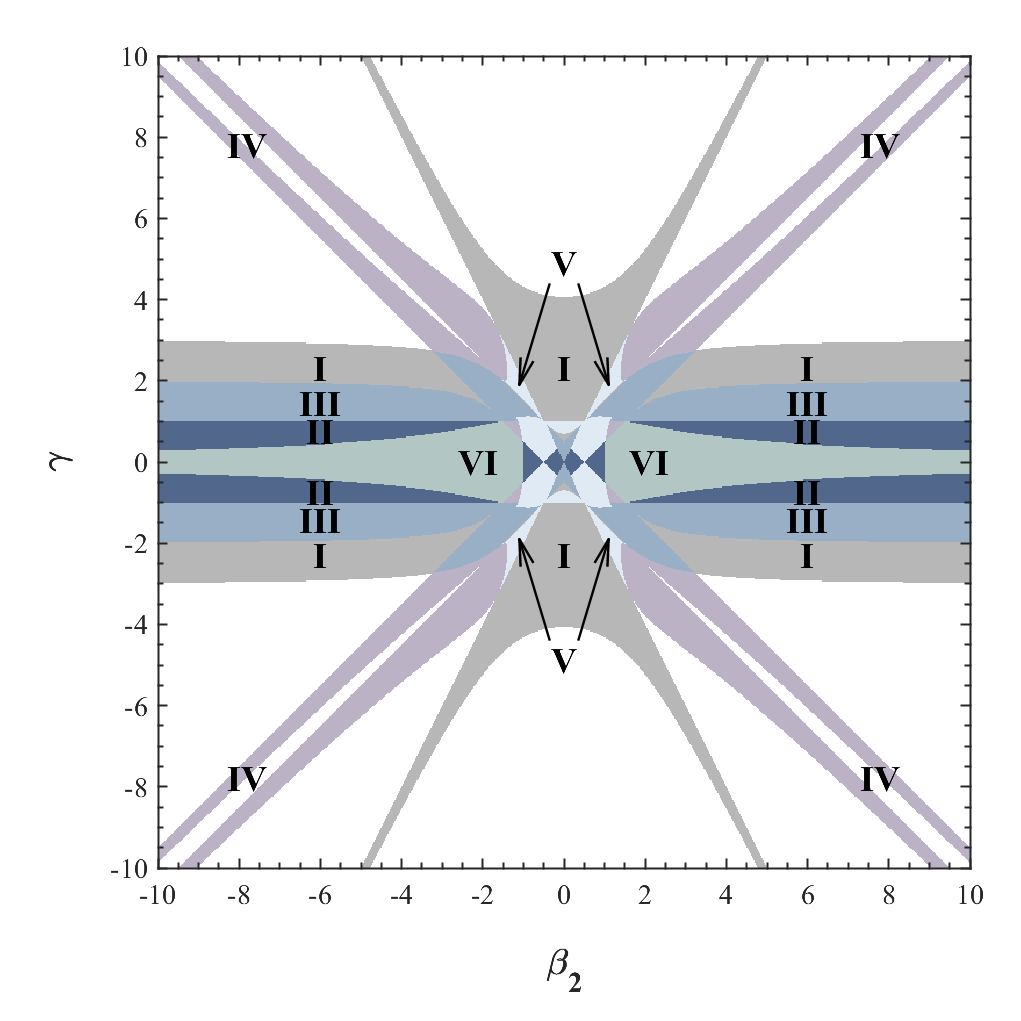} &
			\includegraphics[width=0.23\textwidth]{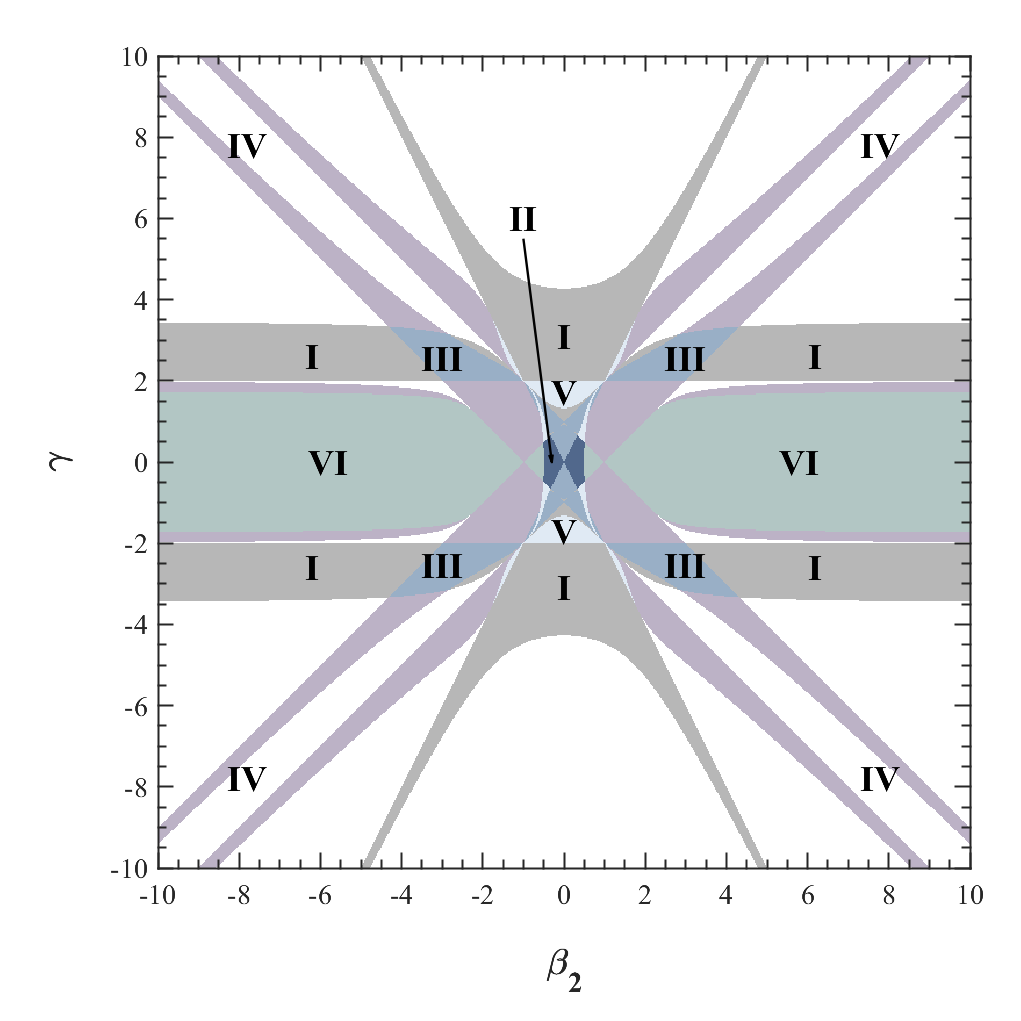} &
			\includegraphics[width=0.23\textwidth]{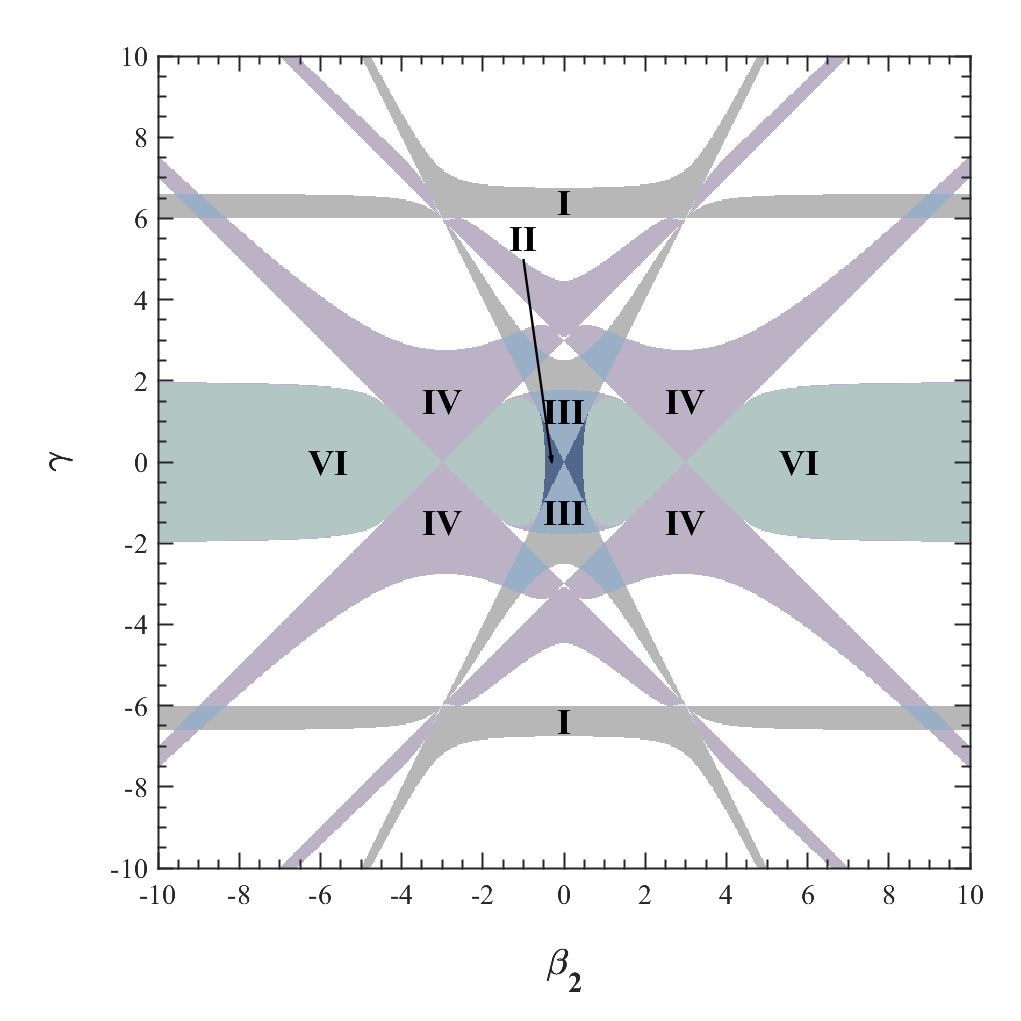} \\
			(a) $\beta_1=0.5$ &
			(b) $\beta_1=1.0$ &
			(c) $\beta_1=3$
		\end{tabular}
		\caption{
			Existence diagrams for different fixed values of $\beta_1$:
			(a) $\beta_1=0.5$, (b) $\beta_1=1.0$, and
			(c) $\beta_1=3$.
		}
		\label{fig:phase-diagrams-general}
	\end{figure*}
For two nonzero carrier wavenumbers with unequal magnitudes, the
admissible eigenvalue branches are governed by the quartic
equation~\eqref{eq:R-quartic-general} together with the
admissibility condition~\eqref{eq:complex-condition-general}.
Unlike the cubic and quadratic reductions considered above, the
quartic discriminant alone does not uniquely determine the root
configuration. For real coefficients, a positive discriminant can
correspond either to four distinct real roots or to two
complex-conjugate pairs. The classification therefore requires both
the root structure of Eq.~\eqref{eq:R-quartic-general} and the
admissibility of the resulting eigenvalue branches.

For $c_1=c_2=1$, $\alpha=0$, and fixed $\beta_1=0.45$, the
existence diagram in the $(\beta_2,\gamma)$ plane is shown in
Fig.~\ref{fig:phase-general-b1045}. Region~I supports only one
eigenvalue-branch pair and hence $d_{\mathrm{nd}}=0$; this pair
lies on the state-transition locus and governs a periodic-wave state.
The other regions are nondegenerate, but their admissible-branch
structures differ substantially. Region~II consists of four pairs
($d_{\mathrm{nd}}=3$), with two on AB branches and two on the
state-transition locus. In region~III, three pairs are admissible
($d_{\mathrm{nd}}=2$), comprising two AB pairs and one
state-transition pair. Regions~IV and V each involve two pairs
($d_{\mathrm{nd}}=1$): both remain on AB branches in region~IV,
whereas both undergo the state transition in region~V. Finally,
region~VI supports four admissible pairs
($d_{\mathrm{nd}}=3$), all associated with AB states.

The topology of the existence diagram depends on the fixed
carrier wavenumber $\beta_1$, as illustrated by the three
representative cases in Fig.~\ref{fig:phase-diagrams-general}.
For $\beta_1=0.5$ [Fig.~\ref{fig:phase-diagrams-general}(a)],
region~II is divided by region~VI, while regions~I and III preserve
their horizontal band-shaped structures. At $\beta_1=1.0$
[Fig.~\ref{fig:phase-diagrams-general}(b)], region~VI occupies a
larger portion of the lateral parameter domain, region~II becomes
more confined to the central part, and the two horizontal bands of
region~I separate from region~VI. For the larger representative
value $\beta_1=3$ [Fig.~\ref{fig:phase-diagrams-general}(c)],
region~V is absent, region~III appears near the center rather than
forming horizontal bands, and region~IV separates into two
X-shaped domains with branch points at $(\beta_1,0)$ and
$(-\beta_1,0)$. These examples demonstrate the pronounced
dependence of the existence topology on $\beta_1$.

\begin{figure}[tbp]
	\centering
	\includegraphics[width=\columnwidth]{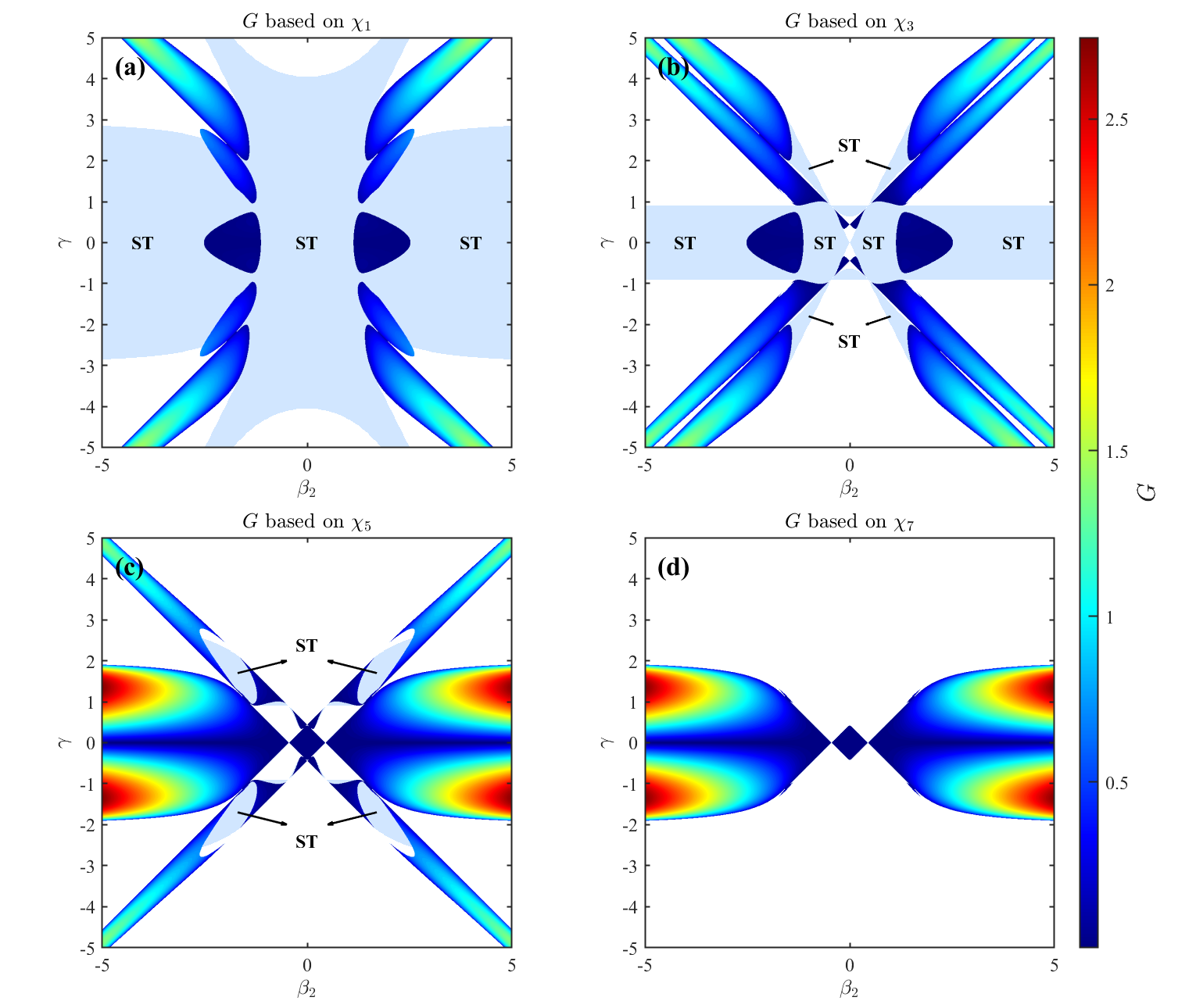}
	\caption{
		Branch-resolved MI gain in the $(\beta_2,\gamma)$ plane for
		two nonzero carrier wavenumbers of unequal magnitudes, with
		$c_1=c_2=1$, $\alpha=0$, and $\beta_1=0.45$.
		One representative from each eigenvalue-branch pair is shown:
		(a) $\chi_1$, (b) $\chi_3$, (c) $\chi_5$, and
		(d) $\chi_7$.
		The regions labeled ``ST'' indicate where the corresponding
		admissible branches obey the state-transition condition.
	}
	\label{fig:MI-general-b1045}
\end{figure}
To further distinguish the individual branch character, we calculate the branch-resolved MI gain for $\beta_1=0.45$, as presented in Fig.~\ref{fig:MI-general-b1045}. These maps determine which admissible branches satisfy the state-transition condition and which retain the AB character in each parameter region. Because the multibranch structure is considerably more involved here, we do not list every branch assignment separately; the classification follows the same admissibility and MI-gain criteria established above. We subsequently use $\beta_1=0.45$ to illustrate the characteristic wave structures in regions~I--VI.

\subsubsection{\label{subsubsec:degenerate-general-phase}
	State-transition periodic wave in region~I}

In region~I, Eq.~\eqref{eq:R-quartic-general} has four distinct real
roots, but only one obeys the admissibility
condition~\eqref{eq:complex-condition-general}. Hence, we have $d_{\mathrm{nd}}=0$ with $N_{\mathrm p}=1$, placing the region
in the degenerate regime.

\begin{figure}[tbp]
	\centering
	\includegraphics[width=\columnwidth]{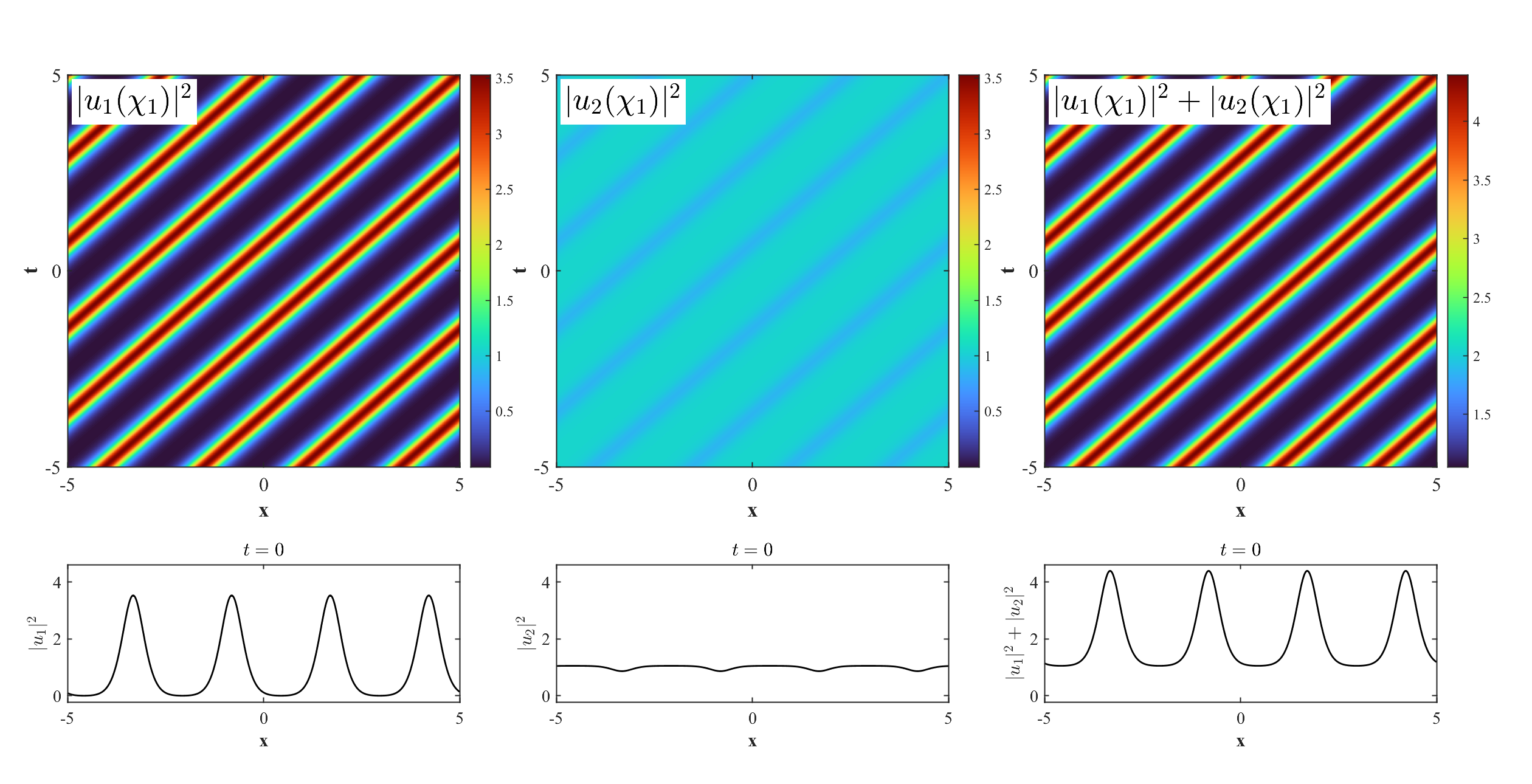}
	\caption{
		State-transition periodic wave at $\mathrm{C}_1=(6,2.5)$ in
		degenerate region~I for $\beta_1=0.45$.
		The representative branch is
		$\chi_a=\chi_1=-1.2500-0.6025i$, with
		$\chi_b=1.2500-0.6025i$.
		It satisfies $\operatorname{Re}(\chi_a)=-\gamma/2$, yielding
		$\Psi=0$ and hence a periodic-wave state.
		The upper row shows, from left to right,
		$|u_1|^2$, $|u_2|^2$, and $|u_1|^2+|u_2|^2$,
		while the lower row gives the corresponding $t=0$ cross sections.
	}
	\label{fig:general-regionI-kai1-ab}
\end{figure}

The unique admissible eigenvalue-branch pair also satisfies the
state-transition condition~\eqref{eq:state-transition-condition},
so that $\Psi=0$ and the corresponding solution reduces to a
periodic-wave state. A representative example at
$\mathrm{C}_1=(6,2.5)$ is exhibited in
Fig.~\ref{fig:general-regionI-kai1-ab}. Although region~I is
degenerate according to the eigenvalue-branch classification, the
two components exhibit distinct intensity modulations: the maxima
of $|u_1|^2$ approximately coincide with the minima of $|u_2|^2$.
The resulting periodic wave displays a nondegenerate-type
mode.

\subsubsection{\label{subsubsec:nondegenerate-general-phase}
	Nondegenerate solutions}

\paragraph*{Region~II.}
\begin{figure}[htbp]
	\centering
	\includegraphics[width=\columnwidth]{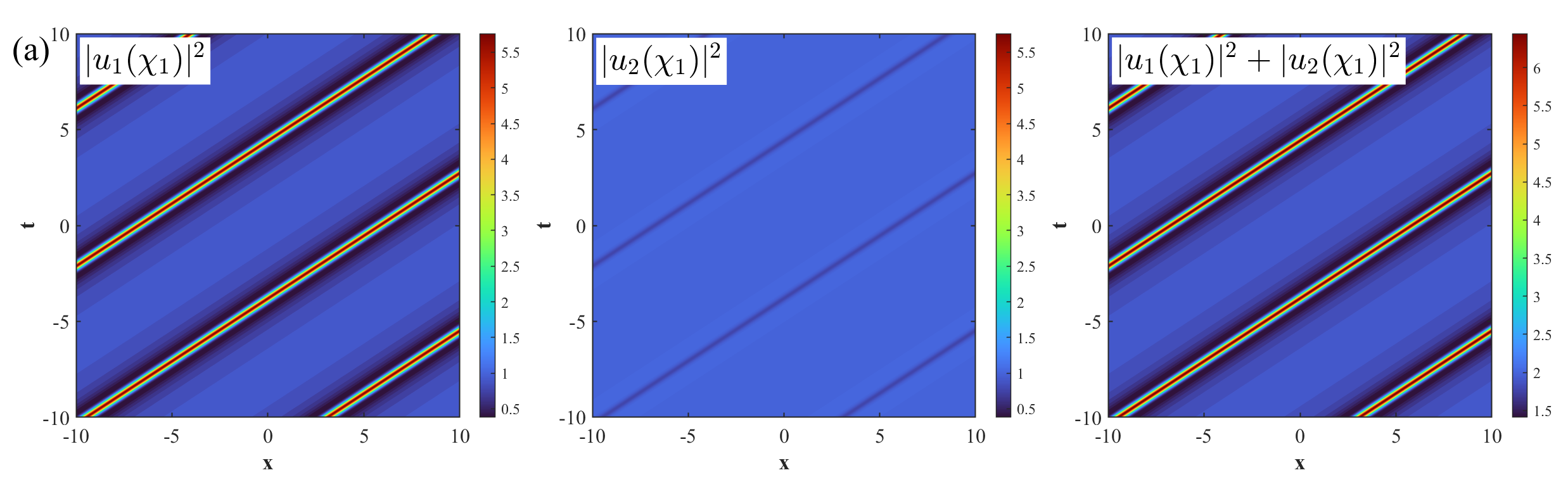}
	\includegraphics[width=\columnwidth]{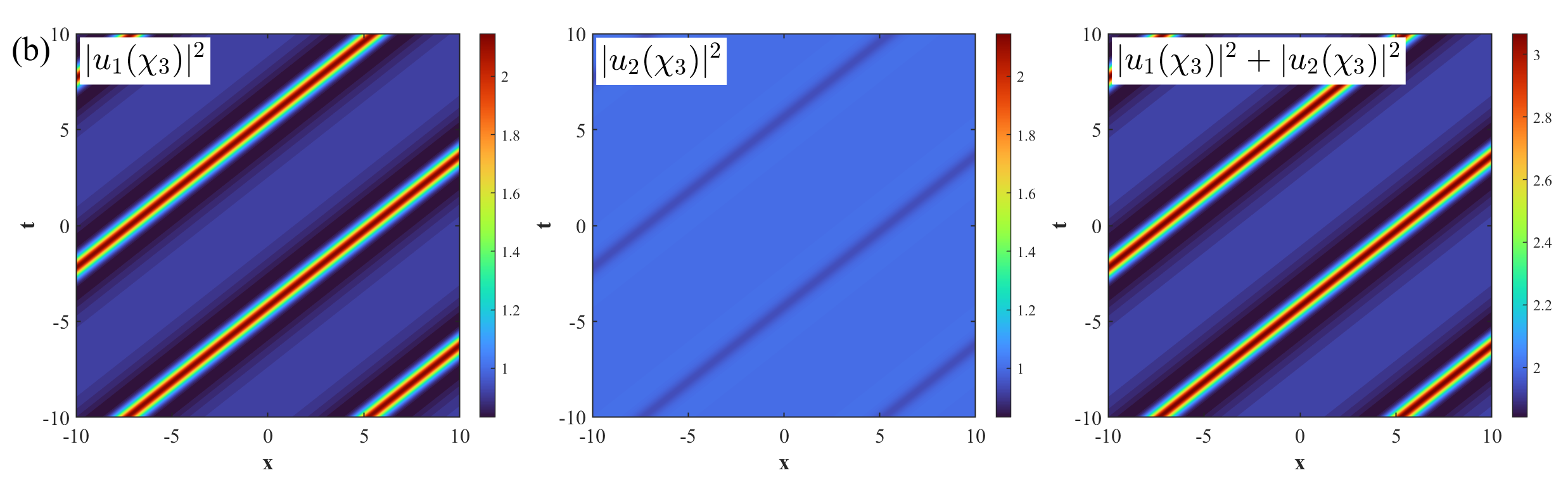}
	\caption{
		State-transition periodic waves at
		$\mathrm{C}_2=(6,0.5)$ in region~II for
		$\beta_1=0.45$.
		(a) $\chi_a=\chi_1=-0.2500+1.0571i$,
		$\chi_b=0.2500+1.0571i$.
		(b) $\chi_a=\chi_3=-0.2500+0.5069i$,
		$\chi_b=0.2500+0.5069i$.
		Both representative branches satisfy
		$\operatorname{Re}(\chi_a)=-\gamma/2$.
		In each row, the panels from left to right show
		$|u_1|^2$, $|u_2|^2$, and $|u_1|^2+|u_2|^2$.
	}
	\label{fig:general-regionII-kai13-ab}
	
\end{figure}
Region~II is characterized by two real roots ($R_1$ and
$R_2$), and one complex-conjugate pair 
($R_4=R_3^\ast$) of Eq.~\eqref{eq:R-quartic-general}.
Both real
roots obeys the admissibility condition and give rise to the pairs
$(\chi_1,\chi_2)$ and $(\chi_3,\chi_4)$. Since the admissible
branches associated with real $R_j$ satisfy
$\operatorname{Re}(\chi_a)=-\gamma/2$, these two pairs lie on the
state-transition locus and yield $\Psi=0$.

\begin{figure}[tbp]
	\centering
	\includegraphics[width=\columnwidth]{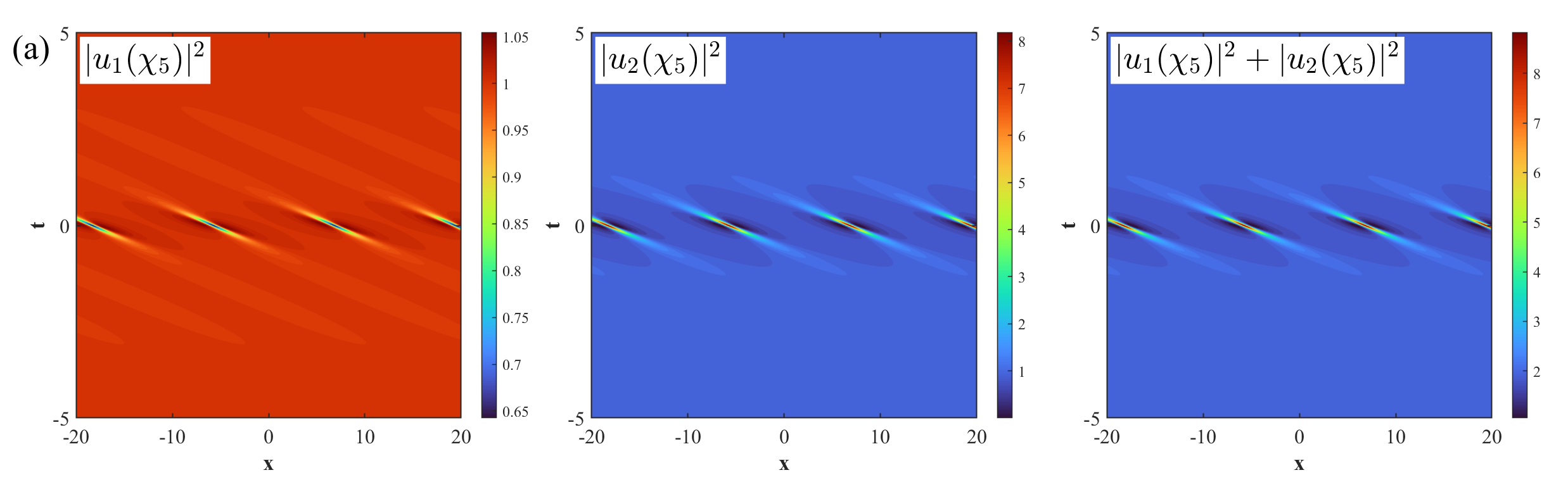}
	\includegraphics[width=\columnwidth]{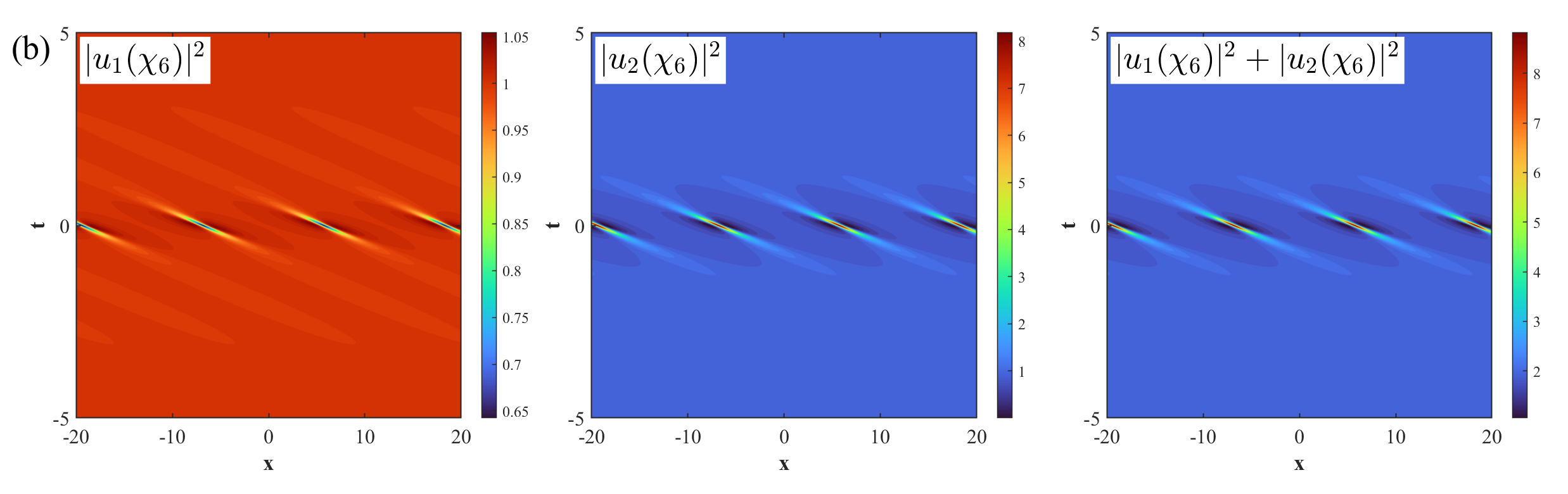}
	\caption{
		Vector ABs at $\mathrm{C}_2=(6,0.5)$ in region~II for
		$\beta_1=0.45$.
		(a) $\chi_a=\chi_5=5.7413-0.9387i$,
		$\chi_b=6.2413-0.9387i$.
		(b) $\chi_a=\chi_6=-6.2413+0.9387i$,
		$\chi_b=-5.7413+0.9387i$.
		Neither branch fulfills the state-transition condition; both
		therefore retain the AB character.
		In each row, the panels from left to right show
		$|u_1|^2$, $|u_2|^2$, and $|u_1|^2+|u_2|^2$.
	}
	\label{fig:general-regionII-kai56-ab}
	
\end{figure}

The complex-conjugate roots generate two further admissible pairs,
$(\chi_5,\chi_8)$ and $(\chi_6,\chi_7)$, which remain on AB
branches. Hence,  one obtain $d_{\mathrm{nd}}=3$ with $N_{\mathrm p}=4$, placing
region~II in the nondegenerate regime. Representative
state-transition periodic waves and ABs at
$\mathrm{C}_2=(6,0.5)$ are illustrated in
Figs.~\ref{fig:general-regionII-kai13-ab} and
\ref{fig:general-regionII-kai56-ab}, respectively.

\begin{figure}[tbp]
	\centering
	\includegraphics[width=0.55\columnwidth]{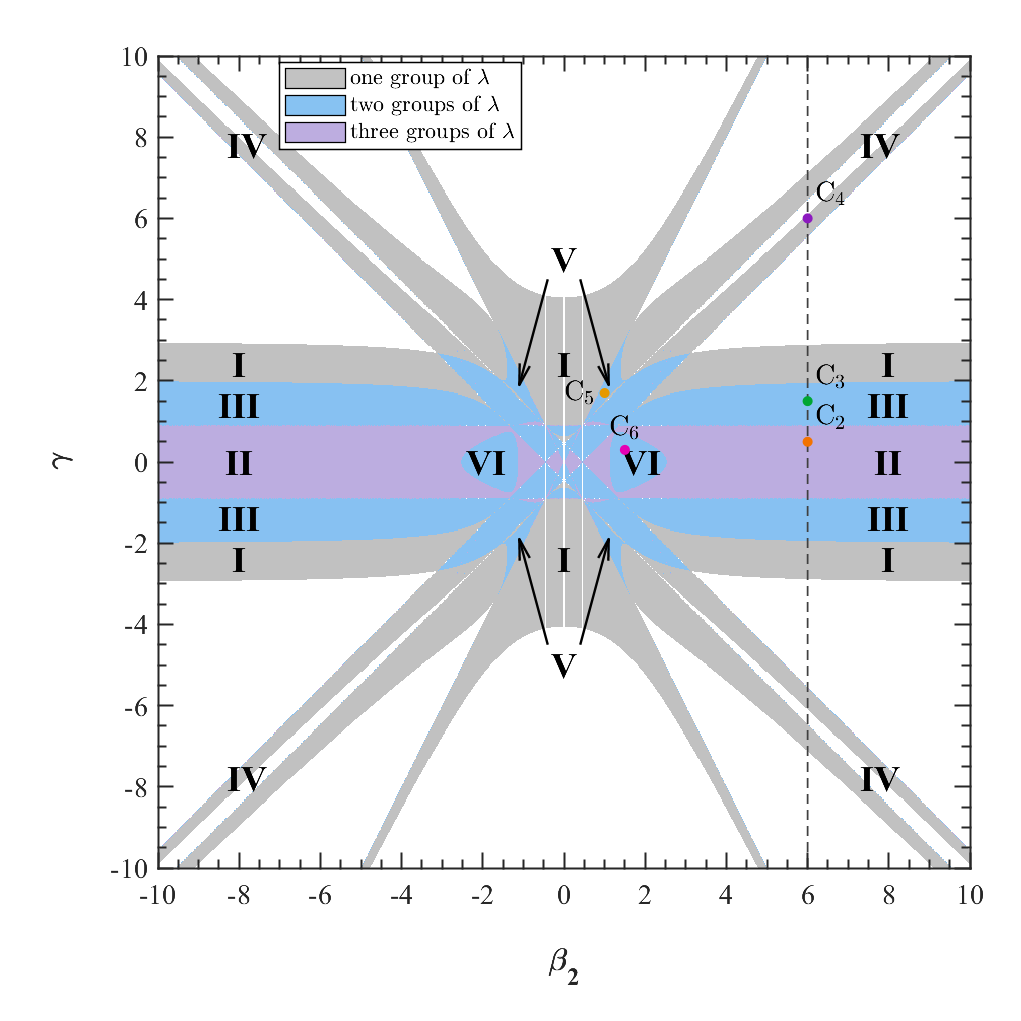}
	\caption{
		Structural-equivalence diagram in the $(\beta_2,\gamma)$ plane
		for $\beta_1=0.45$, with $c_1=c_2=1$ and $\alpha=0$.
		Gray, blue, and purple denote one, two, and three
		structural-equivalence classes, respectively.
		The annotated coordinate points
		$\mathrm{C}_2=(6,0.5)$,
		$\mathrm{C}_3=(6,1.5)$,
		$\mathrm{C}_4=(6,6)$,
		$\mathrm{C}_5=(1,1.7)$, and
		$\mathrm{C}_6=(1.5,0.3)$
		are used in the analyses below.
	}
	\label{fig:lambda-group-general}
\end{figure}
The same symmetry underlying the structural equivalence in
Cases~I and II also applies here, since
Eq.~\eqref{eq:lambda-chi-general} meets
\begin{equation*}
	\lambda(-\chi)=-\lambda(\chi).
\end{equation*}
Accordingly, symmetry-related branches have opposite spectral
parameters and modulus-squared profiles connected by
$(x,t)\mapsto(-x,-t)$. The structural-equivalence diagram in the $(\beta_2,\gamma)$ plane is
shown in Fig.~\ref{fig:lambda-group-general}. At $\mathrm{C}_2$, the
four admissible pairs form three classes: $(\chi_1,\chi_2)$ and
$(\chi_3,\chi_4)$ constitute two distinct state-transition classes,
whereas the branches $\chi_5$--$\chi_8$ form a single AB class.

\paragraph*{Region~III.}

Region~III has the same root configuration as region~II, with two
real roots and one complex-conjugate pair, but a different
admissibility pattern. Only one real root contributes an admissible
pair, $(\chi_1,\chi_2)$, while the complex roots generate the two
additional pairs $(\chi_5,\chi_8)$ and $(\chi_6,\chi_7)$. Hence, one has $d_{\mathrm{nd}}=2$ with $N_{\mathrm p}=3$, placing region~III in
the nondegenerate regime.

\begin{figure}[htbp]
	\centering
	\includegraphics[width=\columnwidth]{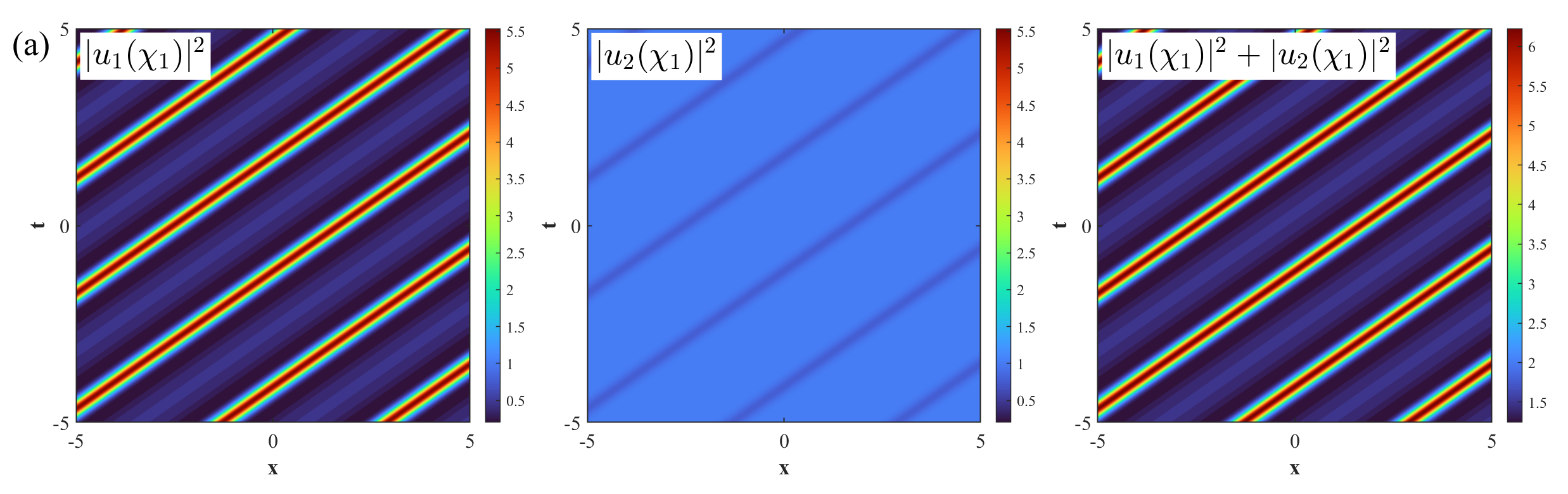}
	
	\includegraphics[width=\columnwidth]{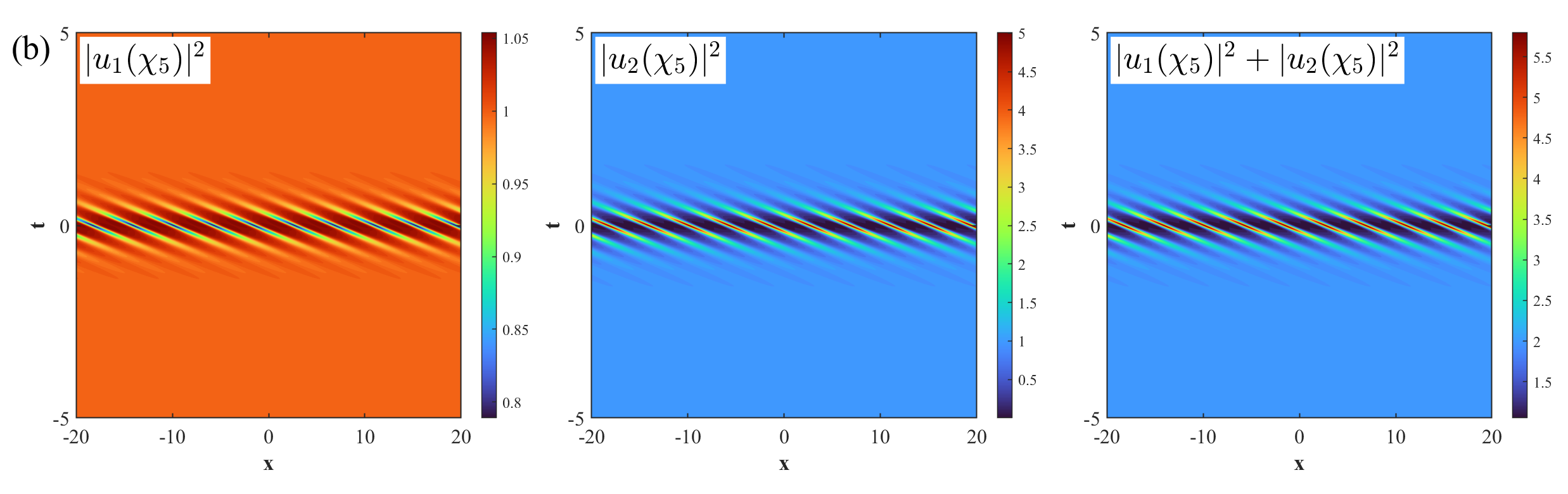}
	
	\includegraphics[width=\columnwidth]{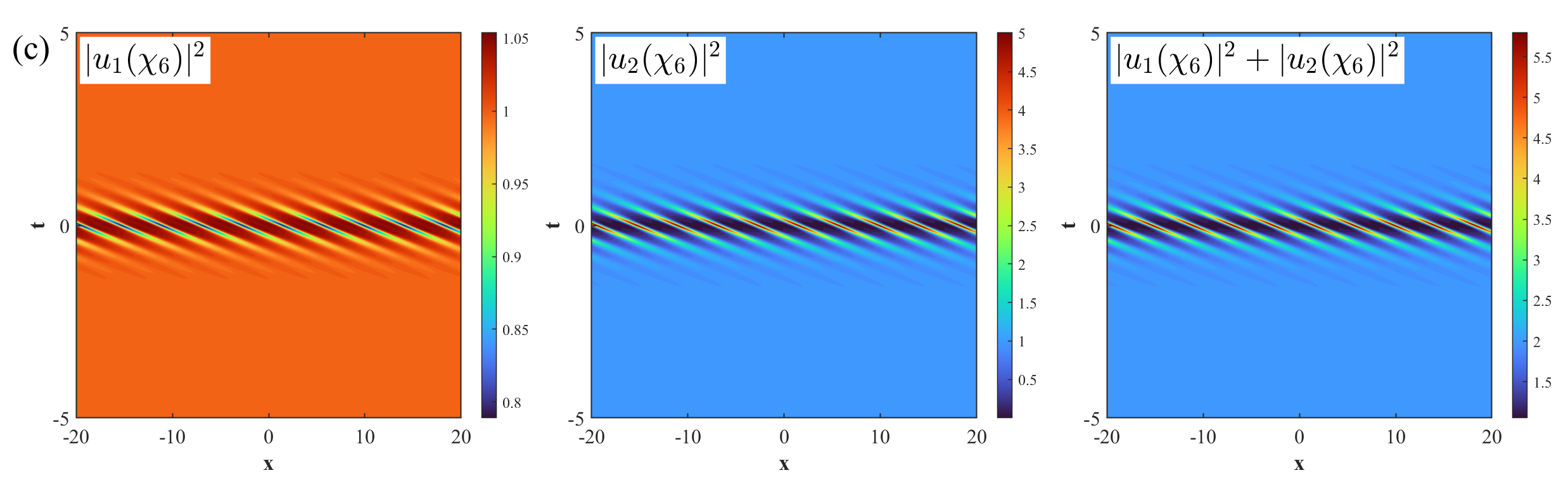}
	\caption{
		Representative wave structures at
		$\mathrm{C}_3=(6,1.5)$ in region~III for
		$\beta_1=0.45$.
		(a) $\chi_a=\chi_1=-0.7500+0.9763i$,
		$\chi_b=0.7500+0.9763i$.
		(b) $\chi_a=\chi_5=5.2409-0.6152i$,
		$\chi_b=6.7409-0.6152i$.
		(c) $\chi_a=\chi_6=-6.7409+0.6152i$,
		$\chi_b=-5.2409+0.6152i$.
		The branch represented by $\chi_1$ lies on the
		state-transition locus, whereas those represented by
		$\chi_5$ and $\chi_6$ correspond to AB states.
		In each row, the panels from left to right show
		$|u_1|^2$, $|u_2|^2$, and $|u_1|^2+|u_2|^2$.
	}
	\label{fig:general-regionIII-all}
\end{figure}

The real-root pair lies on the state-transition locus, while the
two complex-root pairs remain on AB branches. Representative
solutions at $\mathrm{C}_3=(6,1.5)$ are demonstrated in
Fig.~\ref{fig:general-regionIII-all}. The AB branches represented by
$\chi_5$ and $\chi_6$ are symmetry related and thus form a single
structural-equivalence class, consistent with
Fig.~\ref{fig:lambda-group-general}.
\paragraph*{Region~IV.}

Region~IV has two real roots and one complex-conjugate pair,
but the two real roots do not generate admissible branches. The
complex roots instead yield the two admissible eigenvalue-branch
pairs $(\chi_3,\chi_6)$ and $(\chi_4,\chi_5)$. Hence, we arrive at
$d_{\mathrm{nd}}=1$ with $N_{\mathrm p}=2$ , placing region~IV in
the nondegenerate regime. Both pairs remain away from the
state-transition locus and therefore support vector AB states.

\begin{figure}[tbp]
	\centering
	\includegraphics[width=\columnwidth]{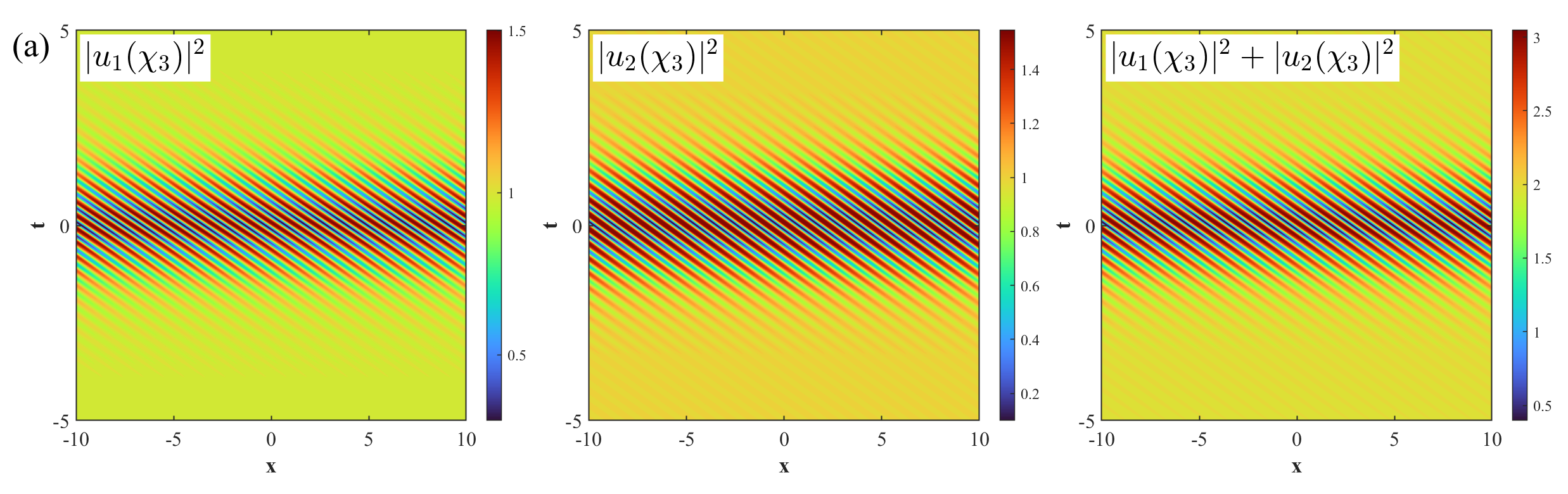}
	\includegraphics[width=\columnwidth]{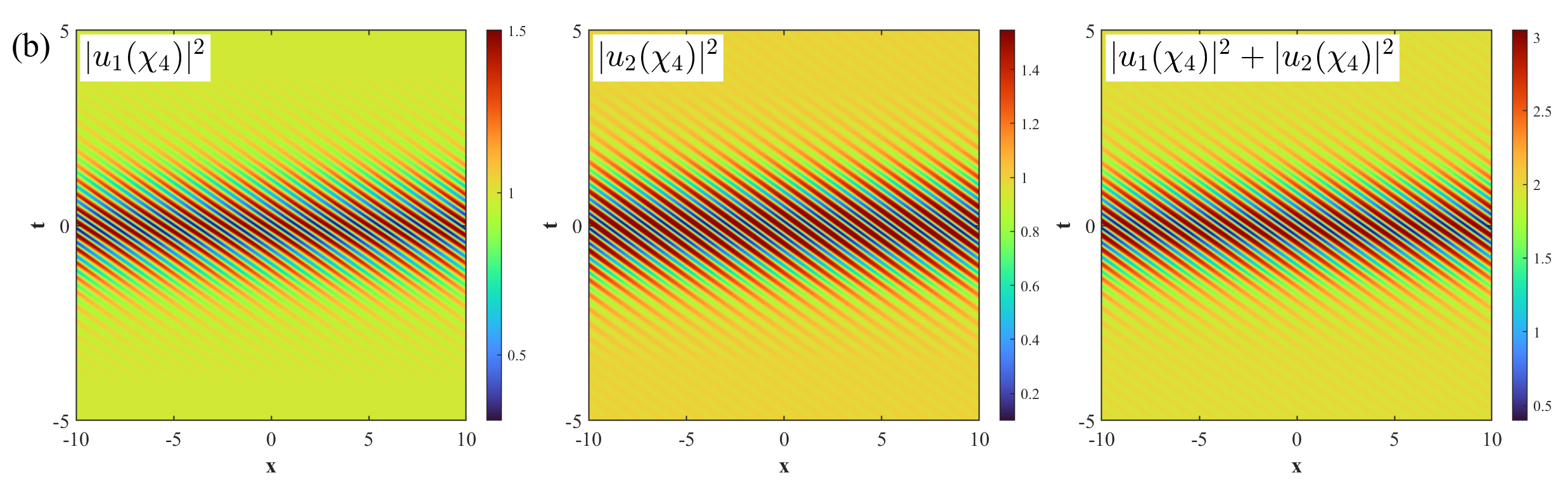}
	\caption{
		Vector ABs at $\mathrm{C}_4=(6,6)$ in region~IV for
		$\beta_1=0.45$.
		(a) $\chi_a=\chi_3=0.2151+0.1153i$,
		$\chi_b=6.2151+0.1153i$.
		(b) $\chi_a=\chi_4=-6.2151-0.1153i$,
		$\chi_b=-0.2151-0.1153i$.
	The two representative branches remain localized and correspond to
	AB states.
		In each row, the panels from left to right show
		$|u_1|^2$, $|u_2|^2$, and $|u_1|^2+|u_2|^2$.
	}
	\label{fig:general-regionIV-kai34-ab}
	
\end{figure}

The profiles of representative solutions at $\mathrm{C}_4=(6,6)$ are presented in
Fig.~\ref{fig:general-regionIV-kai34-ab}. The two AB branches are
symmetry related and therefore belong to the same
structural-equivalence class, in agreement with
Fig.~\ref{fig:lambda-group-general}.

\paragraph*{Region~V.}

In region~V, Eq.~\eqref{eq:R-quartic-general} has four distinct real
roots, two of which satisfy the admissibility condition and generate
the pairs $(\chi_1,\chi_2)$ and $(\chi_3,\chi_4)$. Hence, we obtain $d_{\mathrm{nd}}=1$ with $N_{\mathrm p}=2$, placing region~V in the
nondegenerate regime. Both admissible pairs lie on the
state-transition locus and therefore give periodic-wave states.

\begin{figure}[htbp]
	\centering
	\includegraphics[width=\columnwidth]{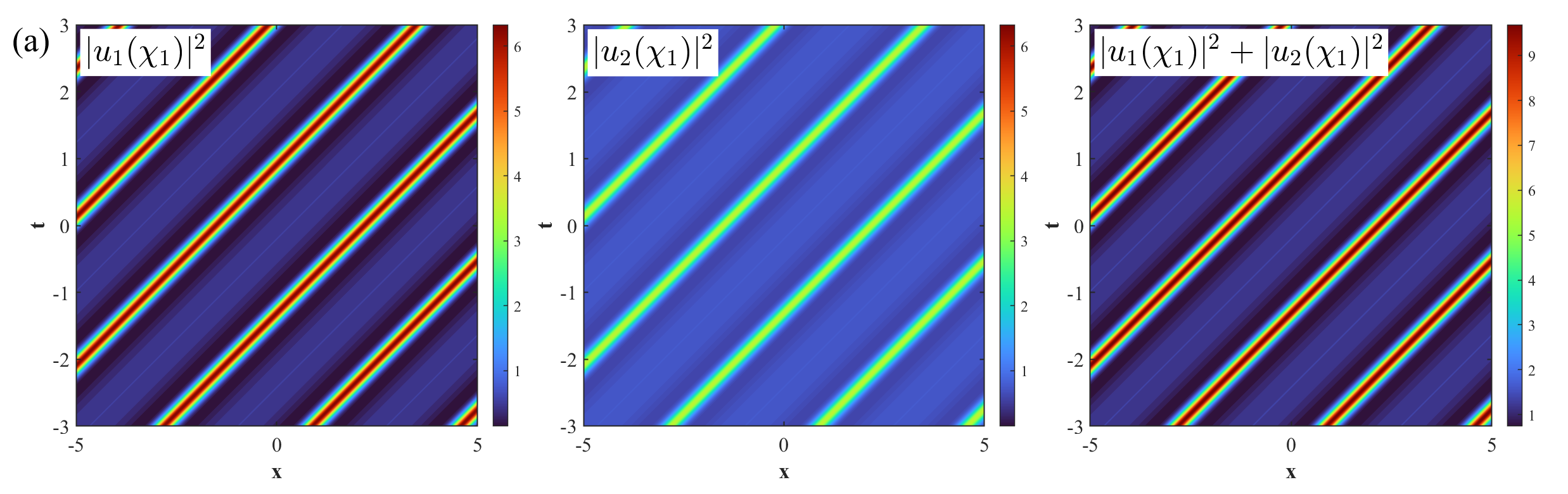}
	
	\includegraphics[width=\columnwidth]{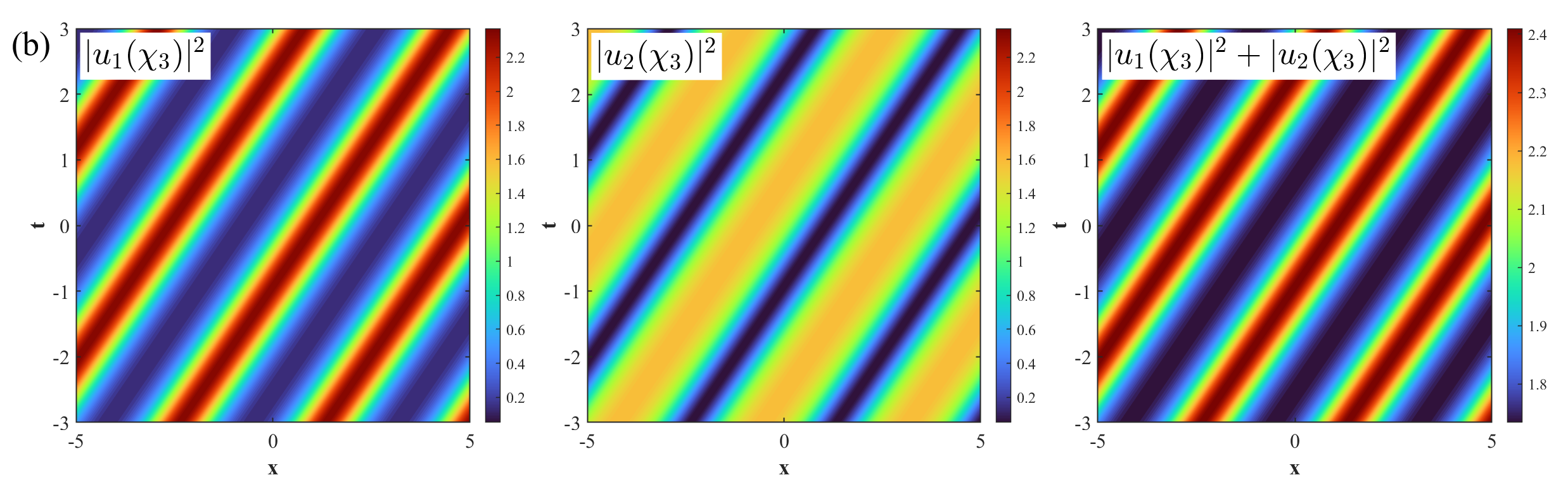}
	\caption{
		State-transition periodic waves at
		$\mathrm{C}_5=(1,1.7)$ in region~V for
		$\beta_1=0.45$.
		(a) $\chi_a=\chi_1=-0.8500-1.3393i$,
		$\chi_b=0.8500-1.3393i$.
		(b) $\chi_a=\chi_3=-0.8500+0.1968i$,
		$\chi_b=0.8500+0.1968i$.
		Each representative branch satisfies the state-transition
		condition and yields a periodic-wave state.
		In each row, the panels from left to right show
		$|u_1|^2$, $|u_2|^2$, and $|u_1|^2+|u_2|^2$.
	}
	\label{fig:general-regionV-kai13-ab}
\end{figure}

Representative modes at $\mathrm{C}_5=(1,1.7)$ are depicted in
Fig.~\ref{fig:general-regionV-kai13-ab}. The two admissible pairs
belong to different structural-equivalence classes, as illustrated in
Fig.~\ref{fig:lambda-group-general}, and thus represent two
inequivalent periodic-wave structures.

\paragraph*{Region~VI.}

Region~VI is characterized by two distinct complex-conjugate pairs of
$R$ roots,
\begin{equation}
	R_2=R_1^\ast,
	\qquad
	R_4=R_3^\ast,
	\qquad
	R_1\neq R_3.
	\label{eq:general-regionVI-R-conj}
\end{equation}
These roots produce the four admissible eigenvalue-branch pairs
$(\chi_1,\chi_4)$, $(\chi_2,\chi_3)$, $(\chi_5,\chi_8)$, and
$(\chi_6,\chi_7)$. Hence, one has
$d_{\mathrm{nd}}=3$ with $N_{\mathrm p}=4$, so region~VI is responsible for the nondegenerate
regime.

\begin{figure}[tbp]
	\centering
	\includegraphics[width=\columnwidth]{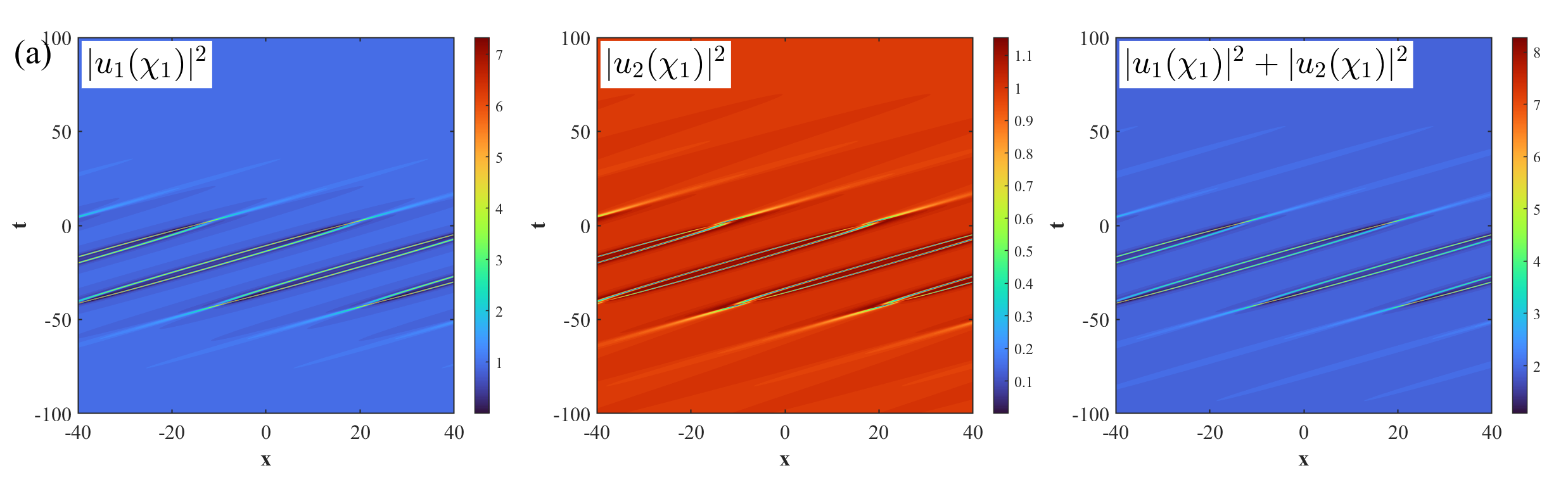}
	
	\includegraphics[width=\columnwidth]{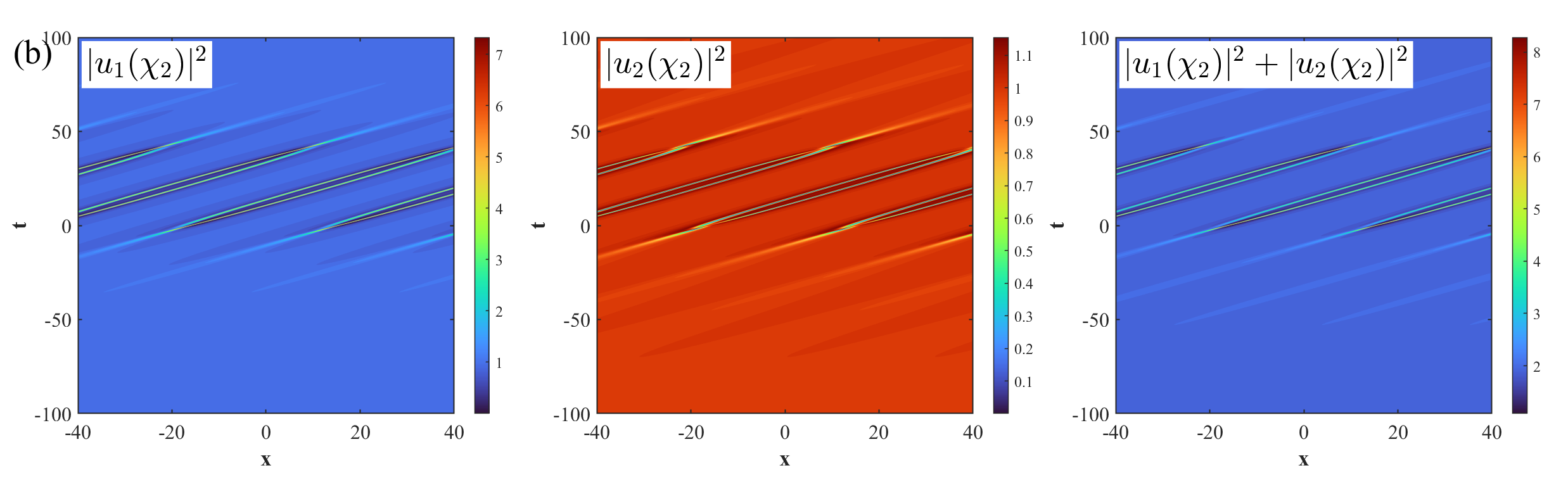}
	
	\includegraphics[width=\columnwidth]{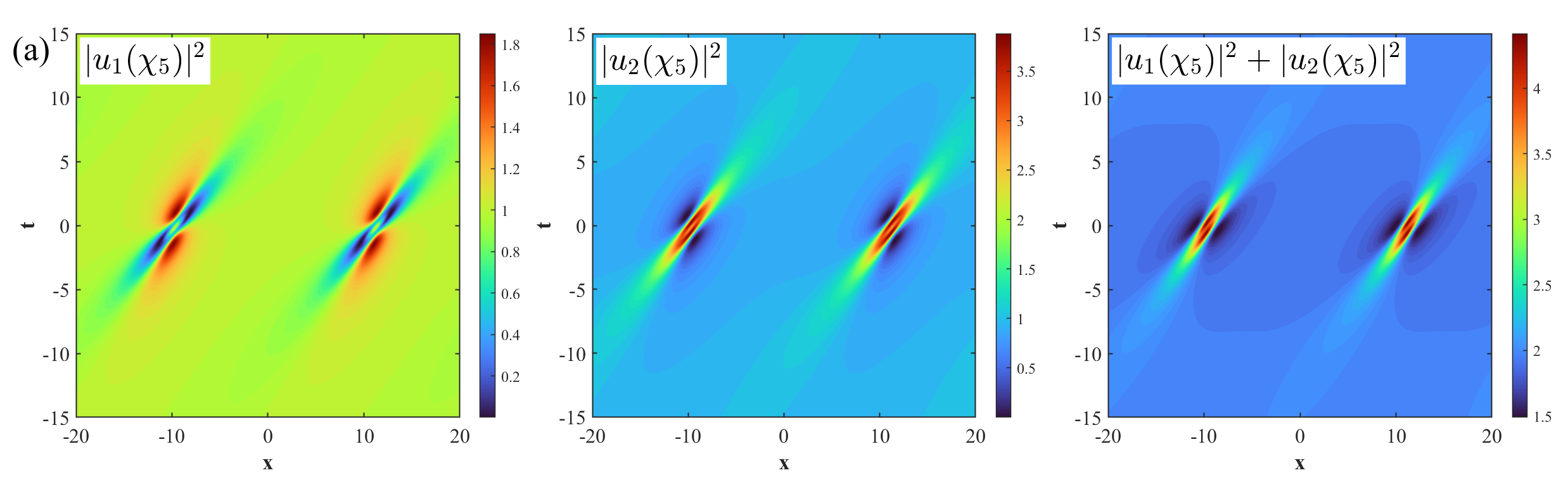}
	
	\includegraphics[width=\columnwidth]{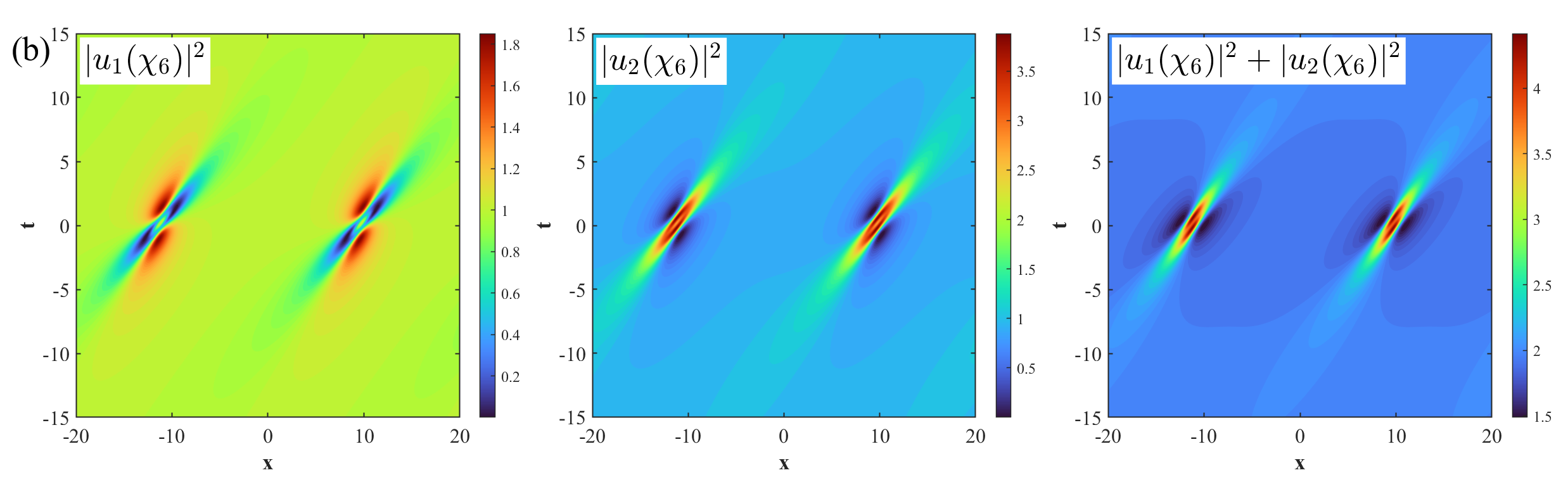}
	\caption{
		Vector ABs at $\mathrm{C}_6=(1.5,0.3)$ in region~VI for
		$\beta_1=0.45$.
		(a) $\chi_a=\chi_1=0.0662-0.8159i$,
		$\chi_b=0.3662-0.8159i$.
		(b) $\chi_a=\chi_2=-0.3662+0.8159i$,
		$\chi_b=-0.0662+0.8159i$.
		(c) $\chi_a=\chi_5=1.1043-0.6758i$,
		$\chi_b=1.4043-0.6758i$.
		(d) $\chi_a=\chi_6=-1.4043+0.6758i$,
		$\chi_b=-1.1043+0.6758i$.
		In each row, the panels from left to right show
		$|u_1|^2$, $|u_2|^2$, and $|u_1|^2+|u_2|^2$.
	}
	\label{fig:general-regionVI-kai1256-ab}
\end{figure}

Because all admissible pairs originate from nonreal $R$ roots, none
lies on the state-transition locus, and region~VI therefore supports
four AB-type eigenvalue-branch pairs. At
$\mathrm{C}_6=(1.5,0.3)$, they form two structural-equivalence
classes, with $\chi_1$--$\chi_4$ in one class and
$\chi_5$--$\chi_8$ in the other. Representative vector ABs from the two
classes are shown in Fig.~\ref{fig:general-regionVI-kai1256-ab}.

\section{\label{sec:numerical}
	Numerical simulations}
\begin{figure}[htbp]
	\centering
	
	\includegraphics[width=0.95\columnwidth]
	{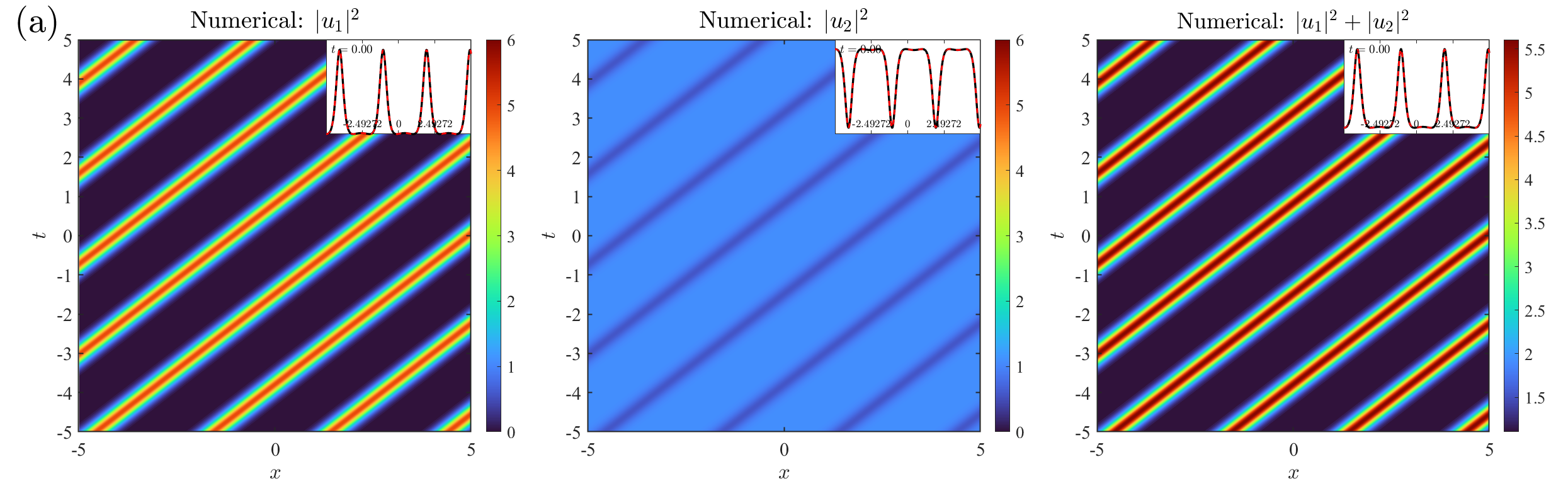}
	
	\vspace{-1mm}
	
	\includegraphics[width=0.95\columnwidth]
	{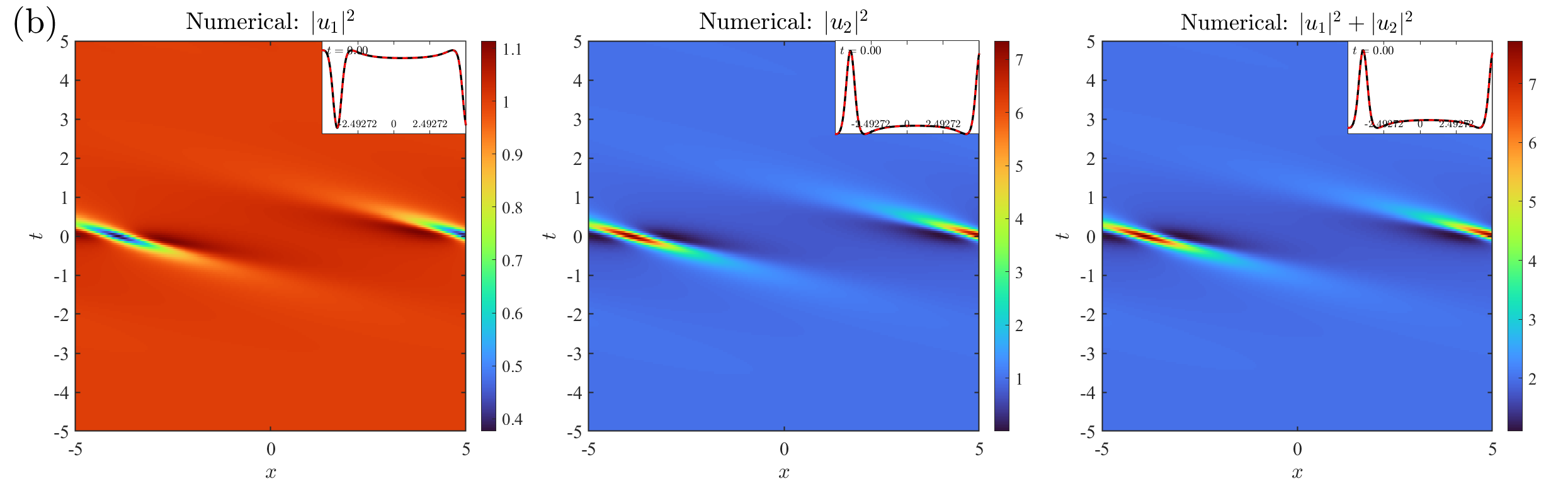}
	
	\vspace{-1mm}
	
	\includegraphics[width=0.95\columnwidth]
	{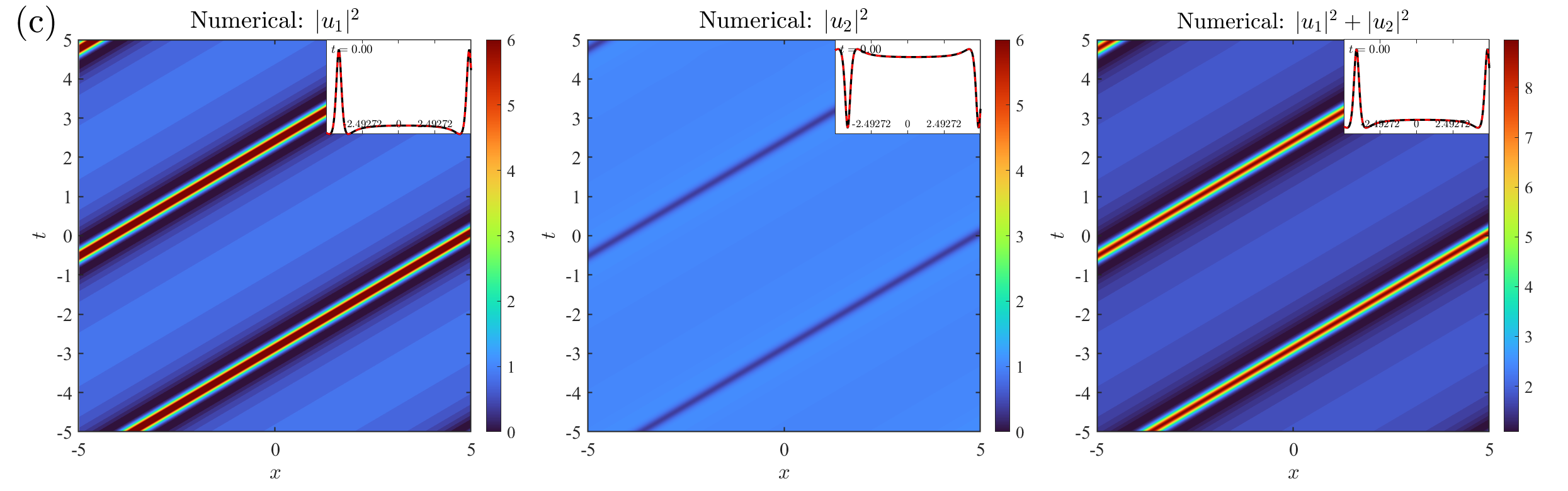}
	
	\caption{
		Numerical evolutions for the single-vanishing-carrier configuration.
		(a) Degenerate state-transition periodic wave.
		(b) Nondegenerate vector AB.
		(c) Nondegenerate state-transition periodic wave.
		In each row, the panels from left to right display the numerical
		evolutions of $|u_1|^2$, $|u_2|^2$, and $|u_1|^2+|u_2|^2$,
		respectively, for $t\in[-5,5]$.
		The inset in each panel gives the corresponding $t=0$ cross section,
		comparing the numerical profile (black solid) with the exact one
		(red dashed).
	}
	\label{fig:numerical-beta1-zero}
\end{figure}
\begin{figure}[tbp]
	\centering
	
	\includegraphics[width=0.95\columnwidth]
	{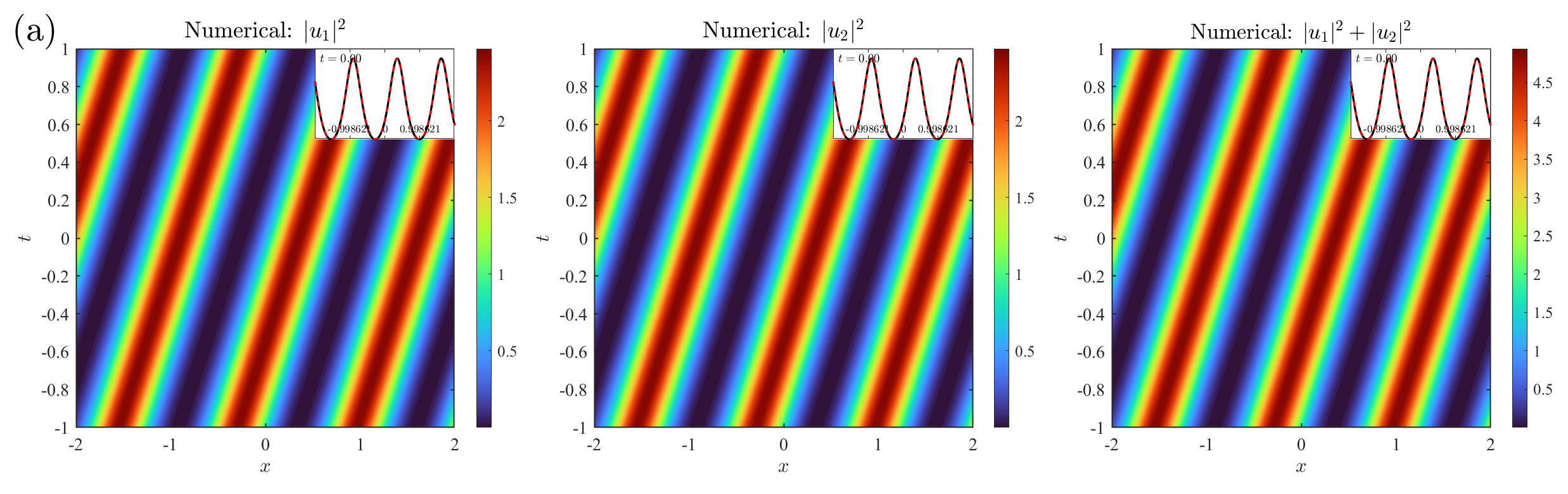}
	
	\vspace{-1mm}
	
	\includegraphics[width=0.95\columnwidth]
	{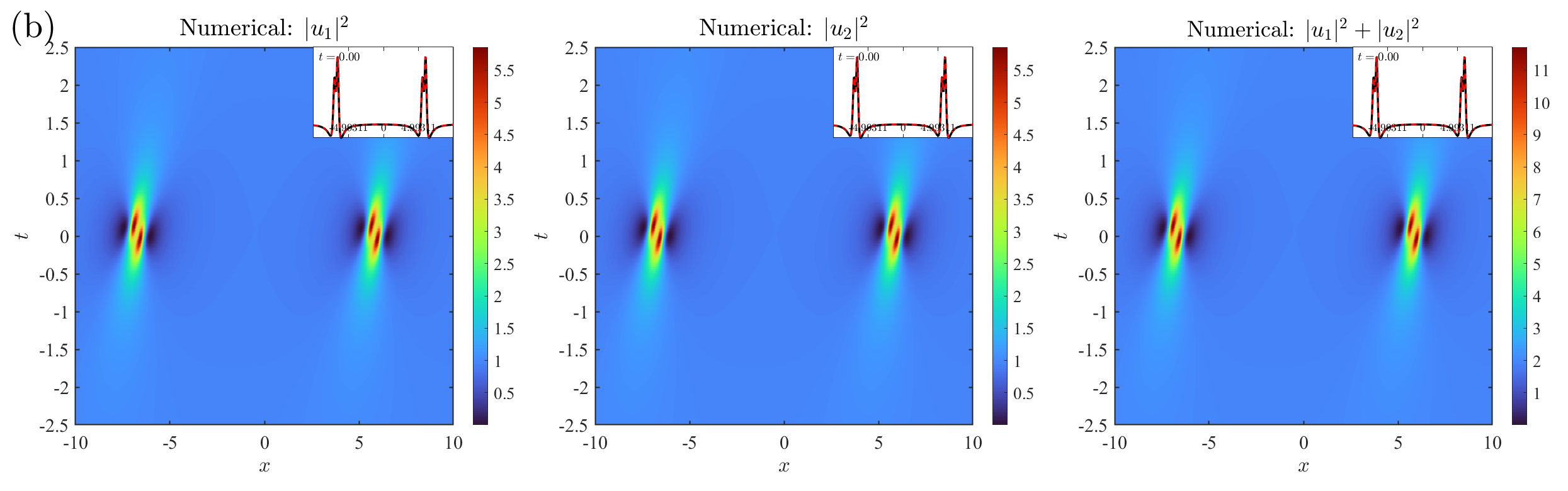}
	
	\vspace{-1mm}
	
	\includegraphics[width=0.95\columnwidth]
	{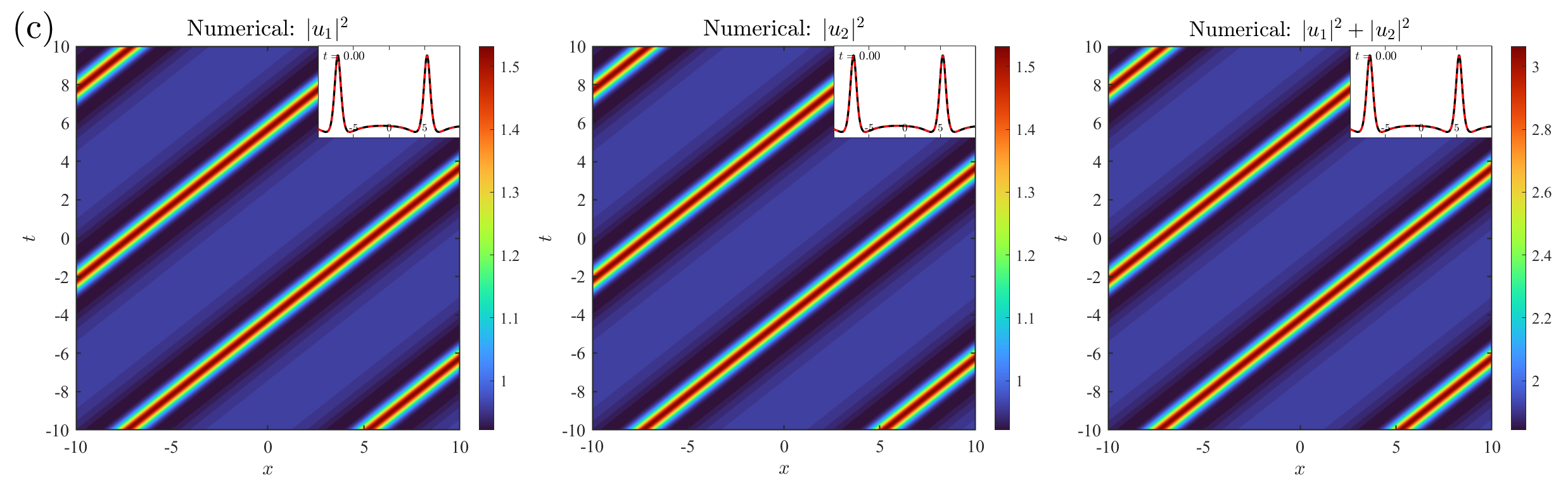}
	
	\caption{
		Numerical evolutions for the opposite-wavenumber configuration.
		(a) Degenerate state-transition periodic wave for $t\in[-1,1]$.
		(b) Nondegenerate vector AB for $t\in[-2.5,2.5]$.
		(c) Nondegenerate state-transition periodic wave for $t\in[-10,10]$.
		In each row, the three panels correspond to
		$|u_1|^2$, $|u_2|^2$, and $|u_1|^2+|u_2|^2$, respectively.
		The upper-right insets compare the numerical $t=0$ profiles
		with the corresponding exact solutions.
	}
	\label{fig:numerical-betaop}
\end{figure}

\begin{figure}[tbp]
	\centering
	
	\includegraphics[width=0.95\columnwidth]
	{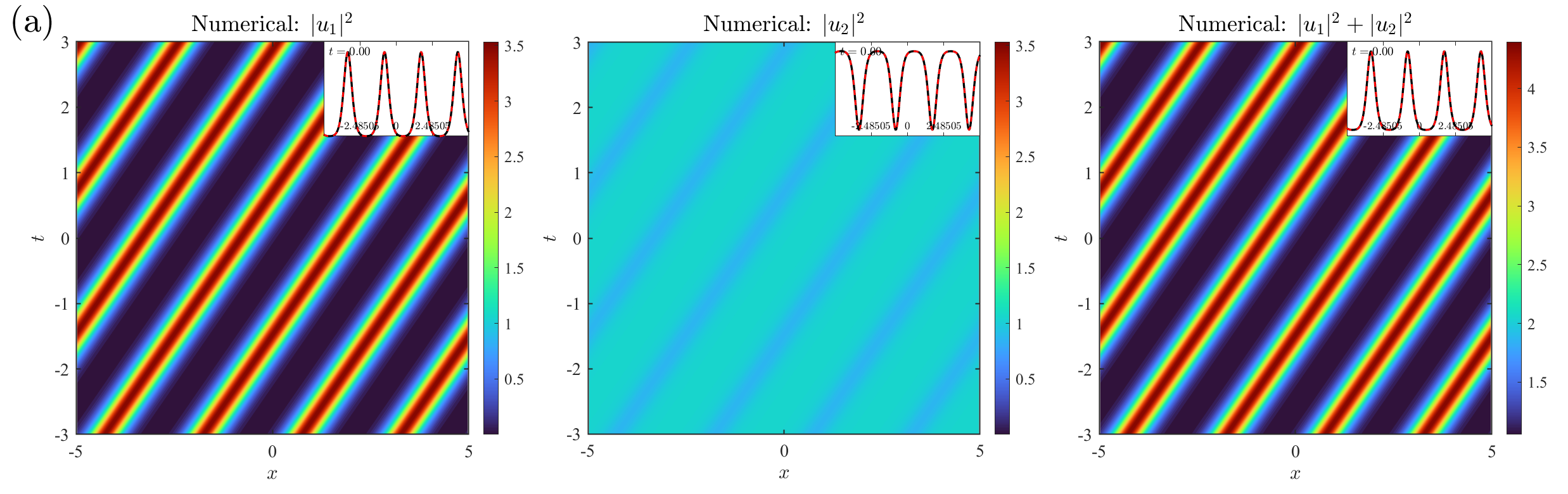}
	
	\vspace{-1mm}
	
	\includegraphics[width=0.95\columnwidth]
	{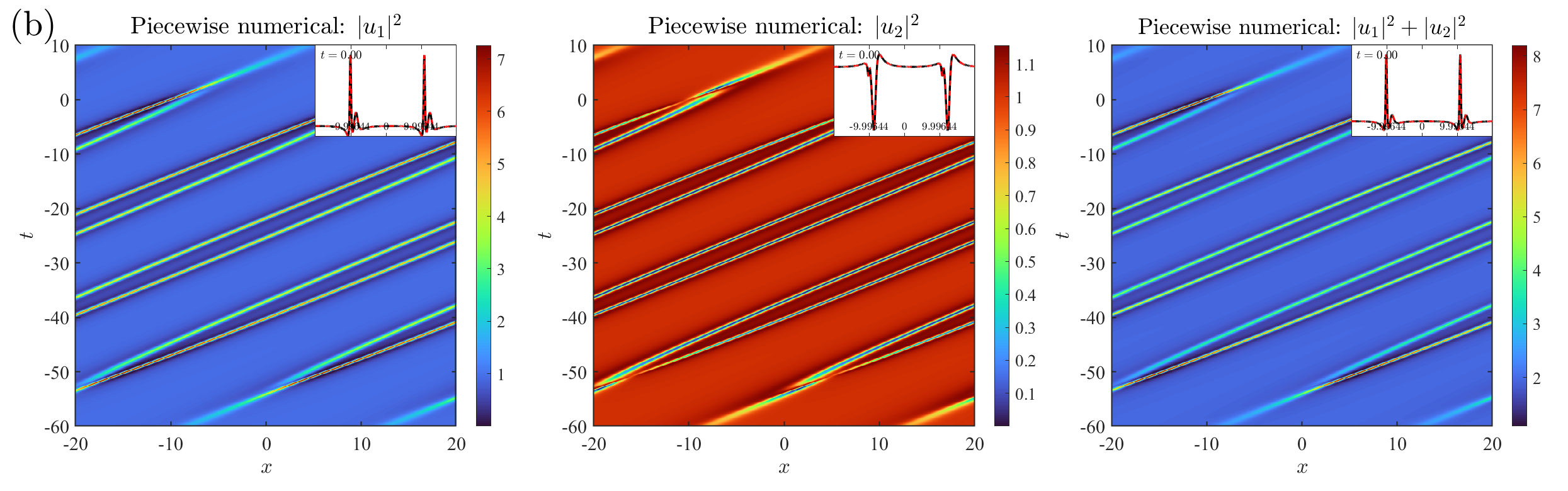}
	
	\vspace{-1mm}
	
	\includegraphics[width=0.95\columnwidth]
	{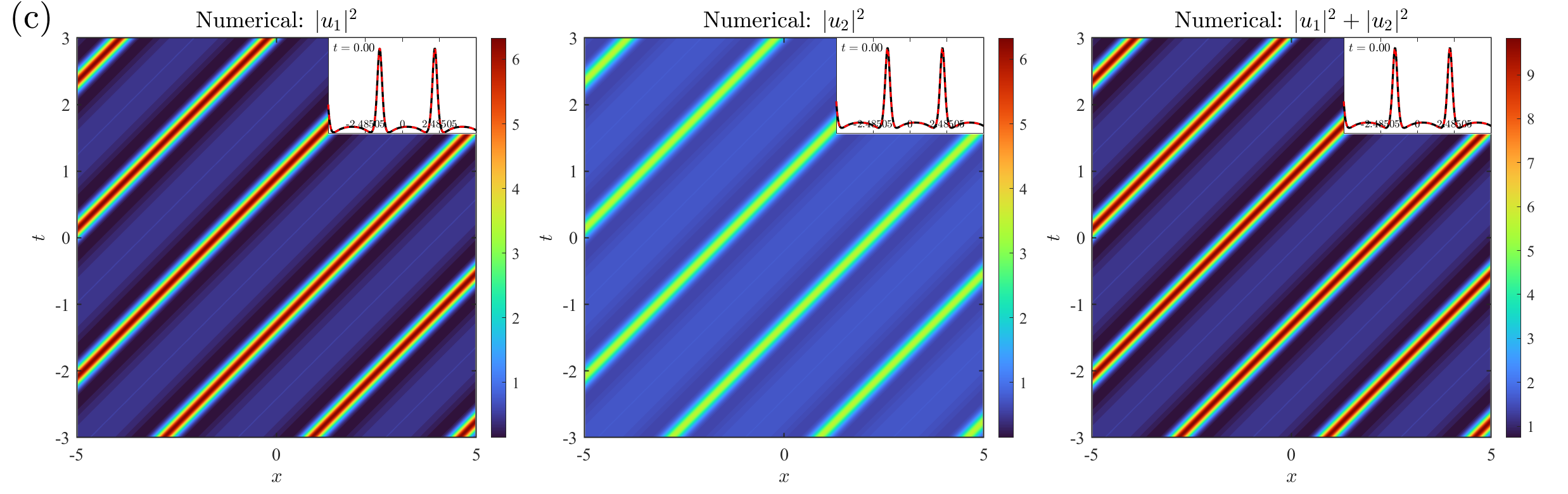}
	
	\caption{
		Numerical results for two nonzero carrier wavenumbers of unequal
		magnitudes.
		(a) Degenerate state-transition periodic wave over $t\in[-3,3]$.
		(b) Nondegenerate vector AB over $t\in[-60,10]$, reconstructed
		piecewise for the long-time evolution.
		(c) Nondegenerate state-transition periodic wave over $t\in[-3,3]$.
		The columns show $|u_1|^2$, $|u_2|^2$, and
		$|u_1|^2+|u_2|^2$ from left to right.
		The insets provide $t=0$ cross-section comparisons between the
		numerical and exact profiles.
	}
	\label{fig:numerical-general}
\end{figure}
To verify the representative exact solutions obtained above, we
numerically integrate the reduced CSS system~\eqref{eq:reduced-system}
using a Fourier pseudospectral discretization together with
exponential time integration. For each carrier-wavenumber
configuration, we consider three types of wave states: a
state-transition periodic wave in the degenerate regime, a vector
AB in the nondegenerate regime, and a state-transition periodic wave
in the nondegenerate regime. Except for the long-time reconstruction
specified below, the corresponding exact solution at the initial
time $t=t_0$ is used as the initial condition and then propagated
numerically over the prescribed interval.

For the single-vanishing carrier configuration,
Fig.~\ref{fig:numerical-beta1-zero} collects the three representative
evolutions associated with
Figs.~\ref{fig:degenerate-regionIII-comparison}(b),
\ref{fig:kai123_ab}(a), and
\ref{fig:kai123_ab}(c), respectively. The opposite-wavenumber results
are summarized in Fig.~\ref{fig:numerical-betaop}, corresponding to
the degenerate state-transition periodic wave in
Fig.~\ref{fig:betaop-regionII-kai3-ab}, the nondegenerate AB in
Fig.~\ref{fig:betaop-regionI-kai12-ab}(a), and the nondegenerate
state-transition periodic wave in
Fig.~\ref{fig:betaop-regionIII-kai13-ab}(a). Finally,
Fig.~\ref{fig:numerical-general} presents the same three types of
wave states for two nonzero carrier wavenumbers of unequal
magnitudes, using the exact solutions shown in
Figs.~\ref{fig:general-regionI-kai1-ab},
\ref{fig:general-regionVI-kai1256-ab}(a), and
\ref{fig:general-regionV-kai13-ab}(a). In all three carrier
configurations, the numerical evolutions reproduce the characteristic
spatiotemporal structures predicted by the corresponding exact
solutions, including both localized breather dynamics and
state-transition periodic-wave patterns. The numerical results are in close agreement with the analytical
predictions, confirming the validity of the exact solutions.

\section{\label{sec:conclusions}
	Conclusions and Discussions}

In this work, we have investigated the vector ABs in the CSS system,  focusing on both the degenerate and nondegenerate regimes. Three representative configurations  were considered: a single vanishing carrier wavenumber,
opposite carrier wavenumbers, and two nonzero carrier wavenumbers of
unequal magnitudes. By introducing the auxiliary variable $R$, the
same spectral-parameter condition was reduced to cubic, quadratic,
and quartic equations for each case, respectively. Their root
structures, combined with the admissibility condition, determine the
relevant eigenvalue branches and the corresponding existence
diagrams of breather solutions.

To quantify the coexistence of distinct eigenvalue-branch pairs, we
introduced the notion of the discrete degree of nondegeneracy
$d_{\mathrm{nd}}=N_{\mathrm p}-1$. Its maximal value depends
on the carrier background:
$d_{\mathrm{nd}}^{\max}=2$ for a single vanishing carrier wavenumber,
$d_{\mathrm{nd}}^{\max}=1$ for opposite carrier wavenumbers, and
$d_{\mathrm{nd}}^{\max}=3$ for two nonzero carrier wavenumbers of
unequal magnitudes. As summarized in
Table~\ref{tab:max-degree-nondegeneracy}, the CSS system exhibits a
richer structures of eigenvalue branches than those in the Manakov system and CH
system. In
particular, the generic unequal-wavenumber case yields the largest
number of coexisting pairs and the highest degree of nondegeneracy
found here.

The exact solutions and branch-resolved MI spectra also reveal a
common state-transition mechanism. For an admissible branch,
$\operatorname{Re}(\chi_a)=-\gamma/2$ implies $\Psi=0$, causing the
localized AB to become a periodic-wave state. The same relation gives
the zero-gain condition of the associated MI branch, thereby linking
the transition of the exact solution directly to the linear stability
of the plane-wave background. Every eigenvalue branch in degenerate
region lies on this state-transition locus, whereas the converse is
not generally true. State-transition branches may also occur in
nondegenerate regions, either together with AB branches or as all
admissible pairs in that region.

The wave-component intensities further show that the eigenvalue-branch
nondegeneracy does not necessarily determine the mode character of
the solution. For the case of single-vanishing carrier, two regions with
the same degenerate eigenvalue-branch structure can exhibit different
mode behaviors: one gives similar spatiotemporal intensity patterns in
the two components, whereas the other shows opposite componentwise
modulations. Conversely, in the opposite-wavenumber case, a degenerate-type mode may arise even in an eigenvalue-branch nondegenerate region. 
This is different from the results in Manakov system, where the modes of two wave components in degenerate region are essentially identical.

Several questions remain open. The underlying spectral problem is
governed by a quintic characteristic equation, which is not generally
solvable in radicals. Rather than treating the full quintic directly,
we impose $\chi_b=\chi_a+K$ together with
$\lambda(\chi_a)=\lambda(\chi_b)$ and introduce
$R=\chi_a\chi_b$. This construction is tailored to the spectral
conditions required for ABs, while alternative treatments of the full
characteristic problem may reveal additional spectral features.
Moreover, the analysis here is restricted to ABs, while nondegenerate KMs
require separate consideration. Only the eigenvalues $\chi_a$ and
$\chi_b$ selected by the Darboux construction have been examined, and
other combinations of characteristic roots or nonlinear
superpositions of multiple wave elements may produce more intricate
vector-wave structures. Finally, although
$d_{\mathrm{nd}}$ quantifies eigenvalue-branch nondegeneracy through
the number of admissible pairs, it does not measure the relative
intensity relation between the two wave components. Developing a new
 physical measure of nondegeneracy therefore
remains an interesting issue for future study.
	\begin{acknowledgments}
		We express our sincere thanks to the referees and to all the
		members of our discussion group for their valuable comments.
		This work is supported by the National Natural Science Foundation
		of China Grant No.~12375002.

Author contributions are as follows: Mengyuan Lan: methodology, analytical calculations, numerical simulation, physics analysis, and writing; Lei Wang: idea of this paper, methodology, analytical calculations, physics analysis, writing, and supervision; Yinchuan Zhao: writing.
	\end{acknowledgments}
	
	\appendix
	
	\section{\label{app:spectral-reduction}Reduction of the spectral problem}
	
In this appendix, we derive the constant-coefficient spectral problem
used in Sec.~\ref{sec:lax-darboux}. We denote
\begin{equation*}
	\rho=c_1^2+c_2^2
\end{equation*}
and introduce
\begin{align*}
		\theta_j
		&=\beta_jx
		+\epsilon\left[
		\beta_j^3-6\beta_j(c_1^2+c_2^2)
		\right]t,\\ 
		E_j^\pm&=\exp(\pm i\theta_j),\quad j=1,2.
\end{align*}

Substituting the plane-wave seeds
\begin{equation*}
	u_1^{[0]}=c_1e^{i\theta_1},
	\qquad
	u_2^{[0]}=c_2e^{i\theta_2}
\end{equation*}
into the Lax pair~\eqref{eq:lax-pair}, the corresponding
background eigenfunction $\Phi_0$ satisfies
\begin{equation*}
	(\Phi_0)_x=U_{\mathrm{bg}}\Phi_0,
	\qquad
	(\Phi_0)_t=V_{\mathrm{bg}}\Phi_0.
\end{equation*}
Here, $U_{\mathrm{bg}}$ and $V_{\mathrm{bg}}$ denote the full Lax
matrices evaluated on the plane-wave background. 

The spatial background matrix is
\begin{equation*}
	U_{\mathrm{bg}}
	=
	i
	\begin{pmatrix}
		\lambda & c_1E_1^+ & c_1E_1^- & c_2E_2^+ & c_2E_2^- \\
		c_1E_1^- & 0 & 0 & 0 & 0 \\
		c_1E_1^+ & 0 & 0 & 0 & 0 \\
		c_2E_2^- & 0 & 0 & 0 & 0 \\
		c_2E_2^+ & 0 & 0 & 0 & 0
	\end{pmatrix}.
\end{equation*}
The temporal background matrix $V_{\mathrm{bg}}$ is obtained by
substituting the same plane-wave seeds into the temporal Lax matrix
$V$.

To remove the background phase factors, we introduce the diagonal
gauge transformation
\begin{equation}
	\Phi_0=\mathcal G\widehat{\Phi}_0,
	\qquad
	\mathcal G
	=
	\operatorname{diag}
	\left(
	1,E_1^-,E_1^+,E_2^-,E_2^+
	\right).
	\label{app:eq:gauge-transformation}
\end{equation}
Differentiating Eq.~\eqref{app:eq:gauge-transformation} with respect
to $x$ and $t$ gives
\begin{equation*}
	(\widehat{\Phi}_0)_x
	=
	\widehat U_{\mathrm{bg}}\widehat{\Phi}_0,
	\qquad
	(\widehat{\Phi}_0)_t
	=
	\widehat V_{\mathrm{bg}}\widehat{\Phi}_0,
\end{equation*}
where
\begin{equation*}
	\widetilde U_0
	=
	\mathcal G^{-1}U_{\mathrm{bg}}\mathcal G
	-\mathcal G^{-1}\mathcal G_x,\quad
	\widetilde V_0
	=
	\mathcal G^{-1}V_{\mathrm{bg}}\mathcal G
	-\mathcal G^{-1}\mathcal G_t.
\end{equation*}
Consequently, all background phase factors cancel, and the transformed
spatial matrix becomes
\begin{equation*}
\widetilde U_0
	=
	iM
	=
	i
	\begin{pmatrix}
		\lambda & c_1 & c_1 & c_2 & c_2 \\
		c_1 & \beta_1 & 0 & 0 & 0 \\
		c_1 & 0 & -\beta_1 & 0 & 0 \\
		c_2 & 0 & 0 & \beta_2 & 0 \\
		c_2 & 0 & 0 & 0 & -\beta_2
	\end{pmatrix}.
\end{equation*}
A direct substitution of the plane-wave background into the temporal
part of the Lax pair, followed by the same gauge transformation, gives
\begin{equation}
\widetilde V_0
	=
	-\epsilon\widetilde U_0^3
	-6\epsilon\rho\,\widetilde U_0.
	\label{app:eq:Vt0}
\end{equation}
Since $\widetilde U_0=iM$, Eq.~\eqref{app:eq:Vt0} can
equivalently be written as
\begin{equation*}
	\widetilde V_0
	=
	i\epsilon\left(M^3-6\rho M\right).
\end{equation*}

	\section{\label{app:conjugation-equivalence}Conjugation equivalence for $\beta_1=\beta_2$ and $\beta_1=-\beta_2$}
	
	Here we prove the intensity equivalence between the equal- and opposite-wavenumber configurations used in Sec.~\ref{subsec:classification-beta-opposite}. We consider the symmetric background
	\begin{equation*}
		c_1=c_2=c,\qquad \rho=c_1^2+c_2^2=2c^2 .
	\end{equation*}
	Let $(u_{1,+},u_{2,+})$ be the Darboux solution for
	\begin{equation*}
		\beta_1=\beta_2=\beta,
	\end{equation*}
	and let $(u_{1,-},u_{2,-})$ be the Darboux solution for
$
		\beta_1=\beta, \beta_2=-\beta .
$
	For $\beta_1=\beta_2=\beta$, the seed backgrounds are
	\begin{equation*}
		u_{10,+}=ce^{i\theta},\qquad
		u_{20,+}=ce^{i\theta},
	\end{equation*}
	where
	\begin{equation*}
		\theta=\beta x+\epsilon(\beta^3-6\rho \beta)t .
	\end{equation*}
	For $\beta_1=\beta$ and $\beta_2=-\beta$, the seeds become
	\begin{equation*}
		u_{10,-}=ce^{i\theta},\qquad
		u_{20,-}=ce^{-i\theta}.
	\end{equation*}
	Hence, we have
	\begin{equation}
		u_{10,-}=u_{10,+},\qquad
		u_{20,-}=u_{20,+}^{\ast}.
		\label{app:eq:seed-conj-equivalence}
	\end{equation}
	
	Since both configurations satisfy $\beta_1^2=\beta_2^2=\beta^2$, they have the same spectral curve,
	\begin{equation}
		\lambda(\chi)
		=\chi-\frac{\rho}{\chi-\beta}
		-\frac{\rho}{\chi+\beta}.
		\label{app:eq:spectral-curve-pm-beta}
	\end{equation}
	Thus, for the same spectral branches and the same Darboux coefficients, the two constructions differ only in the placement of the phase factors associated with the second component.
	
	Let the eigenfunction used in the Darboux transformation be
	\begin{equation}
\Phi=(P_0,P_1,P_2,P_3,P_4)^{\mathrm T}.
	\end{equation}
	For $\beta_1=\beta_2=\beta$ and
	$\beta_1=-\beta_2=-\beta$, the corresponding
	eigenfunction components are related by
	\begin{equation}
		\begin{aligned}
		P_{1,-}=P_{1,+},\quad
		P_{2,-}=P_{2,+},\\
		P_{3,-}=P_{4,+},\quad
		P_{4,-}=P_{3,+}.
		\label{app:eq:eigenvector-component-relation}
		\end{aligned}
	\end{equation}
	Thus changing from the equal-wavenumber case to the opposite-wavenumber case exchanges the two components associated with the second field.
	
	The binary Darboux correction can be written in the bilinear form
	\begin{equation}
		u_1=u_{10}-\mathcal B(P_1,P_2),
		\quad
		u_2=u_{20}-\mathcal B(P_3,P_4),
		\label{app:eq:field-B-form}
	\end{equation}
	where $\mathcal B$ denotes the corresponding Darboux bilinear expression. The Gram matrix appearing in the binary Darboux transformation is unchanged under the exchange \eqref{app:eq:eigenvector-component-relation}, and the bilinear expression satisfies
	\begin{equation}
		\mathcal B(X,Y)=\mathcal B(Y,X)^\ast .
		\label{app:eq:H-conj-symmetry}
	\end{equation}
	Combining Eqs.~\eqref{app:eq:seed-conj-equivalence}--\eqref{app:eq:H-conj-symmetry}, one obtains
	\begin{equation*}
		u_{1,-}=u_{1,+},\qquad
		u_{2,-}=u_{2,+}^{\ast}.
	\end{equation*}
	Consequently,
	\begin{equation*}
		|u_{1,-}|^2=|u_{1,+}|^2,\qquad
		|u_{2,-}|^2=|u_{2,+}|^2,
	\end{equation*}
	and
	\begin{equation*}
		|u_{1,-}|^2+|u_{2,-}|^2=|u_{1,+}|^2+|u_{2,+}|^2 .
	\end{equation*}
	This proves that the two carrier-wavenumber configurations are indistinguishable at the level of component intensities and total intensity.
	\section{\label{app:coefficients-darboux}
		Coefficients in the Darboux exponential forms}
	
	In this appendix, we give the coefficients used in the nine-term binary
	Darboux expression and in the four-term single Darboux expression in
	Sec.~\ref{sec:existence-mi}. We consider the two-mode choice
	associated with the spectral branches $\chi_a$ and $\chi_b$, and set
	\begin{equation*}
		\phi_\mu=e^{\eta_\mu},
		\eta_\mu
		=
		i\chi_\mu x
		+i\epsilon\left(\chi_\mu^3-6\rho\chi_\mu\right)t,\quad
		\mu=a,b .
	\end{equation*}
	For the two-mode solution, we take
	\begin{equation*}
		P_0=\phi_a+\phi_b,
	\end{equation*}
	\begin{equation*}
		P_1=
		u_{10}^{\ast}
		\left(
		\frac{\phi_a}{\chi_a-\beta_1}
		+
		\frac{\phi_b}{\chi_b-\beta_1}
		\right),
	\end{equation*}
	\begin{equation*}
		P_2=
		u_{10}
		\left(
		\frac{\phi_a}{\chi_a+\beta_1}
		+
		\frac{\phi_b}{\chi_b+\beta_1}
		\right),
	\end{equation*}
	\begin{equation*}
		P_3=
		u_{20}^{\ast}
		\left(
		\frac{\phi_a}{\chi_a-\beta_2}
		+
		\frac{\phi_b}{\chi_b-\beta_2}
		\right),
	\end{equation*}
	and
	\begin{equation*}
		P_4=
		u_{20}
		\left(
		\frac{\phi_a}{\chi_a+\beta_2}
		+
		\frac{\phi_b}{\chi_b+\beta_2}
		\right).
	\end{equation*}
	
	Define \begin{equation*}
		D_\Phi
		=
		\sum_{j=0}^{4}|P_j|^2,
		\qquad
		\Delta_\Phi
		=
		P_0^2-2P_1P_2-2P_3P_4.
	\end{equation*}
	Then
	\begin{equation}
		D_\Phi=
		\sum_{\mu,\sigma\in\{a,b\}}
		G_{\mu\sigma}\phi_\mu\phi_\sigma^\ast,
		\label{app:eq:D-G-expand}
	\end{equation}
	where
	\begin{equation}
		\begin{aligned}
			G_{\mu\sigma}
			=
			&1
			+
			\frac{c_1^2}
			{(\chi_\mu-\beta_1)(\chi_\sigma^\ast-\beta_1)}
			+
			\frac{c_1^2}
			{(\chi_\mu+\beta_1)(\chi_\sigma^\ast+\beta_1)}
			\\
			&
			+
			\frac{c_2^2}
			{(\chi_\mu-\beta_2)(\chi_\sigma^\ast-\beta_2)}
			+
			\frac{c_2^2}
			{(\chi_\mu+\beta_2)(\chi_\sigma^\ast+\beta_2)} ,\\
			&
			\mu,\sigma\in\{a,b\}.
		\end{aligned}
		\label{app:eq:Gmusigma-def}
	\end{equation}
	Similarly,
	\begin{equation}
		\Delta_\Phi=
		\sum_{\mu,\nu\in\{a,b\}}
		E_{\mu\nu}\phi_\mu\phi_\nu,
		\label{app:eq:Delta-E-expand}
	\end{equation}
	with
	\begin{equation*}
		E_{\mu\nu}
		=
		1
		-
		\frac{2c_1^2}
		{(\chi_\mu-\beta_1)(\chi_\nu+\beta_1)}
		-
		\frac{2c_2^2}
		{(\chi_\mu-\beta_2)(\chi_\nu+\beta_2)} .
	\end{equation*}
	
	For the binary Darboux case, the reduction-related spectral data are
	\begin{equation*}
		\lambda_1=\lambda,\qquad
		\lambda_2=-\lambda^\ast .
	\end{equation*}
	The corresponding $\Omega$ matrix is
	\begin{equation*}
		\Omega
		=
		\begin{pmatrix}
			\dfrac{D_\Phi}{\lambda-\lambda^\ast}
			&
			-\dfrac{\Delta_\Phi^\ast}{2\lambda^\ast}
			\\[2ex]
			\dfrac{\Delta_\Phi}{2\lambda}
			&
			\dfrac{D_\Phi}{\lambda-\lambda^\ast}
		\end{pmatrix}.
	\end{equation*}
	For compactness, we introduce
	\begin{equation*}
		q_\lambda=\lambda-\lambda^\ast,\quad
		a_\Omega=\frac{1}{q_\lambda},
		\quad
		b_\Omega=-\frac{1}{2\lambda^\ast},
		\quad
		c_\Omega=\frac{1}{2\lambda}.
	\end{equation*}
	Then
\begin{equation*}
	\det\Omega
	=
	a_\Omega^2D_\Phi^2
	+\frac{|\Delta_\Phi|^2}{4|\lambda|^2}.
\end{equation*}
	
	Using Eqs.~\eqref{app:eq:D-G-expand} and
	\eqref{app:eq:Delta-E-expand}, the denominator of the binary solution can be
	expanded as
	\begin{equation*}
		\det\Omega
		=
		\sum_{\mu,\nu,\sigma,\tau\in\{a,b\}}
		C^{(0)}_{\mu\nu\sigma\tau}
		\phi_\mu\phi_\nu\phi_\sigma^\ast\phi_\tau^\ast ,
	\end{equation*}
	where
	\begin{equation*}
		C^{(0)}_{\mu\nu\sigma\tau}
		=
		a_\Omega^2G_{\mu\sigma}G_{\nu\tau}
		+
		\frac{1}{4|\lambda|^2}
		E_{\mu\nu}E_{\sigma\tau}^\ast .
	\end{equation*}
	The numerator coefficients for the $s$th component are
	\begin{equation*}
		C^{(s)}_{\mu\nu\sigma\tau}
		=
		C^{(0)}_{\mu\nu\sigma\tau}
		+
		F^{(s)}_{\mu\nu\sigma\tau},
		\qquad s=1,2 ,
		\label{app:eq:Cs-C0-F}
	\end{equation*}
	where
\begin{equation*}
	\begin{aligned}
		F^{(s)}_{\mu\nu\sigma\tau}
		={}&
		a_\Omega
		\left[
		\frac{G_{\nu\sigma}}
		{\chi_\tau^\ast-\beta_s}
		-
		\frac{G_{\mu\sigma}}
		{\chi_\nu+\beta_s}
		\right]
		\\
		&+
		b_\Omega
		\frac{E_{\sigma\tau}^\ast}
		{\chi_\nu+\beta_s}
		-
		c_\Omega
		\frac{E_{\mu\nu}}
		{\chi_\tau^\ast-\beta_s},\qquad s=1,2.
	\end{aligned}
\end{equation*}
	Therefore, the binary Darboux fields are equivalently
	\begin{equation*}
		u_s
		=
		u_{s0}
		\frac{
			\displaystyle
			\sum_{\mu,\nu,\sigma,\tau\in\{a,b\}}
			C^{(s)}_{\mu\nu\sigma\tau}
			\phi_\mu\phi_\nu\phi_\sigma^\ast\phi_\tau^\ast
		}{
			\displaystyle
			\sum_{\mu,\nu,\sigma,\tau\in\{a,b\}}
			C^{(0)}_{\mu\nu\sigma\tau}
			\phi_\mu\phi_\nu\phi_\sigma^\ast\phi_\tau^\ast
		},
		\qquad s=1,2 .
	\end{equation*}
	
	We now recover the nine-term forms used in
	Eqs.~\eqref{eq:Ns-nine-term-main}
	and~\eqref{eq:D-nine-term-main}.
	Let
	\begin{equation*}
		\delta=\eta_b-\eta_a=\Psi+i\Theta,
		\qquad
		\zeta=e^\delta=e^{\Psi+i\Theta}.
	\end{equation*}
	Then $\phi_b=\phi_a \zeta$ and $\phi_b^\ast=\phi_a^\ast \zeta^\ast$. After extracting
	the common factor $\phi_a^2(\phi_a^\ast)^2$ and multiplying numerator and
	denominator by $e^{-2\Psi}$, the nine coefficients are
	\begin{equation*}
		\begin{aligned}
			\mathcal K^{(s)}_{00}
			&=C^{(s)}_{aaaa},
			\mathcal K^{(s)}_{01}
			=C^{(s)}_{aaab}+C^{(s)}_{aaba},\\
			\mathcal K^{(s)}_{02}
			&=C^{(s)}_{aabb},
			\mathcal K^{(s)}_{10}
			=C^{(s)}_{abaa}+C^{(s)}_{baaa},\\
			\mathcal K^{(s)}_{11}
			&=C^{(s)}_{abab}+C^{(s)}_{abba}
			+C^{(s)}_{baab}+C^{(s)}_{baba},\\
			\mathcal K^{(s)}_{12}
			&=C^{(s)}_{abbb}+C^{(s)}_{babb},
			\mathcal K^{(s)}_{20}
			=C^{(s)}_{bbaa},\\
			\mathcal K^{(s)}_{21}
			&=C^{(s)}_{bbab}+C^{(s)}_{bbba},
			\mathcal K^{(s)}_{22}=C^{(s)}_{bbbb},\quad s=0,1,2 .
		\end{aligned}
	\end{equation*}
	Substituting these coefficients gives
	\begin{align*}
		\mathcal N_s
		={}&
		\mathcal K^{(s)}_{22}e^{2\Psi}
		+\mathcal K^{(s)}_{21}e^{\Psi+i\Theta}
		+\mathcal K^{(s)}_{12}e^{\Psi-i\Theta}+
		\mathcal K^{(s)}_{20}e^{2i\Theta}
\nonumber\\
&+\mathcal K^{(s)}_{11}
				+\mathcal K^{(s)}_{02}e^{-2i\Theta}+
		\mathcal K^{(s)}_{10}e^{-\Psi+i\Theta}
		+\mathcal K^{(s)}_{01}e^{-\Psi-i\Theta}
		\nonumber\\
		&+\mathcal K^{(s)}_{00}e^{-2\Psi}, \qquad s=1,2 ,
	\end{align*}
	and
	\begin{align*}
		\mathcal D
		={}&
		\mathcal K^{(0)}_{22}e^{2\Psi}
		+\mathcal K^{(0)}_{21}e^{\Psi+i\Theta}
		+\mathcal K^{(0)}_{12}e^{\Psi-i\Theta}+
		\mathcal K^{(0)}_{20}e^{2i\Theta}
		\nonumber\\
		&+\mathcal K^{(0)}_{11}
				+\mathcal K^{(0)}_{02}e^{-2i\Theta}+
		\mathcal K^{(0)}_{10}e^{-\Psi+i\Theta}
		+\mathcal K^{(0)}_{01}e^{-\Psi-i\Theta}
		\nonumber\\
		&+\mathcal K^{(0)}_{00}e^{-2\Psi}.
	\end{align*}
	
	We next give the coefficients in the one-fold Darboux expression used when
	$\lambda+\lambda^\ast=0$. In this case,
	\begin{equation}
		u_s
		=
		u_{s0}
		\frac{
			\displaystyle
			\sum_{\mu,\sigma\in\{a,b\}}
			H^{(s)}_{\mu\sigma}
			\phi_\mu\phi_\sigma^\ast
		}{
			\displaystyle
			\sum_{\mu,\sigma\in\{a,b\}}
			H^{(0)}_{\mu\sigma}
			\phi_\mu\phi_\sigma^\ast
		},
		\qquad s=1,2 ,
		\label{app:eq:single-H-form}
	\end{equation}
	where
$
		H^{(0)}_{\mu\sigma}=G_{\mu\sigma},
$
	and
	\begin{equation*}
		H^{(s)}_{\mu\sigma}
		=
		G_{\mu\sigma}
		+
		(\lambda-\lambda^\ast)
		\frac{1}{\chi_\sigma^\ast-\beta_s},
		 \qquad s=1,2 .
	\end{equation*}
	Using again $\phi_b=\phi_a \zeta$ and $\phi_b^\ast=\phi_a^\ast \zeta^\ast$, after
	extracting $\phi_a\phi_a^\ast$ and multiplying numerator and denominator by
	$e^{-\Psi}$, Eq.~\eqref{app:eq:single-H-form} becomes the four-term form
	\begin{equation*}
		u_s
		=
		u_{s0}
		\frac{
			h^{(s)}_{11}e^{-\Psi}
			+h^{(s)}_{12}e^{\Psi}
			+h^{(s)}_{13}e^{i\Theta}
			+h^{(s)}_{14}e^{-i\Theta}
		}{
			h^{(0)}_{11}e^{-\Psi}
			+h^{(0)}_{12}e^{\Psi}
			+h^{(0)}_{13}e^{i\Theta}
			+h^{(0)}_{14}e^{-i\Theta}
		},
		 s=1,2 .
	\end{equation*}
	The four coefficients are
	\begin{equation*}
		\begin{aligned}
			h^{(s)}_{11}&=H^{(s)}_{aa},\quad
			h^{(s)}_{12}=H^{(s)}_{bb},
			\\
			h^{(s)}_{13}&=H^{(s)}_{ba},\quad
			h^{(s)}_{14}=H^{(s)}_{ab},\quad
			s=0,1,2 .
			\end{aligned}
	\end{equation*}
	Explicitly, the denominator coefficients are
	\begin{equation*}
			h^{(0)}_{11}=G_{aa},\quad
			h^{(0)}_{12}=G_{bb},\quad
			h^{(0)}_{13}=G_{ba},\quad
			h^{(0)}_{14}=G_{ab},
	\end{equation*}
	and, for $s=1,2$,
	\begin{equation*}
		\begin{aligned}
			h^{(s)}_{11}
			&=
			G_{aa}
			+
			\frac{\lambda-\lambda^\ast}{\chi_a^\ast-\beta_s},\quad
			h^{(s)}_{12}=
			G_{bb}
			+
			\frac{\lambda-\lambda^\ast}{\chi_b^\ast-\beta_s},
			\\
			h^{(s)}_{13}
			&=
			G_{ba}
			+
			\frac{\lambda-\lambda^\ast}{\chi_a^\ast-\beta_s},\quad
			h^{(s)}_{14}
			=
			G_{ab}
			+
			\frac{\lambda-\lambda^\ast}{\chi_b^\ast-\beta_s}.
		\end{aligned}
	\end{equation*}
	\section{Linear stability analysis}
	\label{app:mi-analysis}
	
	In this appendix, we study the MI of the plane-wave backgrounds considered in Sec.~\ref{sec:existence-mi}. The linear stability analysis
	is carried out directly for the system~\eqref{eq:reduced-system} and applies to all three
	carrier-wavenumber configurations considered in this work.
	We further relate the zero-MI-gain condition to the state-transition one.
	
	The vector plane-wave solutions of system~\eqref{eq:reduced-system} are given by Eq.~\eqref{eq:seed-solution}.
	 We introduce small perturbations $p_j(x,t)$ through
	\begin{equation}
		u_j(x,t)
		=
		\left[c_j+p_j(x,t)\right]e^{i\theta_j},\quad
		|p_j|\ll c_j,
		j=1,2.
		\label{eq:mi-perturbation}
	\end{equation}
	Substituting the perturbation ansatz in
	Eq.~\eqref{eq:mi-perturbation} into the
	system~\eqref{eq:reduced-system} and retaining only terms linear
	in the perturbations, we obtain
	\begin{equation}
		\begin{aligned}
			p_{j,t}
			&+\epsilon\big[
			p_{j,xxx}
			+3i\beta_j p_{j,xx}
			+3(2\rho-\beta_j^2)p_{j,x}
			\\
			&+3c_j\left(\partial_x+2i\beta_j\right)\delta S
			\big]
			=0,
			\qquad j=1,2,
		\end{aligned}
		\label{eq:mi-linearized}
	\end{equation}
	where
	\begin{equation*}
		\rho=c_1^2+c_2^2,
		\qquad
		\delta S
		=
		\sum_{k=1}^{2}c_k\left(p_k+p_k^*\right).
	\end{equation*}
	
	We take Fourier-mode perturbations of the form
	\begin{equation}
		p_j
		=
		f_j e^{i(\gamma x-\omega t)}
		+g_j^*e^{-i(\gamma x-\omega^*t)},
		\qquad
		j=1,2,
		\label{eq:mi-ansatz}
	\end{equation}
	where $\gamma\in\mathbb{R}$ is the modulation wavenumber,
	$\omega$ is the generally complex modulation frequency, and
	$f_j$ and $g_j$ are constant perturbation amplitudes.
	
	Substitution of Eq.~\eqref{eq:mi-ansatz} into the linearized
	system~\eqref{eq:mi-linearized} gives, for $j=1,2$,
	\begin{align}
		\omega f_j
		&=
		d_j^+f_j
		+3\epsilon c_j
		\left(\gamma+2\beta_j\right)\mathcal{P},
		\label{eq:mi-fj}
		\\
		\omega g_j
		&=
		d_j^-g_j
		+3\epsilon c_j
		\left(\gamma-2\beta_j\right)\mathcal{P},
		\label{eq:mi-gj}
	\end{align}
	where
	\begin{align}
		d_j^+
		&=
		\epsilon\gamma
		\left(
		6\rho-3\beta_j^2
		-3\beta_j\gamma-\gamma^2
		\right),
		\label{eq:mi-d+}
		\\
		d_j^-
		&=
		\epsilon\gamma
		\left(
		6\rho-3\beta_j^2
		+3\beta_j\gamma-\gamma^2
		\right),
		\label{eq:mi-d-}
	\end{align}
	and
	\begin{equation*}
		\mathcal{P}
		=
		c_1(f_1+g_1)+c_2(f_2+g_2).
	\end{equation*}
	
	Introducing the perturbation vector
	\begin{equation*}
		\mathbf{Z}
		=
		(f_1,g_1,f_2,g_2)^{\mathrm T},
	\end{equation*}
	Eqs.~\eqref{eq:mi-fj} and \eqref{eq:mi-gj} can be written as
	the eigenvalue problem
	\begin{equation*}
		\omega\mathbf{Z}
		=
		\mathcal{M}_{\mathrm{MI}}\mathbf{Z}.
	\end{equation*}
	The linearized matrix can be expressed compactly as
	\begin{equation}
		\mathcal{M}_{\mathrm{MI}}
		=
		\boldsymbol{\Lambda}_{\mathrm{MI}}
		+3\epsilon\,\mathbf{r}\mathbf{s}^{{\mathrm T}},
		\label{eq:mi-matrix-compact}
	\end{equation}
	where
	\begin{align*}
		\boldsymbol{\Lambda}_{\mathrm{MI}}
		&=
		\operatorname{diag}
		\left(
		d_1^+,d_1^-,d_2^+,d_2^-
		\right),
		\\
		\mathbf{r}
		&=
		\left(
		c_1(\gamma+2\beta_1),
		c_1(\gamma-2\beta_1),
		c_2(\gamma+2\beta_2),
		c_2(\gamma-2\beta_2)
		\right)^{\mathrm T},
		\\
		\mathbf{s}
		&=
		(c_1,c_1,c_2,c_2)^T.
	\end{align*}
	
	Equivalently, the explicit form of the linearized matrix is
	\begin{equation*}
		\mathcal{M}_{\mathrm{MI}}
		=
		\begin{pmatrix}
			d_1^++\Gamma_1^+c_1
			&
			\Gamma_1^+c_1
			&
			\Gamma_1^+c_2
			&
			\Gamma_1^+c_2
			\\
			\Gamma_1^-c_1
			&
			d_1^-+\Gamma_1^-c_1
			&
			\Gamma_1^-c_2
			&
			\Gamma_1^-c_2
			\\
			\Gamma_2^+c_1
			&
			\Gamma_2^+c_1
			&
			d_2^++\Gamma_2^+c_2
			&
			\Gamma_2^+c_2
			\\
			\Gamma_2^-c_1
			&
			\Gamma_2^-c_1
			&
			\Gamma_2^-c_2
			&
			d_2^-+\Gamma_2^-c_2
		\end{pmatrix},
	\end{equation*}
	where
	\begin{equation*}
		\Gamma_j^\pm
		=
		3\epsilon c_j
		\left(\gamma\pm2\beta_j\right),
		\qquad
		j=1,2.
	\end{equation*}
	
	The linear dispersion relation is determined by
	\begin{equation*}
		\det\left(
		\omega I-\mathcal{M}_{\mathrm{MI}}
		\right)
		=0.
	\end{equation*}
	Exploiting the rank-one structure in
	Eq.~\eqref{eq:mi-matrix-compact}, the determinant can be written
	explicitly as
	\begin{equation}
		\begin{aligned}
			&
			\prod_{j=1}^{2}
			(\omega-d_j^+)(\omega-d_j^-)
			-3\epsilon
			\sum_{j=1}^{2}c_j^2
			\Bigg\{
			\Big[
			(\gamma+2\beta_j)(\omega-d_j^-)\\
			&+
			(\gamma-2\beta_j)(\omega-d_j^+)
			\Big]
			\prod_{\substack{k=1\\k\neq j}}^{2}
			(\omega-d_k^+)(\omega-d_k^-)
			\Bigg\}
			=0.
		\end{aligned}
		\label{eq:mi-dispersion-explicit}
	\end{equation}
	Equation~\eqref{eq:mi-dispersion-explicit} is a quartic
	equation for the modulation frequency $\omega$. The plane-wave
	background is modulationally unstable when at least one root
	has a nonzero imaginary part.
	
We now relate the MI frequency $\omega$ to the characteristic roots used in the Darboux construction. For the vector ABs considered here, $\alpha=0$, and hence $K=\gamma$. We further assume a nonzero modulation wavenumber, $\gamma\neq0$, and introduce the auxiliary variable $\widehat R$ through
\begin{equation}
	\omega
	=
	\epsilon\gamma
	\left(
	6\rho-\gamma^2-3\widehat R
	\right),
	\label{eq:mi-omega-Rhat}
\end{equation}
Substituting this expression into the dispersion relation~\eqref{eq:mi-dispersion-explicit}, which is polynomial in $\omega$, relates it to the characteristic equations obtained for the three carrier-wavenumber settings: Eq.~\eqref{eq:char-beta1-zero} for $\beta_1=0$, Eq.~\eqref{eq:char-equal-square} for $\beta_1=-\beta_2$, and Eq.~\eqref{eq:char-compact} for two unequal nonzero wavenumbers. These three cases are discussed below.
	
	When $\beta_1=0$,
	Eqs.~\eqref{eq:mi-d+} and \eqref{eq:mi-d-} reduce to
	\begin{equation*}
		d_1^+=d_1^-
		=
		\epsilon\gamma
		\left(
		6\rho-\gamma^2
		\right)
		\equiv d_1.
	\end{equation*}
	Accordingly, the linear dispersion relation factorizes as
\begin{equation}
	\det\left(
	\omega I-\mathcal{M}_{\mathrm{MI}}
	\right)
	=
	(\omega-d_1)\mathcal{F}_{\mathrm I}(\omega)
	=0,
	\label{eq:mi-factor-beta1zero}
\end{equation}
	where
	\begin{equation}
		\begin{aligned}
			\mathcal{F}_{\mathrm I}(\omega)
			={}&
			(\omega-d_1)
			(\omega-d_2^+)
			(\omega-d_2^-)
			\\
			&-
			6\epsilon\gamma c_1^2
			(\omega-d_2^+)
			(\omega-d_2^-)
			\\
			&-
			3\epsilon c_2^2
			(\gamma+2\beta_2)
			(\omega-d_1)(\omega-d_2^-)
			\\
			&-
			3\epsilon c_2^2
			(\gamma-2\beta_2)
			(\omega-d_1)(\omega-d_2^+).
		\end{aligned}
		\label{eq:mi-cubic-beta1zero}
	\end{equation}
	The isolated root $\omega=d_1$ is real and therefore carries
	no MI gain. The remaining three frequency branches are
	determined by
	\begin{equation*}
		\mathcal{F}_{\mathrm I}(\omega)=0.
	\end{equation*}
	
	Substituting Eq.~\eqref{eq:mi-omega-Rhat} into
	Eq.~\eqref{eq:mi-cubic-beta1zero} gives
	\begin{equation*}
			\mathcal{F}_{\mathrm I}\bigl(\omega(\widehat R)\bigr)
			=
			-27\epsilon^3\gamma^3
			\Big[
			\widehat R^3
			+a_2\widehat R^2
			+a_1\widehat R
			+a_0
			\Big],
	\end{equation*}
	where $a_2$, $a_1$, and $a_0$ are exactly those defined in
	Eq.~\eqref{eq:R-cubic-coefficients} with $K=\gamma$.
	Since $\epsilon\gamma\neq0$, the condition
	$\mathcal{F}_{\mathrm I}(\omega)=0$ is therefore equivalent to
	\begin{equation}
		\widehat R^3
		+a_2\widehat R^2
		+a_1\widehat R
		+a_0
		=0.
		\label{eq:mi-cubic-equivalence}
	\end{equation}
The polynomial equation~\eqref{eq:mi-cubic-equivalence}
is identical to the characteristic-root equation
\eqref{eq:R-cubic-beta1zero} with $K=\gamma$.
Therefore, $\widehat R$ can be identified with the spectral
variable $R$, giving
\begin{equation*}
	\widehat R
	=
	R
	=
	\chi_a\chi_b.
\end{equation*}
Since $\chi_b=\chi_a+\gamma$,
Eq.~\eqref{eq:mi-omega-Rhat} immediately yields
\begin{equation}
	\omega(\chi_a;\gamma)
	=
	-\epsilon\gamma
	\left(
	3\chi_a^2
	+3\gamma\chi_a
	+\gamma^2
	-6\rho
	\right).
\label{eq:mi-omega-chi}
\end{equation}
	
	When $\beta_1=-\beta_2=\beta$, Eqs.~\eqref{eq:mi-d+} and \eqref{eq:mi-d-} take the form
	\begin{align*}
		d_1^+=d_2^-
		&=
		\epsilon\gamma
		\left(
		6\rho-3\beta^2-3\beta\gamma-\gamma^2
		\right)
		\equiv d^+,
		\\
		d_1^-=d_2^+
		&=
		\epsilon\gamma
		\left(
		6\rho-3\beta^2+3\beta\gamma-\gamma^2
		\right)
		\equiv d^-.
	\end{align*}
	Accordingly, the dispersion relation can be written as
	\begin{equation}
		\det\left(
		\omega I-\mathcal{M}_{\mathrm{MI}}
		\right)
		=
		(\omega-d^+)(\omega-d^-)
		\mathcal{F}_{\mathrm {II}}(\omega)
		=0,
		\label{eq:mi-factor-opposite}
	\end{equation}
	with
	\begin{equation}
		\begin{aligned}
			\mathcal{F}_{\mathrm {II}}(\omega)
			={}&
			(\omega-d^+)(\omega-d^-)
			\\
			&-
			3\epsilon\rho
			(\gamma+2\beta)(\omega-d^-)
			\\
			&-
			3\epsilon\rho
			(\gamma-2\beta)(\omega-d^+).
		\end{aligned}
		\label{eq:mi-quadratic-opposite}
	\end{equation}
	The two explicit factors give the real frequencies
	$\omega=d^+$ and $\omega=d^-$, neither of which produces MI gain.
	The nontrivial MI branches are therefore specified by
	$\mathcal{F}_{\mathrm {II}}(\omega)=0$.
	
	Inserting Eq.~\eqref{eq:mi-omega-Rhat} into
	Eq.~\eqref{eq:mi-quadratic-opposite} gives
	\begin{equation}
			\mathcal{F}_{\mathrm {II}}\bigl(\omega(\widehat R)\bigr)
			=
			9\epsilon^2\gamma^2
			\Big[
			\widehat R^2
			+2(\rho-\beta^2)\widehat R
			+
			\beta^2
			\left(
			\beta^2-\gamma^2+2\rho
			\right)
			\Big].
		\label{eq:mi-map-opposite}
	\end{equation}
	Because $\epsilon\gamma\neq0$, the condition
	$\mathcal{F}_{\mathrm {II}}(\omega)=0$ becomes
	\begin{equation}
		\widehat R^2
		+2(\rho-\beta^2)\widehat R
		+\beta^2
		\left(
		\beta^2-\gamma^2+2\rho
		\right)
		=0.
		\label{eq:mi-quadratic-equivalence}
	\end{equation}
	Equation~\eqref{eq:mi-quadratic-equivalence} coincides with
	the characteristic-root equation~\eqref{eq:R-quadratic-betaop}
	for $K=\gamma$. We may therefore set
	\begin{equation*}
		\widehat R=R=\chi_a\chi_b,
	\end{equation*}
	which, together with $\chi_b=\chi_a+\gamma$, recovers
	Eq.~\eqref{eq:mi-omega-chi}.

For $\beta_1\beta_2\neq0$ and $\beta_1^2\neq\beta_2^2$,
the MI dispersion relation remains quartic in general. We write the corresponding polynomial in Eq.~\eqref{eq:mi-dispersion-explicit} as
\begin{equation*}
	\mathcal{F}_{\mathrm {III}}(\omega)
	=
	\det\left(
	\omega I-\mathcal{M}_{\mathrm{MI}}
	\right).
\end{equation*}
Using Eq.~\eqref{eq:mi-omega-Rhat}, this expression becomes
\begin{equation}
	\mathcal{F}_{\mathrm {III}}\bigl(\omega(\widehat R)\bigr)
	=
	81\epsilon^4\gamma^4
	\left(
	\widehat R^4
	+b_3\widehat R^3
	+b_2\widehat R^2
	+b_1\widehat R
	+b_0
	\right),
	\label{eq:mi-map-general}
\end{equation}
where $b_3$, $b_2$, $b_1$, and $b_0$ are given in
Eq.~\eqref{eq:R-quartic-general} for $K=\gamma$.
Because $\epsilon\gamma\neq0$, the condition
$\mathcal{F}_{\mathrm {III}}(\omega)=0$ reduces to
\begin{equation}
	\widehat R^4
	+b_3\widehat R^3
	+b_2\widehat R^2
	+b_1\widehat R
	+b_0
	=0.
	\label{eq:mi-quartic-equivalence}
\end{equation}
This is precisely the characteristic-root equation
\eqref{eq:R-quartic-general} evaluated at $K=\gamma$.
Thus,
$
	\widehat R=R=\chi_a\chi_b,
$
and, together with $\chi_b=\chi_a+\gamma$, the relation
\eqref{eq:mi-omega-chi} follows.

	Accordingly, each admissible characteristic-root pair
	corresponds to a linear MI branch, while the additional
	real-frequency branches in the first two configurations
	have zero MI gain and are unrelated to the characteristic-root equations.
	\raggedbottom
	Writing $\chi_a=\chi_a^r+i\chi_a^i$, the imaginary part
	of Eq.~\eqref{eq:mi-omega-chi} is
	\begin{equation*}
		\operatorname{Im}\omega
		=
		-3\epsilon\gamma\chi_a^i
		\left(
		2\chi_a^r+\gamma
		\right).
	\end{equation*}
	For all three carrier-wavenumber configurations, the MI gain of an admissible characteristic-root branch is
	\begin{equation}
		G(\chi_a;\gamma)
		=
		3
		\left|
		\epsilon\gamma\chi_a^i
		\left(
		2\chi_a^r+\gamma
		\right)
		\right|.
		\label{eq:mi-gain-chi}
	\end{equation}
	The branch-resolved MI-gain diagrams in Sec.~\ref{sec:existence-mi} are constructed using Eq.~\eqref{eq:mi-gain-chi}.
	The frequency--root relation obtained above also appears
	directly in the exponential phases of the Darboux solution.
	Indeed, using $\chi_b-\chi_a=\gamma$, one finds
	\begin{equation*}
		\begin{aligned}
			\eta_b-\eta_a
			={}&
			i(\chi_b-\chi_a)x+
			i\epsilon
			\left[
			\chi_b^3-\chi_a^3
			-6\rho(\chi_b-\chi_a)
			\right]t
			\\
			={}&
			i\gamma x
			+i\epsilon\gamma
			\left(
			3\chi_a^2
			+3\gamma\chi_a
			+\gamma^2
			-6\rho
			\right)t
			\\
			={}&
			i(\gamma x-\omega t),
		\end{aligned}
	\end{equation*}
	where $\omega$ is precisely the linear MI frequency given by
	Eq.~\eqref{eq:mi-omega-chi}.
	
	Writing $\eta_b-\eta_a=\Psi+i\Theta$, the localization
	variable is therefore
	\begin{equation*}
		\Psi
		=
		\operatorname{Im}(\omega)t
		=
		-3\epsilon\gamma\chi_a^i
		\left(
		2\chi_a^r+\gamma
		\right)t.
	\end{equation*}
	For a nonzero modulation wavenumber and an admissible complex
	branch, $\gamma\neq0$ and $\chi_a^i\neq0$. Hence, for all three carrier-wavenumber configurations,
	the zero-gain condition is given by
	\begin{equation}
		\chi_a^r
		=
		-\frac{\gamma}{2}.
		\label{eq:mi-zero-gain}
	\end{equation}
	Thus, the zero-gain condition of the linear MI spectrum
	coincides with the state-transition condition of the Darboux
	solutions in each of the three carrier-wavenumber
	configurations. When Eq.~\eqref{eq:mi-zero-gain} holds,
	the localization variable $\Psi$ vanishes identically, and the
	corresponding AB branch reduces to a periodic-wave state.
	
	\bibliography{refs_1}

	\end{document}